\documentstyle[12pt]{article}
\topmargin - 0.8cm

\catcode`@=11
\@addtoreset{equation}{section}
\catcode`@=12
\newcommand{\fatg}[1]{\mbox{\boldmath $#1$}}
\begin{document}
\title{\vskip-1.7cm \bf Tunnelling geometries I. Analyticity,
unitarity and
instantons in quantum cosmology}
\author{A.O.Barvinsky$^{1,2}$\ and A.Yu.Kamenshchik$^{2}$ }
\date{}
\maketitle
\hspace{-8mm}$^{1}${\em
Theoretical Physics Institute, Department of Physics, \ University of
Alberta, Edmonton, Canada T6G 2J1}
$^{*}$
\\ $^{2}${\em Nuclear Safety
Institute, Russian Academy of Sciences , Bolshaya Tulskaya 52, Moscow
113191, Russia}
\begin{abstract}
We present a theory of tunnelling geometries originating from the
no-boundary
quantum state of Hartle and Hawking, which describes in the language
of
analytic continuation the nucleation of the Lorentzian Universe from
the
Euclidean spacetime. We reformulate the no-boundary wavefunction in
the
manifestly unitary representation of true physical variables and
calculate it
in the one-loop approximation. For this purpose a special technique
of complex
extremals is developed, which reduces the formalism of complex
tunnelling
geometries to the real ones, and also the method of collective
variables is
applied, separating the macroscopic collective degrees of freedom
from the
microscopic modes that are treated perturbatively. The quantum
distribution
function of Lorentzian universes, defined on the space of such
collective
variables, incorporates the probabilty conservation and boils down to
the
partition function of quasi-DeSitter gravitational instantons
weighted by their
Euclidean effective action. These instantons represent closed compact
manifolds
obtained by the procedure of doubling the Euclidean spacetime which
nucleates
the Lorentzian universes. The over-Planckian behaviour of their
distribution is
determined by the anomalous scaling of the theory on such instantons,
which
serves as a criterion for the high-energy normalizability of the
no-boudary
wavefunction and the validity of the semiclassical expansion. Thus
the
presented formalism combines the covariance of the Euclidean
effective action
with unitarity and analyticity in the Lorentzian spacetime and can be
regarded
as a step towards the unification of Euclidean and Lorentzian
versions of
quantum gravity and its third quantization.
\end{abstract}
\\
$^{*}$ Present address\\

\baselineskip6.8mm
\section{Introduction}
\hspace{\parindent}
The history of quantum cosmology reveals in the last decade the whole
cycle of
the developement characteristic of the theories standing at the
advanced
frontiers of physics and lacking a good experimental status. Beeing
founded in
the pioneering works of Dirac, Wheeler, DeWitt \cite{DeW} and the
others in
late fifties and early sixties, this theory underwent a long period
of a lull
in order to recover in early eighties due to practically simultaneous
invention
of the inflationary cosmology and two, conceptually similar,
proposals  for its
quantum state -- no-boundary proposal of Hartle and Hawking
\cite{HH,H} and the
tunnelling proposal of Vilenkin \cite{Vilenkin}. A productive
application of
these proposals in the theory of the inflationary Universe was
followed by a
great uprise of interest in the Euclidean quantum gravity -- the
corner stone
of the no-boundary wavefunction -- in connection with the ideas of
the topology
change, baby universe production and the theory of the cosmological
constant of
Coleman \cite{Gid-S,Coleman,Hawk-w}, which could be regarded as a
first visible
step towards the third quantization of gravity. However, an extensive
accumulation of applications in this field (see the bibliography in
\cite{Halliwell:bibliography}), very useful from the viewpoint of the
theory of
the early Universe -- the subject acquiring now a well established
experimental
status, could not resolve the principal difficulties of this theory.
As a
result, it has again entered a slugish stage, to an essential extent,
caused by
the fact that the fundamental ideas of gravitational tunnelling
phenomena
remained at the heuristic level and, actually, were never pushed
beyond the
tree-level approximation.

In short, the difficulties of quantum gravity, arising in the above
context
\footnote
{We don't discuss here the other fundamental difficulty of quantum
gravity
theory -- its perturbative nonrenormalizability, which might be
resolved in the
framework of the fundamental string theory manifesting itself
basically at the
over-Planckian energy scale. We focuse at the problems equally
inherent to
gravity theory both in the ultraviolet and infrared limits, arising,
for
example, in the theory of the cosmological constant of Coleman.
},
can be formulated as a controversy between covariance and unitarity.
In search
for completely covariant formulation of quantum gravity, Hartle and
Hawking
postulated the path integral over Euclidean (positive-signature)
four-geometries and matter fields, which under certain boundary
conditions
semiclassically generates the usual Lorentzian spacetimes and
propagating
quantum fields. It is obvious, that this construction does not
guarantee the
unitarity of the theory, the more so unitarity implies a well-defined
notion of
time and Lorentzian metric signature missing in such a Euclidean
formalism.
This feature of the covariant Euclidean quantum gravity makes it
formulated at
the half-heuristic level lacking the fundametal principles which
close it as a
self-contained physical theory: the selection of physical quantum
states beeing
incomplete, their physical inner product not defined, etc.

As a counterpart to this covariant approach there exists the
quantization of
the theory reduced to physical (ADM) variables, featuring
particularly selected
time, manifest unitarity and a Hilbert space of quantum states
\cite{ADM,Kuchar:bubble,BPon,B:GenSem,BarvU}. It is generally
believed that
this reduced phase-space quantization is not equivalent to the
original
Dirac-Wheeler-DeWitt scheme (or its Euclidean extension of Hartle and
Hawking)
-- the statement usually relying on the obvious inequivalence of the
ADM
quantizations in different gauges with some {\it ad hoc} operator
realizations
of the physical Hamiltonian and observables. Ultimately, such
statements might
be true, because ADM quantization turns out to be intrinsically
inconsistent in
view of the problem of Gribov copies \cite{BarvU,BKr,Operd}. This
problem
underlies a purely mathematical motivation for the secondary
quantization of a
relativistic particle, which leads to a more fundamental concept of
the quantum
field theory, and apparently compells physicists for the third
quantization of
gravity \cite{Kuchar:third,Gid-S,Coleman,Hawk-w}. However, at the
present stage
of quantum gravity theory such statements do not give much insight
into the
nature of its Euclidean version and do not help to close it as a
physical
theory.

On the contrary, the extension of the ADM quantization to the
Euclidean regime
can yield the missing principles of constructing and interpreting the
Euclidean
quantum gravity and, also, clearly indicates the points where this
method fails
and what kind of new physical phenomena follow from this failure.
This approach
is very promising in the framework of the semiclassical expansion,
for
perturbatively there exist very strong theorems on the equivalence of
covariant
and reduced phase space quantizations \cite{Faddeev,Fr-V,BF:Ann},
confirmed at
the operatorial level in the one-loop approximation \cite{BKr}.
Despite its
limitations, the asymptotic $\hbar$-expansion is universally
applicable to
theories of general type and, apparently, gives a key also to the
phenomena
which can be completely described only at the non-perturbative level.
As a good
analogy of this approach, one can list a method of many-trajectory
representation in quantum field theory \cite{traject}, describing the
former in
terms of sets of first quantized relativistic particles, the first
quantized
string theory as an approximation to the full string field theory,
etc.

With this paper we begin the series of publications devoted to a
partial
implementation of this strategy. Our starting point, partly
formulated in
\cite{BarvU,BKK}, will be a manifestly unitary gravity theory in
Lorentzian
spacetime, quantized in terms of physical variables. This theory
produces by
certain analytic continuation the Euclidean quantum gravity -- the
auxiliary
mathematical tool which serves us to describe the classically
forbidden states
of the gravitational field \cite{Sakharov}, just as the concept of
the
imaginary time \cite{Banks-Bender-Wu}, Euclidean spacetime and
instantons is
usually used for the description of the decay and tunnelling
phenomena in
quantum mechanical and field models. Semiclassically, the necessity
of such an
analytic continuation originates from the fact that families of
classical
solutions may have caustics in the configuration space of the theory.
In order
to extend these solutions beyond caustics, into the "shadow" domains,
one has
to continue them analytically into the complex plane of time -- the
fact which
follows (at least semiclassically) from the two-fold analyticity of
the path
integral. The gauge properties and the nature of the caustic surfaces
for the
Wheeler-DeWitt equations will be considered in the third paper of
this series
\cite{tunnelIII}
\footnote
{This paper demonstrates, in particular, that an important class of
such
caustics exactly corresponds to the occurrence of Gribov copies in
the
quantization of gravity as a gauge theory with local diffeomorphism
invariance.
In connection with the discussion of \cite{BKr} this serves as a
compelling
argument that the quantum penetration through these caustics belongs
to the
phenomena of the universe creation in the third quantization of
gravity theory,
described in the language of the secondary quantization.
},
while in the first two papers (this one and \cite{tunnelII}) we
develope
elements of the general theory and the one-loop approximation for
quantum
systems penetrating beyond these caustics, which we shall call the
tunnelling
geometries.

A very simple picture of tunnelling geometry, demonstrating the
purposes of
this paper, is presented on Fig.1. In Einstein's gravity theory with
the
positive cosmological constant $\Lambda=3H^2$, there exists a well
known
DeSitter solution with the Lorentzian (indefinite-signature) metric
	\begin{eqnarray}
	&&ds^2_{L}=-dt^2+
	a^2_{L}(t)\,c_{ab}\,dx^{a}dx^{b},         \label{eqn:1.1}\\
	&&a_{L}(t)=
	\frac{1}{H}\,{\rm cosh}\,(Ht)            \label{eqn:1.2}
	\end{eqnarray}
describing the expansion of a spatially homogeneous spherical
hypersurface,
having a round three-dimensional metric $a^2_{\!L}(t)\,c_{ab}$ with
the scale
factor $a_{L}(t)$. As a counterpart to this construction there exists
the
solution of Euclidean Einstein equations with the positive-signature
DeSitter
metric
	\begin{eqnarray}
	&&ds^2=g^{D\!S}_{\mu\nu}dx^{\mu}dx^{\nu}=
	d\tau^2+a^2(\tau)\,c_{ab}\,dx^{a}dx^{b},
\label{eqn:1.3}\\
	&&a\,(\tau)=
	\frac{1}{H}\,{\rm sin}\,(H\tau),
\label{eqn:1.4}
	\end{eqnarray}
which corresponds to the geometry of the four-dimensional sphere of
radius
$R=1/H$ with spherical three-dimensional sections labelled by the
latitude
angle $\theta=H\tau$. The both metrics are related by the analytic
continuation
into the complex plane of the Euclidean "time" $\tau$
\cite{Mottola,Laflamme}
	\begin{eqnarray}
	\tau=\frac{\pi}{2H}+it,\;\;\;
	a_{L}(t)=a\,(\pi/2H+it),
\label{eqn:1.5}
	\end{eqnarray}
which is a Wick rotation with respect to the point $\tau=\pi/2H$ in
this plane.
This analytic continuation can be interpreted as a quantum nucleation
of the
Lorentzian DeSitter spacetime from the Euclidean hemisphere and shown
on Fig.1
as a matching of the two manifolds (\ref{eqn:1.1}) - (\ref{eqn:1.4})
across the
equatorial section $\tau=\pi/2H\;(t=0)$ -- the bounce surface
$\Sigma_{B}$ of
zero extrinsic curvature.

This quantum tunnelling from the classically forbidden Euclidean
domain with
$a\leq 1/H$ served as a heuristic basis for the Hartle-Hawking
no-boundary and
Vilenkin tunnelling proposals, which recently constituted the whole
epoch in
the developement of quantum cosmology. The two major difficulties
with this
simple picture are the following. First, this idea was never pushed
beyond the
tree-level approximation. In the best case, the quantum fields were
considered
on the classical background by using the well-known fact that the
Wheeler-DeWitt equations generate the Schrodinger
\cite{DeW,Rubakov-L,Banks}
(or the heat \cite{Rubakov}) equation for quantized matter, when the
collective
variables of the tunnelling background are completely classical.
Therefore, in
this approach no quantum properties of this background can be
analyzed and no
question of unitarity in the sector of these variables can be posed.

The second difficulty is that, even in the tree-level approximation,
the above
simple picture is applicable only to a limited class of problems
called {\it
real\,} tunnelling geometries \cite{Gibbons-Hartle,Gibbons-Pohle}. In
these
problems the metric and matter fields are completely real both on the
Lorentzian and Euclidean parts of spacetime and smoothly match across
the
nucleation surface which must necessarily have a vanishing extrinsic
curvature
and vanishing normal derivative of matter fields. This requirement of
reality
is very strong and in many physically interesting situations cannot
be
satisfied. The most important example is the Hawking model of chaotic
inflation
\cite{H} driven by the effective Hubble constant $H\,(\varphi)$ which
is
generated by the inflaton scalar field $\varphi$. In this model the
geometry of
the spacetime is approximately  described by the above equations
(\ref{eqn:1.1}) - (\ref{eqn:1.4}) with $H=H(\varphi)$. As it follows
from
equations of motion, for large $H(\varphi)$ the scalar field is
approximately
conserved in time (which justifies the inflationary ansatz
(\ref{eqn:1.2}) with
constant $H$), but its derivative never exactly vanishes for
solutions
satisfying appropriate boundary conditions. Therefore, no nucleation
surface
exists at which exactly {\it real\,} Euclidean solution can be
smoothly matched
to the {\it real\,} Lorentzian one \cite{Lyons}.

Obviously, this is a general case of complex tunnelling in contrast
to a rather
narrow class of real tunnelling geometries. This situation is
characteristic
not only of quantum gravity, but also applicable to the theory of
instantons in
non-linear gauge theories. The curious fact is that not much
attention has been
paid in literature to the underbarrier penetration phenomena,
described by
complex-valued instantons, and filling in this omission in the body
of quantum
theory might lead to interesting physical applications extending
beyond quantum
gravity and cosmology
\footnote
{One of the authors (A.O.B.) enjoied discussion of this point with
S.Coleman.
}.

Thus, in this paper we shall undertake several unassuming steps
towards the
resolution of the above difficulties and omissions in the theory of
tunnelling
geometries. To begin with, we present in Sect.2 the Euclidean quantum
gravity
as an analytic continuation from its Lorentzian counterpart and
reformulate in
this language the no-boundary proposal of Hartle and Hawking. In
Sect.3 we
discuss this analytic continuation for the wavefunction $\Psi(q,t)$
in the
representation of true physical variables $q$. As it was discussed
above, this
representation is intrinsically non-covariant, but this obvious
disadvatage of
the ADM wavefunction is justified by the transparency of its
interpretation
based on a simple inner product
          \begin{equation}
	  <\!\Psi_{1}\,|\,\Psi_{2}\!>=\int dq\,
	  \Psi_{1}^{*}\,(q,t)\,\Psi_{2}(q,t).
\label{eqn:1.6}
	  \end{equation}
The one-loop approximation for $\Psi(q,t)$ also looks simple in this
representation for it involves the functional determinant of the wave
operator
only on the subspace of physical modes. We use a reduction method of
the second
paper of this series \cite{tunnelII} to convert this determinant to a
special
form which, together with (1.6), will be used for the proof of the
conservation
of the total probability -- the corner stone of unitarity.

Sect.4 deals with the separation of the physical configuration space
into
collective macroscopic variables, describing the tunnelling spacetime
background, and the microscopic modes treated perturbatively on this
background
-- the only constructive way of handling the realistic models with
the infinite
amount of the degrees of freedom.

Sect.5 contains the method generalizing the semiclassical expansion
to the case
of complex solutions of dynamical equations. Under good convexity
properties of
the Euclidean action this method boils down to the perturbation
expansion in
the imaginary part of the extremal, which can be used asymptotically
in
$\hbar\rightarrow 0$ even for large imaginary corrections. This
method, in
particular, explains the secondary role of complex instantons in
quantum
theory, which turn out to be exponentially suppressed in comparison
with the
real ones. The application of this method to the Hartle-Hawking
wavefunction in
Sect.6 shows that the complex nature of the underlying tunnelling
geometry does
not prevent from interpreting this wavefunction as a special
Euclidean vacuum
of the microscopic physical modes.

Sect.7 occupies the central place in the paper for it contains the
derivation
of the quantum distribution function for tunnelling Hartle-Hawking
geometries
on the space of collective variables. This distribution, on one hand,
demonstrates unitarity and, on the other hand, reduces to the
partition
function of closed gravitational instantons of spherical topology
weighted by
their Euclidean {\it effective\,} action including all one-loop
corrections.
This algorithm, based on the reduction technique of the next paper in
this
series \cite{tunnelII}, establishes the link between the
non-covariant but
manifestly unitary Lorentzian theory with its covariant Euclidean
counterpart
-- one of the main purposes of this paper.

Sect.8 contains the first important application of this quantum
distribution,
briefly reported in \cite{BKam:norm,BarvU}, -- the over-Planckian
behaviour of
the partition function of the inflationary cosmologies in the Hawking
model of
the chaotic inflation \cite{Linde,H}. In contrast to the tree-level
approximation, which always generates the Hartle-Hawking wavefunction
unnormalizable at high energies \cite{H-Page,Vilenkin:tun-HH}, the
corresponding quantum distribution is determined asymptotically by
the
anomalous scaling of the Euclidean theory on the gravitational
instanton of the
above type. Thus, depending on the value of this scaling parameter
determined
by the particle content of the model, this distribution can either
enhance the
contribution of the over-Planckian energy scales, or suppress them
and justify
the use of the semiclassical expansion.

\section{Euclidean quantum gravity as an analytic continuation from
its
Lorentzian counterpart}
\hspace{\parindent}
Let us denote the collection of three-dimensional metric
$g_{ab}\,({\bf x})$
and matter fields $\phi({\bf x})$ in the canonical quantization of
gravity by
     \begin{equation}
     \mbox{\boldmath $q$}=
     (\,g_{ab}\,({\bf x}),\,\phi\,({\bf x})\,).    \label{eqn:2.1}
     \end{equation}
Then, what is usually called the transition amplitude from the
configuration
$\mbox{\boldmath $q$}_{-}$ at spatial hypersurface $\Sigma_{-}$ to
the
configuration $\fatg{ q}_{+}$ at $\Sigma_{+}$ is given by the path
integral
over Lorentzian spacetime geometries and histories of matter fields
$\fatg{g}=(\,g_{\mu\nu} ({\bf x},t),\,\phi\,({\bf x},t)\,)$
interpolating
between the data on $\Sigma_{-}$ and $\Sigma_{+}$
      \begin{equation}
      \fatg{K}(\mbox{\boldmath $q$}_{+},\mbox{\boldmath $q$}_{-}\!)=
      \int D\fatg{\mu}[\,\fatg{g}\,]\,\,
      {\rm e}^{\!\!\phantom{0}^{\textstyle
      \frac{i}{\hbar}\fatg{S}\,[\,\fatg{g}\,]}}.
\label{eqn:2.2}
      \end{equation}
Here $\fatg{S}\,[\,\fatg{g}\,]$ is the gravitational action in
spacetime domain
sandwiched between the hypersurfaces $\Sigma_{-}$ and $\Sigma_{+}$,
which takes
the form
       \begin{equation}
       \fatg{S}\,[\,\fatg{g}\,]=
       \int_{t_{-}}^{t_{+}}dt\,\fatg{\cal{L}}\,
       (\mbox{\boldmath $q$},
       \dot{\mbox{\boldmath $q$}},N)
\label{eqn:2.3}
       \end{equation}
of the time integral with the Lagrangian
$\fatg{\cal{L}}\,(\mbox{\boldmath
$q$},\dot{\fatg{ q}}, N)$ when spacetime is foliated by the
$t$-parameter
family of spacelike hypersurfaces, so that $\fatg{g}=(\mbox{\boldmath
$q$}\,(t),N(t))$ is decomposed into spatial 3-metric and matter
fields
$\mbox{\boldmath $q$}\,(t)$ (\ref{eqn:2.1})  and lapse and shift
functions
$N(t)=(N^{\perp}({\bf x},t),\,N^{a}({\bf x},t)\,)$. In contrast to
variables
$\mbox{\boldmath $q$}\,(t)$, which enter the gravitational Lagrangian
together
with their space and time derivatives, the time derivatives of lapse
and shift
functions $N$ do not appear in (\ref{eqn:2.3}) -- a well-known fact
accounting
for the constrained nature of the gravitational theory.

The integration measure $D\fatg{\mu[\,g\,]}$ in (\ref{eqn:2.2})
includes the
Faddeev-Popov gauge fixing procedure for local diffeomorphism
invariance of the
theory with the corresponding ghost contribution and implies
integration over
the histories $\fatg{g}=(\mbox{\boldmath $q$}\,(t),\,N(t))$ matching
the fixed
fields $\mbox{\boldmath $q$}\,(t_{\pm})=\mbox{\boldmath $q$}_{\pm}$
at
$\Sigma_{\pm}$. Lapse and shift functions $N$, which have the nature
of
Lagrange multiplyers generating the gravitational constraints, are
integrated
over at the boundary surfaces and, therefore, do not enter the
arguments of the
transition kernel (\ref{eqn:2.2}) in accordance with the BRST
symmetry
truncated to the Dirac quantization of gravity
\cite{Henneaux,BarvU,BKr}
\footnote
{For the formulation of boundary conditions on the full set of
integration
variables and the boundary-value problems for their propagators in
the Feynman
diagrammatic technique corresponding to (\ref{eqn:2.2}) see
\cite{B:loopexp}
}.
As a result this kernel satisfies with respect to both of its
arguments the
system of the Wheeler-DeWitt equations, which are just the
operatorial version
of the gravitational constraints, and does not depend on time
coordinates
$t_{\pm}$ labeling the initial and final hypersurfaces.

\subsection{Two-fold analyticity of the path integral}
\hspace{\parindent}
This mathematical construction allows to realize the qualitative
ideas of
tunnelling geometries, sketched above, in the following language of
analytic
continuation. Under the assumption that the field variables
$\fatg{g}=
(\mbox{\boldmath $q$}\,(t),\,N(t))$ are analytic functions of time,
one can
deform the contour of integration over $t$ in (\ref{eqn:2.3}) into
its complex
plane without changing the value of the action. In view of the formal
$t_{\pm}$-independence of the kernel (\ref{eqn:2.2}) this can be done
even
without keeping the end points of this contour fixed: they can also
be
arbitrarily shifted into the complex plane of time. Such an
assumption on the
analyticity of the field variables in the integrand of the path
integral is
certainly of dubious nature, because generally the path integration
goes in the
class of non-smooth fields: the histories of \mbox{\boldmath
$q$}-variables can
be not differentiable, which fact is well known from the
skeletonization of the
path integral on the lattice, while the histories of Lagrange
multiplyers
$N(t)$ can be even discontinuous (at least in the class of unitary
nonrelativistic gauges of the Faddeev-Popov gauge fixing procedure
\cite
{BarvU}). However, the contribution of such non-smooth histories is
inessential
in the $\hbar$-expansion which we consider here.

As a consequence of such analytic continuation, the integration field
variables
generally become complex-valued, which means that the integral
(\ref{eqn:2.2})
must be taken over some complex contour \fatg{C} in the configuration
space of
the theory. The possibility of doing it without changing the value of
the
transition kernel is again provided by the analyticity of the
integrand, but
this time, of the functional integral with respect to its functional
argument
$\fatg{g}$. Due to this two-fold analyticity in {\it time} and {\it
configuration-space} variables the path integral representation
(\ref{eqn:2.2})
can be identically rewritten as
      \begin{equation}
      \fatg{K}(\mbox{\boldmath $q$}_{+},
      \mbox{\boldmath $q$}_{-}\!)=
      \int_{\fatg{C}} D\fatg{\mu[\,\Phi\,]}\,
      \,{\rm e}^{\!\!\phantom{0}^{\textstyle
      -\frac{1}{\hbar}
      \fatg{\cal I}\,[\,\fatg{\Phi}(z)\,]}},
\label{eqn:2.4}
      \end{equation}
where $z,\,\fatg{\Phi}(z)$ and $\fatg{\cal I}\,[\,\fatg{\Phi}(z)\,]$
are the
results of the above two-fold analytic continuation of
correspondingly the
time, configuration-space fields and their gravitational action
(\ref{eqn:2.3})
       \begin{eqnarray}
       t&\rightarrow&z,\;\;\;
       \fatg{g}(t)\rightarrow\fatg{\Phi}(z)=
       (\,\mbox{\boldmath $q$}\,(z),
       \,N_{E}(z)\,),                             \nonumber \\
       -i\fatg{S}\,[\,\fatg{g}\,]
       &\rightarrow &
       \fatg{\cal I}\,[\,\fatg{\Phi}(z)\,]=
       \int_{C}dz\,\,\fatg{\cal L}_{E}
       (\mbox{\boldmath $q$}\,(z),
       d\mbox{\boldmath $q$}\,(z)/dz,N_{E}(z)\,).
\label{eqn:2.5}
       \end{eqnarray}
Here the contour of integration $C$ can generally represent an
arbitrary
continuous curve in the complex plane of time variable, and the
complex
Lagrangian $\fatg{\cal L}_{E}(\mbox{\boldmath $q$},d\mbox{\boldmath
$q$}/dz,N_{E})$ is related to the original gravitational Lagrangian
by the
equation
        \begin{eqnarray}
	&&\fatg{\cal L}_{E}
	(\mbox{\boldmath $q$},
	d\mbox{\boldmath $q$}/d\tau,N_{E})=
	-\fatg{\cal L}(\mbox{\boldmath $q$},\,
	id\mbox{\boldmath $q$}/d\tau,N),
\label{eqn:2.6}\\
	&&N\equiv(N^{\perp},N^{a})=
	(N^{\perp}_{E},\,iN^{a}_{E})
\label{eqn:2.7}
	\end{eqnarray}
for arbitrary functions $\mbox{\boldmath $q$}=\mbox{\boldmath
$q$}(\tau)$ and
$N_{E}= (N^{\perp}_{E}(\tau),\,N^{a}_{E}(\tau))$.

The notations in the eqs.(\ref{eqn:2.5})-(\ref{eqn:2.7}) obviously
imply that
the quantities labelled by subscript $E$ denote the objects in the
Euclidean
spacetime having $\tau$ as a time variable. In particular, equation
(\ref
{eqn:2.6}) means that in order to get the Euclidean Lagrangian from
the
Lorentzian one we have to perform the Wick rotation which basically
boils down
to multiplying the velocities and lapse functions by $i$. This
procedure makes
a formal transition from the Lorentzian metric
         \begin{equation}
	 ds^{2}=-N^{2}dt^{2}+
	 g_{ab}\,(dx^{a}+N^{a}dt)\,(dx^{b}+N^{b}dt)
\label{eqn:2.8}
	 \end{equation}
to the metric of Euclidean spacetime
         \begin{equation}
	 ds^{2}_{E}=N^{2}d\tau^{2}+
	 g_{ab}\,(dx^{a}+N^{a}_{E}\,d\tau)\,
	 (dx^{b}+N^{b}_{E}\,d\tau),                   \label{eqn:2.9}
         \end{equation}
foliated by the surfaces of constant $\tau$.
\footnote
{Strictly speaking, Wick rotation implies multiplying by {\normalsize
$\pm i$}
the time components of those matter fields {\normalsize
$\mbox{\boldmath $q$}$}
which have tensor indices, so that {\normalsize $\mbox{\boldmath
$q$}$}'s in
the left-hand side of (\ref{eqn:2.6}) should be also labelled by $E$,
but for
brevity we disregard this subtlety. Actually, in what follows we
shall not
encounter this problem because we shall work with the theory in the
representation of true physical variables which usually comprise the
subset of
spatial tensor fields and do not include time components.
}
Comparison of equations (\ref{eqn:2.7}) and (\ref{eqn:2.5}) shows
that one can
regard $z$ as a variable in a complex plane of the Euclidean time
$\tau$ so
that the real part of $z$ can be identified with $\tau$ itself, while
its
imaginary part coincides with $t$. Correspondingly, in the integral
(\ref{eqn:2.5}) over arbitrary contour $C$ the element of integration
is
$dz=d\tau+idt$.

To clarify further the analytic continuation of the above type, let
us consider
the following two choices of such a contour $C$. One choice is
obvious and can
be called the Lorentzian one:
        \begin{equation}
	C_{L}:\{\,z=it,\,{\rm Im}\,t=0,\,
	t_{-}\leq{t}\leq{t_{+}}\}.               \label{eqn:2.10}
	\end{equation}
It generates exactly the Lorentzian gravitational action
(\ref{eqn:2.3})
provided the restriction of the fields $\fatg{\Phi}(z)$ to this
contour gives
real-valued configuration space variables of the Lorentzian gravity
theory
$\fatg{g}(t)$:
        \begin{eqnarray}
	-i\fatg{S}\,[\,\fatg{g}\,(t)\,]=
	\left.\fatg{\cal I}\,[\,\fatg{\Phi}\,(z)\,]
	^{\phantom {0}}
	_{\phantom {0}}\!\!
	\right|_{C_{L}},\;\;\;
	\left.\fatg{\Phi}\,(z)
	^{\phantom {0}}
	_{\phantom {0}}\!\!
	\right|_{C_{L}}=\fatg{g}\,(t).
\label{eqn:2.11}
        \end{eqnarray}
With such a contour and the integration over real $\fatg{g}\,(t)$ one
gets from
(\ref{eqn:2.4}) the original Lorentzian path integral representation
of the
transition kernel.

Another choice is the Euclidean contour $C_{E}$:
        \begin{equation}
	C_{E}:\{\,z=\tau,\,{\rm Im}\,\tau=0,\,
	\tau_{-}\leq{\tau}\leq{\tau_{+}}\},
\label{eqn:2.12}
	\end{equation}
which gives rise to the Euclidean action of the gravitational and
matter
variables $\fatg{g}(\tau)$ corresponding to the metric
(\ref{eqn:2.9}) and
generates the basic path integral of the Euclidean quantum gravity
        \begin{eqnarray}
	&&\fatg{I}[\,\fatg{g}(\tau)\,]=
	\fatg{\cal I}\,[\,\fatg{\Phi}(z)\,]\,
	|_{\!\!\!\!\textstyle\phantom{0}_{C_{E}}},\;\;
	\fatg{\Phi}(z)\,|_{\!\!\!\!\textstyle
	\phantom{0}_{C_{E}}}=
	\fatg{g}(\tau),
\label{eqn:2.13}\\
      &&\fatg{K}(\mbox{\boldmath $q$}_{+},
      \mbox{\boldmath $q$}_{-}\!)=
      \int D\fatg{\mu[\,g\,]}\,
      \,{\rm e}^{\!\!\phantom{0}^{\textstyle
      -\frac{1}{\hbar}\fatg{I}\,
      [\,\fatg{g}(\tau)\,]}}.               \label{eqn:2.14}
      \end{eqnarray}

In view of the formal time independence of $\fatg{K}(\mbox{\boldmath
$q$}_{+},\mbox{\boldmath $q$}_{-}\!)$ and the analyticity
assumptions, all
three path integrals (\ref{eqn:2.2}), (\ref{eqn:2.4}) and
(\ref{eqn:2.14}) are
equal, provided the {\it configuration space} integration contours in
(\ref{eqn:2.2}) and (\ref{eqn:2.14}) either coincide or turn out to
be
continuously deformable to the contour \fatg{C} in (\ref{eqn:2.4})
without
crossing the domains of non-analyticity of the integrand. Actually
the integral
(\ref{eqn:2.4}) is parametrized only by the boundary fields
\fatg{q_{\pm}} and
by the contour \fatg{C} (or, more precisely, by the equivalence class
of the
contours deformable in the above sense to each other). Recent studies
of this
property occupied one of the central places in the theory of quantum
gravity
and resulted in a major achievement of modern quantum cosmology --
the
no-boundary proposal for the quantum state of the Universe
\cite{HH,H}.

\subsection{The no-boundary proposal}
\hspace{\parindent}
The formulation of the no-boundary proposal begins with the analysis
of the
\mbox{\boldmath $q$}-dependence of the transition kernel
(\ref{eqn:2.2}). It
has two arguments $\mbox{\boldmath $q$}_{\pm}$ associated with two
hypersurfaces $\Sigma_{\pm}$ and describes the dynamical transition
from
$\Sigma_{-}$ to $\Sigma_{+}$ via the Lorentzian spacetime realizing
the
cobordism of these two spacelike surfaces. In contrast to transition
amplitudes, the wavefunction has one argument associated with the
spacelike
hypersurface to which the quantum state is ascribed. The idea of
Hartle and
Hawking \cite{HH,H} to obtain the one-argument wavefunction from the
two-argument transition kernel (\ref{eqn:2.2}) or (\ref{eqn:2.14})
consists in
shrinking the spacetime section $\Sigma_{-}$ to a point, demanding
the
regularity of all integration fields in its vicinity and integrating
over all
physical fields at this point compatible with the regularity
condition. However
this set of requirements cannot be implemented in spacetime with the
Lorentzian
signature: shrinking $\Sigma_{-}$ to a point implies that $t$ plays
the role of
a radial coordinate in its neighbourhood, which is timelike in
contrast to
spacelike concentric spheres of constant $t$ around the origin $t=0$.
Such a
4-geometry is singular at the origin $t=0$ which, thus, cannot be
considered as
a regular internal point treated on equal footing with the other
points of
spacetime. Therefore, the above transition to the one-argument path
integral
has to be performed with such integration 4-geometry which can be
regular
around the hypersurface tending to a point.

In the Hartle-Hawking no-boundary proposal the cosmological
wavefunction is
constructed by integrating over Euclidean geometries and matter
fields
$\fatg{g}$ on spacetime $\fatg{M}$ which has a topology of a compact
4-dimensional ball $\fatg{\cal B}^{4}$ bounded by a 3-dimensional
hypersurface
$\Sigma_{+}$ with the boundary fields $\mbox{\boldmath $q$}_{+}$:
      \begin{eqnarray}
      \fatg{\Psi}(\mbox{\boldmath $q$}_{+})=
      \int D\fatg{\mu[\,g\,]}\,
      \,{\rm e}^{\!\!\phantom{0}^{\textstyle
      -\frac{1}{\hbar}\fatg{I}\,[\,\fatg{g}\,]}}.
\label{eqn:2.15}
      \end{eqnarray}
The Euclidean gravitational action $\fatg{I}\,[\,\fatg{g}\,]$ in this
equation
can be regarded as a particular case of (\ref{eqn:2.13})
corresponding to the
Euclidean time contour $C_{E}$ (\ref{eqn:2.12}) with $\tau_{-}=0$
       \begin{equation}
       \fatg{I}\,[\,\fatg{g}\,]=
       \int_{0}^{\tau_{+}}d\tau\,\,\fatg{\cal L}_{E}
       (\,\mbox{\boldmath $q$},
       d\mbox{\boldmath $q$}/d\tau,N_{E}),
\label{eqn:2.16}
       \end{equation}
because the manifold $\fatg{M}$ can be viewed as originating from the
tube-like
spacetime $\Sigma \times [\tau_{-},\tau_{+}]$ by the procedure of the
above
type: shrinking one of its boundaries $\Sigma_{-}$ to a point and
inhabiting it
by a positive-signature metric and matter fields (see Fig.2). In such
a
manifold the role of a radial coordinate is played by the Euclidean
time $\tau$
with the origin at $\tau_{-}=0$ -- point-like remnant of the boundary
$\Sigma_{-}$ of vanishing size.

Therefore the no-boundary path integral (\ref{eqn:2.15}) is also a
particular
case of the kernel (\ref{eqn:2.4}) with the no-boundary prescription
replacing
the specification of $\mbox{\boldmath $q$}_{-}$. This is the
topological part
of the definition of the Hartle-Hawking wave function. The rest part
of this
definition is the choice of the integration contour $\fatg{C}$. Since
the work
\cite{Gibbons-H-P}, revealing the indefiniteness of the Euclidean
gravitationa
action in the sector of a conformal mode, it is known that this
integration
cannot run over real 4-geometries: to make the path integral formally
convergent one should rotate the integration contour for this mode
into the
complex plane.

Unfortunately, at present, there is no theory which could have
uniquely fixed
this contour in the no-boundary proposal \cite{Hal-Hartle}. Its
choice can be
constrained by a number of compelling but disjoint arguments,
including
convergence of the path integral, the recovery of quantum field
theory in a
semiclassical curved spacetime, the enforcement of quantum
constraints, etc.
\cite{Hal-Hartle}, but still has essential freedom, which was
confirmed by
considerations of several minisuperspace models \cite{Hal-Louko}. The
general
conclusion claimed by the authors of \cite{Hal-Hartle} was that the
program of
finding the unique integration contour on the basis of quantizing the
true
physical degrees of freedom \cite{Hartle-Sch,Schleich} has not been
successfully implemented in the theory of spatially closed
cosmologies, in
contrast to asymptotically-flat gravitational systems subject to very
powerful
positive-energy and positive-action theorems \cite{Schoen-Y,Witten}.
This
conclusion can be, probably, understood by taking into account that
the
quantization of physical variables is intrinsically incomplete beyond
the
semiclassical expansion, for it suffers from the problem of Gribov
copies
\cite{BarvU,BKr}, which in canonical quantum gravity manifest
themselves as
different aspects of a notorious "problem of time"
\cite{Kuchar:time,Wald}. The
presence of the Gribov problem serves as a compelling argument
\cite{BarvU,BKr}
in favour of the third-order quantization of gravity which was
intensively
discussed recently in connection with the ideas of the wormhole
physics,
baby-universe production and Coleman's "big-fix" mechanism
\cite{Gid-S,Coleman,Hawk-w}. Then, the above ambiguity in integration
contour
of the path integral can be interpreted as corresponding to different
choices
of  Green's functions and vacua of the third-quantized gravity theory
\footnote
{I am grateful to Bruce Campbell for this observation.}.

Here we shall not select the path integration contour from the
variety of all
possible equivalence classes of the above type. We suppose that this
choice is
already done by this or that selection rule, so that we have at our
disposal
one such class within which we can freely deform the contour. In
particular, we
suppose that we can pass it through the saddle point $\fatg{g}$ of
the
Euclidean gravitational action, which gives the following dominant
contribution
      \begin{equation}
      \fatg{\Psi}(\mbox{\boldmath $q$}_{+})\sim
      {\rm e}^{\!\!\phantom{0}^{\textstyle
      -\frac{1}{\hbar}\fatg{I}\,[\,\fatg{g}\,]}}.
\label{eqn:2.17}
      \end{equation}
to the path integral within the $\hbar$-expansion. Our purpose now
will be to
see how the complex nature of the time contour $C$ in the definition
of the
complexified action and configuration-space contour $\fatg{C}$ will
show up in
the calculation of the no-boundary wavefunction (\ref{eqn:2.15}) and
thus
provide a mathematical ground for a simple picture of the tunnelling
geometry
given in Introduction.

\subsection{Nucleation of the Lorentzian Universe from the Euclidean
spacetime
of the no-boundary type}
\hspace{\parindent}
The saddle point $\fatg{g}=(\mbox{\boldmath $q$}\,(\tau),N(\tau))$ of
the
Euclidean action in (\ref{eqn:2.15}) is a solution of the
boundary-value
problem
       \begin{eqnarray}
       &&\left.\frac{\delta\fatg{I}[\,\fatg{g}\,]}
       {\delta\fatg{g}}\;
       \right|_{\fatg{M}}=0,
\label{eqn:2.18}\\
       \nonumber\\
       &&\;\mbox{\boldmath $q$}\,
       |_{\textstyle\phantom{0}\!\!
       _{\fatg{\partial M}}}
       \equiv\mbox{\boldmath $q$}\,(\tau_{+}\!)=
       \mbox{\boldmath $q$}_{+},\;
       \fatg{g}\,|_{\textstyle\phantom{0}\!\!
       _{\fatg{M}}}={\rm reg},
\label{eqn:2.19}
       \end{eqnarray}

            which is a system of elliptic differential equations with
the
3-metric and matter fields $\mbox{\boldmath $q$}_{+}$ prescribed at
the
boundary and lapse and shift functions $N$ determined by necessary
gauge
conditions fixing the coordinatization of $\fatg{M}$. In a sherical
coordinate
system with $\tau$ playing the role of geodetic radius in the
vicinity of
$\tau_{-}=0$ the no-boundary regularity condition (\ref{eqn:2.19})
reduces in
the main to the following behaviour of the 4-metric
        \begin{equation}
	ds^{2}=d\tau^{2}+
	\tau^{2}c_{ab}\,dx^{a}dx^{b}+O(\tau^{3}),\;\;
	\tau\rightarrow 0,
\label{eqn:2.20}
        \end{equation}
and sufficiently smooth differentiability of matter fields.

When the argument $\mbox{\boldmath $q$}_{+}$ of the wavefunction
corresponds to
a 3-geometry of "small size", that is close to the above asymptotic
expression,
one is granted, for obvious reasons, to have a real-valued solution
of the
classical Euclidean equations (\ref{eqn:2.18})-(\ref{eqn:2.19}). For
a
pure-gravity theory with the cosmological constant $\Lambda=3H^{2}$,
considered
in Introduction, in the case of the round 3-metric $\mbox{\boldmath
$q$}_{+}=a^{2}c_{ab}$ with a scale factor $a\leq 1/H$ this solution
coincides
with the 4-geometry (\ref{eqn:1.3}) - (\ref{eqn:1.4}). However, when
the
3-geometry $\mbox{\boldmath $q$}_{+}$ is big enough, such a
real-valued
solution with the positive-signature metric may not exist, as it
happens in the
above example for a scale factor $a\geq 1/H$. This is a manifestation
of the
fact that, generally, the solutions of Einstein equations have
caustics in
superspace of $\fatg{q}$ and cannot regularly be continued into its
"shadow"
domains. But such a solution, which we shall denote by
$\fatg{\Phi}\,(z)$, can
exist when the segment of the Euclidean contour (\ref{eqn:2.12}) is
replaced,
via the procedure of analytic continuation, with some contour in the
complex
plane of the Euclidean time $z=\tau+it$, $C_{+}:\{z=z\,(\sigma),
\,0\leq\sigma\leq 1,\,z\,(0)=0, \,z\,(1)=z_{+}\}$, starting at zero
value of
the Euclidean "radius" and ending at some point
$z_{+}=\tau_{B}+it_{+}$. In
principle, this contour can represent an arbitrary curve in the
complex
z-plane, but for reasons of convenience and good physical
interpretation it
makes sense to break this curve into the union of two straight
segments
        \begin{equation}
	 C_{+}=C_{E}\cup C_{L},
\label{eqn:2.22}
	\end{equation}
which are particular examples of the Euclidean (\ref{eqn:2.12}) and
Lorentzian
(\ref{eqn:2.10}) contours (see Fig.3):
        \begin{eqnarray}
	 &&C_{E}:\{\,z=\tau,\,{\rm Im}\,\tau=0,\,
	 0\leq{\tau}\leq{\tau_{B}}\}, \\
\label{eqn:2.23}
	 &&C_{L}:\{\,z=\tau_{B}+it,\,{\rm Im}\,t=0,\,
	 0\leq{t}\leq{t_{+}}\}.
\label{eqn:2.24}
	 \end{eqnarray}

When the solution of classical equations on $C_{E}$ and $C_{L}$ is
{\it
real-valued} and represents respectively the real metrics
(\ref{eqn:2.9}) and
(\ref{eqn:2.8}) and correspondingly related matter fields, then the
above two
segments can be ascribed to real Euclidean and Lorentzian sections of
one
complex spacetime. These two sections are analytically matched
together across
the bounce surface $\tau=\tau_{B}$ and form one spacetime manifold of
a
combined Euclidean-Lorentzian signature. At the "moment" $\tau_{B}$
the
Euclidean solution suffers a bounce or tunnells into the Lorentzian
regime and,
thus, gives rise to the "beginning of time" \cite{Gibbons-Pohle}.
This is
precisely the situation of the DeSitter Lorentzian spacetime
(\ref{eqn:1.1}) -
(\ref{eqn:1.2}) with the Hubble parameter $H=\sqrt{\Lambda/3}$,
nucleating from
the Euclidean four-dimensional hemisphere (\ref{eqn:1.3}) -
(\ref{eqn:1.4}) of
radius $R=1/H$ at its equator -- the spatial section at
$\tau_{B}=\pi/2H$ (see
Fig.1).

The above case of the so-called real tunnelling geometries has been
intensively
studied since the invention of the no-boundary proposal.
Semiclassically its
interpretation goes as follows. Due to the complexity of the contour
$C_{+}$
the exponential of (\ref{eqn:2.17}) should be replaced by the complex
functional $\fatg{\cal I}[\,\fatg{\Phi}(z)\,]$ computed at the saddle
point
$\fatg{\Phi}\,(z)$. In view of reality of $\fatg{\Phi}\, (z)$ at both
Euclidean
and Lorentzian segments of the full contour $C_{+}$, this action
functional has
a complex structure
         \begin{equation}
	 \fatg{\cal I}\,[\,\fatg{\Phi}\,(z)\,]=
	 \fatg{I}-i\fatg{S}                         \label{eqn:2.25}
	 \end{equation}
with the real and imaginary parts contributed respectively by the
Euclidean and
Lorentzian domains of the total spacetime. For the same reason these
real
fields satisfy in these domains correspondingly the Euclidean and
Lorentzian
equations of motion, so that $\fatg{S}$ turns out to be the classical
Hamilton-Jacobi function of the system. This fact, when
(\ref{eqn:2.25}) is
substituted into the expression (\ref{eqn:2.17}), prompts to
interprete the
resulting wavefunction $\fatg{\Psi}(\mbox{\boldmath $q$}_{+})\sim
{\rm
exp}\{-\fatg{I}/ \hbar+i\fatg{S}/\hbar\}$ as describing the family of
semiclassical Lorentzian universes weighted by the exponentiated
action of the
corresponding Euclidean domain responsible for their tunnelling or
birth from
"nothing"\cite{HH,H}. The analyses of this weight
\cite{Gibbons-Hartle} shows
that under certain positive-energy assumptions for matter fields a
real
tunnelling solution can nucleate only a single connected Lorentzian
spacetime
with the most probable topology ${\bf R}\times S^{3}$ and the
DeSitter metric.
The rest of interpretation for the no-boundary wavefunction is
usually based on
incorporating the old method of deriving the quantum field theory of
matter
fields in curved spacetime from the semiclassically approximated
Wheeler-DeWitt
equations \cite{DeW,Rubakov-L,Banks,Halliwell-Hawking}. This method
shows that
the quantum state of matter fields on the background of such a
tunnelling
geometry coincides with the Euclidean DeSitter invariant vacuum
\cite{Halliwell-Hawking,Laflamme,Gibbons-Pohle} which generates in
the theory
of the inflationary Universe the large scale cosmological structure
compatible
with observations \cite{Linde,large-scale}.

As was discussed in Introduction, there are basically two
difficulties with the
results of the above type: the limitations of the semiclassical
interpretation
and the restriction of the whole scheme to real tunnelling geometries
and
matter fields. The first difficulty invalidates the attempts to
interprete
beyond the tree level those modes of the gravitational field which
are treated
as classical in the $\hbar$-expansion of the Wheeler-DeWitt equation.
In
particular, for the Hawking model of chaotic inflation it does not
allow to get
the no-boundary quantum state as a normalizable wavefunction of the
inflaton
scalar field generating the effective cosmological constant. The
second
difficulty restricts the applicability of the above theory to the
class of toy
models with a special type of "inert" matter fields, because any
realistic
matter (including the chaotic inflation model of Hawking) generally
gives
tunnelling geometries which are complex both in the Euclidean (that
is on
$C_{E}$) and Lorentzian (on $C_{L}$) regimes.

As a remedy against the first difficulty (and as a starting point for
generalizing the above theory to the case of complex tunnelling
geometries), we
shall consider the construction of the no-boundary wavefunction in
the
representation of true physical variables. For problems exploiting
within the
$\hbar$-perturbation theory only local properties of the
configuration space,
this approach seems to be very promising \cite{BarvU,BKr}: under a
proper
operator realization it turns out to be equivalent to Dirac and BFV
(BRST)
quantization, allows to construct the conserved physical inner
product in the
space of physical states and, thus, provides the unitarity of the
theory.
Therefore, we shall use it here for the one-loop calculation of the
no-boundary
wavefunction.

\section{No-boundary wavefunction in the representation of physical
variables}
\subsection{ADM reduction and path-integral quantization}
\hspace{\parindent}
The perturbative construction of the no-boundary wavefunction of
physical
variables repeats with slight modifications the general formalism
presented
above. The starting point of this procedure is the ADM reduction of
the
gravitational theory to dynamically independent degrees of freedom
\cite{ADM}
adjusted to the case of spatially closed cosmology
\cite{Kuchar:bubble,BPon,BarvU}. It consists in imposing the gauge
conditions
on the original phase-space variables  $\mbox{\boldmath $q$}$ and
$\fatg{p}$
($\fatg{p}$ forms the set of canonical momenta conjugated to
$\mbox{\boldmath
$q$}$). These gauge conditions fix the local invariance of the theory
with
respect to general coordinate diffeomorphisms and allow one to
disentangle the
physical degrees of freedom as follows. The full system of the
first-class
gravitational constraints and imposed gauges can be solved for
$(\mbox{\boldmath $q$},\fatg{p})$ in terms of the dynamically
independent
canonical coordinates $q=q^{i}$ and their conjugated momenta
$p=p_{i}$, which
we shall label by a condensed index $i$. The requirement of the
conservation of
gauge conditions in time yields also the lapse and shift functions
$N$ as
functions of $(q,p)$ and thus specifies a concrete foliation of
spacetime by
spacelike hypersurfaces. Substituting $\mbox{\boldmath
$q$}=\mbox{\boldmath
$q$}\,(q,p),\,\fatg{p}= \fatg{p}\,(q,p)$ into the canonical action of
gravity
theory produces its reduced phase space canonical action in terms of
the
unconstrained variables $(q,p)$. This action contains the
nonvanishing, but
generally time-dependent, Hamiltonian of these physical variables and
by a
standard procedure of the Legendre transform from $p$ to $\dot
q=dq/dt$
generates the Lagrangian ${\cal L}(q,dq/dt,t)$ and the corresponding
Lagrangian
action
        \begin{equation}
        S[\,q(t)\,]=\int^{t_{+}}_{t_{-}}
	dt\,{\cal L}(q,dq/dt,t).                    \label{eqn:3.1}
        \end{equation}

According to the general theory of constrained systems
\cite{Faddeev,Fr-V} the
path integral (\ref{eqn:2.2}) over the full configuration space of
fields
$\fatg{g}=(\mbox{\boldmath $q$},N)$ with the correct integration
measure,
including the gauge fixing and contribution of ghosts, can be
identically
rewritten as a path integral over reduced phase space variables
$(q,p)$ of
their exponentiated canonical action
\footnote
{For a kernel {\normalsize $\fatg{K}(\mbox{\boldmath
$q$}_{+},\mbox{\boldmath
$q$}_{-}\!)$} of transition between the hypersurfaces {\normalsize
$\Sigma_{\pm}$} this relation holds up to certain operatorial factors
associated with {\normalsize $\Sigma_{\pm}$}, which are responsible
for the
unitary map between the Dirac-Wheeler-DeWitt quantization and reduced
phase
space quantization. This property is discussed in much detail in the
series of
author's papers \cite{B:GenSem,BarvU,BKr} both at the level of
operatorial and
path-integral quantizations.}.
The ADM reduction to $(q,p)$ is very complicated due to the
nonlinearity of the
gravitational constraints. Therefore it generally leads to the
physical
Hamiltonian which is a non-polynomial function of $p$, and the
corresponding
integration over momenta in this integral has a non-gaussian nature.
This
property complexifies the transition from the phase-space path
integral to its
Lagrangian version, but in the main it boils down to the expression
(\ref{eqn:2.2}) with the covariant action $\fatg{S}\,[\,\fatg{g}\,]$
replaced
by its reduced version $S\,[\,q\,]$ and the new integration measure
$D\mu[\,q\,]$. This measure accumulates the result of this nontrival
integration over momenta and can be calculated as a power series in
$\hbar$
beginning with the following one-loop contribution:
        \begin{eqnarray}
	&&D\mu[\,q\,]=\prod_{t}dq(t)\;
	[\,{\rm det}\,a\,]^{1/2}\!(t)
	+O\,(\,\hbar\,),\;\;\;
	dq=\prod_{i}dq^{i},
\label{eqn:3.2}\\
	&&{\rm det}\,a={\rm det}\,a_{ik}, \;\;\;
	a_{ik}=\frac{\partial^{2}{\cal L}}
	{\partial\dot q^{i}\partial\dot q^{k}}.
\label{eqn:3.4}
	\end{eqnarray}
Here the determinant of the Hessian matrix $a_{ik}$ is understood
with respect
to condensed indices $i$ and $k$. They include, depending on the
representation
of field variables, either continuous labels of spatial coordinates
or discrete
quantum numbers labelling some complete infinite set of functions on
a spatial
section of spacetime. Therefore the above determinant is functional,
but its
functional nature is restricted to a spatial slice of constant time
$t$. The
product over time points of ${\rm det}\,a(t)$ can be regarded as a
determinant
of higher functional dimensionality associated with the whole
spacetime if we
redefine $a_{ik}$ as a time-ultralocal operator $\fatg{a}=
a_{ik}\delta(t-t')$.
We shall denote such functional determinants for both ultralocal and
differential operators in time by Det. Therefore, in view of the
ultralocality
of $\fatg{a}$, the contribution of the one-loop measure becomes
        \begin{equation}
	\prod_{t}\,[\,{\rm det}\,a\,]^{1/2}(t)=
	[\,{\rm Det}\,\fatg{a}\,]^{1/2}=
	{\rm exp}\left \{\frac{1}{2}
	\int_{t_{-}}^{t_{+}}\!dt\,
	\delta\,(0)\;{\rm ln\,det}\,a(t)\right\}.
\label{eqn:3.6}
	\end{equation}
Since this expression contains the coincidence limit of the
one-dimensional
delta-function $\delta\,(0)$, the local measure contributes to the
path
integral a pure power divergence. As was shown in \cite{FV:Bern}
within the
four-dimensional treatment of spacetime covariant differential
operators, this
contribution identically cancels the strongest (quartic) divergences
of
one-loop Feynman diagrams -- the property which will be demonstrated
in
\cite{tunnelII} within the canonical framework.

One more modification inherent to the reduced phase space
quantization is that
the transition kernel (\ref{eqn:2.2}) and the Hartle-Hawking
wavefunction
(\ref{eqn:2.15}) have in the representation of physical variables $q$
the
explicit dependence on time. This follows from the fact that in the
ADM
reduction the role of time is played by some functional combinations
of the
initial phase-space coordinates $\mbox{\boldmath $q$}$ (or momenta
$\fatg{p}$),
so that the arguments $\mbox{\boldmath $q$}_{\pm}$ of
$\fatg{K}(\mbox{\boldmath
$q$}_{+},\mbox{\boldmath $q$}_{-}\!)$ after the reduction give rise
to time
variables $t_{\pm}$. Simultaneously with disentangling time from
phase space of
the theory, the ADM procedure recovers the nonvanishing physical
Hamiltonian,
and its operatorial version governs the Schrodinger evolution of the
transition
kernel and the wavefunction \cite{BPon,BarvU,BKr}.

\subsection{Analytic continuation technique}
\hspace{\parindent}
The reformulation of the analytic continuation technique in terms of
physical
variables is rather straightforward. Mainly it repeats the equations
(\ref{eqn:2.2}) - (\ref{eqn:2.6}) and (\ref{eqn:2.10}) -
(\ref{eqn:2.25} ) with
the original configuration variables $\fatg{g}=(\mbox{\boldmath
$q$},\,N)$ ,
their Lagrangian $\fatg{\cal{L}}\,(\mbox{\boldmath
$q$},\dot{\mbox{\boldmath
$q$}},N)$ and  action $\fatg{S}\,[\,\fatg{g}\,]$ beeing replaced by
the
corresponding physical space counterparts $q=q^{i},\;{\cal
L}\,(q,dq/dt,t)$ and
$S[\,q(t)\,]$. In contrast to bold letters for the objects in the
original
formulation we shall use the usual letters for their analogues in the
ADM
quantization. In particular the physical transition kernel
$K(q_{+},t_{+}|q_{-},t_{-}\!)$ and the wavefunction
$\Psi(q_{+},t_{+}\!)$ will
replace $\fatg{K}(\mbox{\boldmath $q$}_{+}, \mbox{\boldmath
$q$}_{-}\!)$ and
$\fatg{\Psi}\,(\mbox{\boldmath $q$}_{+})$
\footnote
{The precise relation between the unitary transition kernels and
wavefunctions
in the ADM and the Dirac-Wheeler-DeWitt quantization schemes, which
establishes
their local equivalence, is considered in \cite{B:GenSem,BarvU,BKr}}.

The analytic continuation of both time $t$ and configuration space
variables
$q(t)$ into their complex planes $z$ and $\Phi(z)$ can be done
similarly to
eqs.(\ref{eqn:2.5}) -- the procedure that generates the complex
action of
physical variables
         \begin{equation}
         {\cal{I}}[\,\Phi\,(z)\,]=\int_{C}dz\,
         {\cal{L}}_{E}(\Phi\,(z),d\Phi\,(z)/dz,z)
\label{eqn:3.7}
         \end{equation}
with the Euclidean Lagrangian ${\cal{L}}_{E}(\Phi,d\Phi/dz,z)$. The
latter is
related to its Lorentzian version of the eq.(\ref{eqn:3.1}) by the
definition
analogous to (\ref{eqn:2.6}). There is now one subtlety in this
definition
originating from the explicit dependence of the physical Lagrangian
${\cal
L}(q,dq/dt,t)$ on time. Since it is no longer translationally
invariant in the
complex plane of $z=\tau+it$, one should specify the point in this
plane with
respect to which the Wick rotation generates the Euclidean
Lagrangian. In
principle, the choice of this point is a part of the gauge fixing
procedure to
the same extent as the choice of time is a part of ADM reduction.
Therefore one
can expect that the physical results and the invariant unitary
dynamics of the
theory are independent of this choice. But there exists a physically
distinguished definition of the Wick rotation associated with the
no-boundary
proposal, which looks as follows.

According to the Hartle-Hawking construction the contour of
integration $C_{+}$
(\ref{eqn:2.22}) in the action functional runs from the point $z=0$
to some
complex point $z_{+}$. In the Dirac-Wheeler-DeWitt quantization this
final
point does not explicitly enter the wavefunction, beeing
semiclassically
determined by its argument $\mbox{\boldmath $q$}_{+}$ from the
solution of the
boundary-value problem (\ref{eqn:2.18})-(\ref{eqn:2.19}) in some
coordinate
gauge. On the contrary, in the ADM quantization the physical
wavefunction
explicitly depends on time, and according to the above method of
analytic
continuation real and imaginary ranges of this argument can be
associated
respectively with the classically allowed and forbidden transitions
of the
system. This interpretation urges us to identify
the above final point of the integration contour
$z_{+}=\tau_{B}\!+it_{+}$ with
the complex time argument of the physical wavefunction
$\Psi(q_{+},z_{+})$
\footnote
{This does not, certainly, mean that there is no gauge-fixing
ambiguity in the
determination of $z_{+}$ analogous to the coordinate gauge for the
classical
equations (\ref{eqn:2.18})-(\ref{eqn:2.19}). The construction of time
in the
ADM reduction relies on some particular gauge which is implicitly
equivalent to
the coordinate gauge for these equations.
}.
Breaking the contour $C_{+}$ into the union (\ref{eqn:2.22}) of the
Euclidean
$C_{E}$ and Lorentzian $C_{L}$ segments heuristically implies that
the
no-boundary quantum state at the moment $t_{+}$ of the physical time
is a
result of the underbarrier penetration along $C_{E}$ followed by the
unitary
evolution along the segment of real time $C_{L}$. An obvious choice
of the
center of Wick rotation, which follows from this picture, is the
intersection
point $z=\tau_{B}$ of these two segments. Under this choice the
natural
definition of the Euclidean Lagrangian of complexified physical
variables
$\Phi(z)$ looks like
         \begin{equation}
	 {\cal L}_{E}(\,\Phi\,(z),\,d\Phi\,(z)/dz,\,z\,)=
	 -{\cal L}\,(\,\Phi\,(z),
	 \,id\Phi\,(z)/dz,\,(z-\tau_{B})/i\,),
\label{eqn:3.8}
         \end{equation}
because it generates from the universal complex action functional
(\ref{eqn:3.7}) the Lorentzian action (\ref{eqn:3.1}) on the contour
$C_{L}$
          \begin{equation}
	  iS\,[\,q(t)\,]=-{\cal{I}}\,[\,\Phi\,(z)\,]\,
	  |_{\!\!\!\!\textstyle\phantom{0}_{C_{L}}},\;\;
	  q\,(t)=\Phi\,(z)\,|_{\!\!\!\!
	  \textstyle\phantom{0}_{C_{L}}}
\label{eqn:3.9}
	  \end{equation}
and the Euclidean physical action on the contour $C_{E}$
          \begin{eqnarray}
	  &&I\,[\,\phi(\tau)\,]={\cal{I}}\,[\,\Phi(z)\,]\,
	  |_{\!\!\!\!\textstyle\phantom{0}_{C_{E}}},\;\;
	  \phi\,(\tau)=\Phi(z)\,|_{\!\!\!\!
	  \textstyle\phantom{0}_{C_{E}}}\\
\label{eqn:3.10}
	  &&I\,[\,\phi(\tau)\,]=
	  \int\limits_{0}^{\tau_{+}}d\tau\,
	  {\cal{L}}_{E}(\phi,d\phi/d\tau,\tau).
\label{eqn:3.11}
	  \end{eqnarray}
Here we denoted the values of the complex physical field $\Phi\,(z)$
on the
contours $C_{L}$ and $C_{E}$ respectively by $q\,(t)$ and
$\phi\,(\tau)$, so
that these functions are related by the following analytic
continuation from
the real axis of the Euclidean time $\tau$:
          \begin{equation}
          q\,(t)=\phi\,(\tau_{B}\!+it).
\label{eqn:3.12}
          \end{equation}

Now we can write the general expression for the no-boundary
wavefunction of
physical variables as a path integral
          \begin{eqnarray}
          \Psi\,(q_{+},z_{+})=
          \int D\mu\,[\,\Phi(z)\,]\,
          \,{\rm e}^{\!\!\phantom{0}^{\textstyle
          -\frac{1}{\hbar}{\cal I}\,[\,\Phi(z)\,]}}.
\label{eqn:3.13}
          \end{eqnarray}
with the complex action (\ref{eqn:3.7}) and a local measure
(\ref{eqn:3.2})
defined on the contour $C=C_{+}$ joining the points $z=0$ and $z_{+}$
in the
complex plane of time. The integration here goes over physical fields
matching
$q_{+}$ at the boundary $z=z_{+}$ of complex spacetime manifold
$\fatg{M}=\Sigma\times C_{+}$ and satisfying the no-boundary
regularity
conditions in the center of it $z=0$:
           \begin{equation}
	   \Phi\,(z_{+})=q_{+},\;\;\;\Phi\,(z)\,|_
	   {\!\!\!\phantom{0}_{\textstyle
	   z\rightarrow 0}}={\rm reg}.
\label{eqn:3.15}
	   \end{equation}

The wavefunction (\ref{eqn:3.13}) can be regarded as a result of the
analytic
continuation into the complex plane of $\tau_{+}$ of the Euclidean
path
integral
           \begin{equation}
	   \Psi\,(q_{+},\tau_{+})=
           \int D\mu\,[\,\phi(\tau)\,]\,
           \,{\rm e}^{\!\!\phantom{0}^{\textstyle
           -\frac{1}{\hbar}I\,[\,\phi(\tau)\,]}}
\label{eqn:3.16}
           \end{equation}
which corresponds to $z_{+}=\tau_{+}$ and the choice of the contour
$C_{+}=C_{E}$ in (\ref{eqn:3.13}). When $z_{+}=\tau_{B}\!+it_{+}$ in
$\Psi\,(q_{+},z_{+}\!)$ it makes sense to identify this function with
the {\it
Lorentzian} quantum state of the system $\Psi_{L}(q_{+},t_{+}\!)$
evolving in
the real physical time $t_{+}$ and originating from
$\Psi\,(q_{+},\tau_{+}\!)$
by this analytic continuation
            \begin{equation}
	    \Psi_{L}(q_{+},t_{+}\!)=
	    \Psi(q_{+},\tau_{B}\!+it_{+}\!).
\label{eqn:3.17}
	    \end{equation}
So our further strategy will consist in the calculation of the
Euclidean
wavefunction (\ref{eqn:3.16}) by the method of semiclassical
expansion and its
analytic continuation into the Lorentzian regime.

\subsection{One-loop approximation}
\hspace{\parindent}
The semiclassical expansion of (\ref{eqn:3.16}) is rather
straightforward and
begins in the one-loop approximation with the expression
           \begin{equation}
           \Psi\,(q_{+},\tau_{+}\!)=\left(\,\frac{{\rm
	   Det}\,\fatg{F}\,[\,\phi\,]}{{\rm
	   Det}\,\fatg{a}\,[\,\phi\,]}\,\right)^{-1/2}
	   {\rm e}^{\!\!\phantom{0}^{\textstyle
           -\frac{1}{\hbar}I\,[\,\phi\,]}}\,
	   [\;1+O\,(\hbar)\;].
\label{eqn:3.18}
	   \end{equation}
Here the dominant tree-level contribution originates from the
classical
extremal $\phi$ satisfying the boundary-value problem for physical
fields
similar to (\ref{eqn:2.18}) - (\ref{eqn:2.19})
       \begin{eqnarray}
       &&\left.\frac{\delta I\,[\,\phi\,]}
       {\delta\phi\,(\tau)}\;\right|_{\fatg{M}}=0,
\label{eqn:3.19}\\
       \nonumber \\
       &&\;\phi\,(\tau)\,|_{\textstyle\phantom{0}\!\!
       _{\fatg{M}}}={\rm reg},\;\;\;\;\phi\,(\tau)\,
       |_{\textstyle\phantom{0}\!\!_{\fatg{\partial M}}}
       \equiv\phi(\tau_{+})=q_{+},                  \label{eqn:3.20}
       \end{eqnarray}
while the preexponential factor is a combination of the local measure
(\ref{eqn:3.6}) and contribution of the gaussian functional
integration over
quantum disturbancies around this extremal. The latter is determined
by the
functional determinant of $\fatg{F}$ -- the kernel of the quadratic
part of the
action, which is a differential operator with respect to $\tau$
       \begin{equation}
       \fatg{F}\equiv\fatg{F}\,(d/d\tau\!)\,
       \delta(\tau-\tau^{\prime})=
       \frac{\delta^{2}I\,[\,\phi\,]}
       {\delta\phi(\tau)\,
       \delta\phi(\tau^{\prime})}.                   \label{eqn:3.21}
       \end{equation}
Because of the usual structure of the Lagrangian ${\cal{L}}_{E}
(\phi,d\phi/d\tau,\tau)$, containing at most first-order time
derivatives of
field variables, this matrix-valued differential operator
$\fatg{F}\,(d/d\tau\!)= \fatg{F}_{ik}\,(d/d\tau\!)$ is of the second
order and
has the form
       \begin{equation}
       \fatg{F}\,(d/d\tau)=-\frac{d}{d\tau}\,
       a\,\frac{d}{d\tau}
       -\frac{d}{d\tau}\,b+b^{T}\frac{d}{d\tau}+c,
\label{eqn:3.22}
       \end{equation}
where the coefficients $a=a\,_{ik},\;b=b\,_{ik}$ and $c=c\,_{ik}$ are
the
(functional) matrices acting in the space of field variables
$\phi\,(\tau)=\phi\,^{k}(\tau)$, and the superscript $T$ denotes
their
(functional) transposition $(b^{T})\,_{ik}\equiv b\,_{ki}$. These
coefficients
can be easily expressed as mixed second-order derivatives of the
Euclidean
Lagrangian with respect to $\phi\,^{i}$ and
$\dot\phi\,^{i}=d\phi\,^{i}/d\tau$.
In particular, the matrix of the second order derivatives $a\,_{ik}$
is given
by the Euclidean version of the Hessian matrix (\ref{eqn:3.4}):
$a\,_{ik}=\partial^{2}{\cal
L}_{E}/\partial\dot\phi\,^{i}\partial\dot\phi\,
^{k}$.

\subsection{The choice of gauge and nature of physical variables}
\hspace{\parindent}
The general scheme of the above type bears inalienable ambiguity in
the choice
of gauge for physical variables $q\,^{i}$. This choice specifies the
way these
variables are disentangled from the initial phase space of
$\mbox{\boldmath
$q$}$ and $\fatg{p}$ and fixes the physical internal time $t$ for
their
dynamics. This ambiguty shows up in the physical wavefunction and its
analytic
continuation $\Psi(q_{+},z)$ into the complex plane of $t$, which is
not
gauge-invariant object in contrast to the Dirac-Wheeler-DeWitt
wavefunction
$\fatg{\Psi(q)}$. There exist several physically reasonable
requirements which
can restrict the excessive freedom in the choice of the gauge-fixing
procedure.
Note, first of all, that the Wick rotation (\ref{eqn:3.8}) generally
leads to a
complex Euclidean Lagrangian in view of the explicit time dependence
in ${\cal
L}(q,dq/dt,t)$. Therefore it makes sense to define such an ADM
reduction that
makes ${\cal{L}}_{E}\,(\phi,d\phi/d\tau,\tau)$ real valued at real
values of
the Euclidean time $\tau$ and $\phi\,(\tau)$.

Another property, which will be important for our further
derivations, is the
requirement of the boundedness from below for the Euclidean action of
physical
variables (\ref{eqn:3.11}). It provides the formal (mode by mode)
convergence
of the Euclidean path integral (\ref{eqn:3.16}) over real fields
$\phi\,(\tau)$, establishes the normalizability of the wavefunction
$\Psi(q,t)$
on the real section of the $q-$configuration space (and, therefore,
the
possibility to regard physical variables as Hermitian operators) and
serves as
a ground for a special technique of complex extremals which we
develope in
Sect.5. This property of the {\it physical } Euclidean action can be
a
consequence of both a special reduction procedure and the properties
of the
original {\it covariant} gravitational action which, as is known,
suffers from
the problem of indefiniteness in the conformal sector that can
invalidate the
attempts to construct a good Euclidean action.

In this paper we shall not consider an ADM reduction which guarantees
all the
properties of the above type. We simply assume that the positivity of
the
quadratic form of the physical action near its extremum can be
provided and,
then, develope the consequnces of this assumption. Actual reduction
procedure
and its application to the Hawking model of chaotic inflation will be
considered in much detail in \cite{tunnelIII}. Just to give the idea
of how
this reduction works, we give here only its basic ingredients and
formulate the
no-boundary regularity conditions (\ref{eqn:3.20}).

The basic approximation to the chaotic inflationary
(spatially-closed)
cosmology consists in the minisuperspace model described by the
metric
(\ref{eqn:1.1}) - (\ref{eqn:1.2}) with the scale factor $a_{\!L}(t)$
driven by
the effective Hubble constant $H$ which is generated by the spatially
homogeneous (inflaton) scalar field $\varphi,\,H=H(\varphi)$. All the
other
spatially inhomogeneous fields of all possible spins are treated as
perturbations on this background and, therefore, the full set of the
initial
phase-space coordinates of the theory can be represented as
$\mbox{\boldmath
$q$}=(a,\,\varphi,\,\phi({\bf x}),\,\psi({\bf x}),\,A_{a}({\bf x}),\,
\psi_{a}({\bf x}),\,h_{ab}\,({\bf x}),...)$
\footnote
{Time-components of vector $A_{\mu}$, gravitino $\psi_{\mu}$,
graviton
$h_{\mu\nu}$, etc. form in the canonical formalism the set of
Lagrange
multiplyers (including lapse and shift functions) and, therefore, do
not enter
the set of phase-space coordinates.}.

A physically meaningful reduction from $\mbox{\boldmath $q$}$ to $q$
goes
separately in the minisuperspace sector of the full superspace
$(a,\,\varphi)$
and the sector of spatially inhomogeneous modes. There is only one
Hamiltonian
constraint which is effectively imposed on spatially homogeneous
modes
$(a,\,\varphi)$, because the momentum constraints are identically
satisfied in
the minisuperspace model of Bianchi IX type with the round
three-dimensional
metric $c_{ab}$ \cite{Ryan}. Therefore, there can be only one
physical degree
of freedom among these two minisuperspace variables $(a,\,\varphi)$.
The second
of these variables has to be fixed by the gauge condition which
simultaneously
disentangles time (or, in other words, identifies the second variable
with the
internal time). It is useful to choose the inflaton scalar field
$\varphi$ as
this physical degree of freedom and interprete the approximate
solution of
classical equations of motion (\ref{eqn:1.2}) as a gauge
         \begin{equation}
	 a(t)=\frac{1}{H\,(\varphi)}\,
	 {\rm cosh}\,H\,(\varphi)\,t,                \label{eqn:3.23}
	 \end{equation}
which thus plays the role of the parametrization of the initial
phase-space
coordinates in terms of the physical ones in the minisuperspace
sector of the
theory.

This gauge is very convenient because it corresponds to the choice of
cosmic
time with the lapse $N=1$ in classical solutions \cite{tunnelIII}
and, what is
most important for our purposes, provides the reality of the
Euclidean
Lagrangian (\ref{eqn:3.8}). Indeed, choosing
$\tau_{B}=\pi/2H\,(\varphi)$ and
analytically continuing the gauge (\ref{eqn:3.23}) along the contour
$C_{+}=C_{E}\cup C_{L}$ onto the real axis of the Euclidean time
$\tau$, one
finds that the Euclidean scale factor remains real and coincides with
the
Euclidean solution (\ref{eqn:1.4}). Therefore, the Lagrangian of
physical
variables on the background of such an Euclidean spacetime hemisphere
$0\leq\tau\leq\pi/2H\,(\varphi)$ is also real and even
positive-definite in the
sector of transverse-traceless graviton modes \cite{BKK}.

The ADM reduction in the sector of other fields can be performed in
very many
different ways. In the linearization approximation it mainly boils
down to
selecting the transverse $(T)$, transverse-traceless $(TT)$, etc.
modes of
spatial components of the corresponding tensor fields, so that the
full set of
physical variables can be written as
        \begin{equation}
	q^{i}=(\varphi,\,\phi({\bf x}),\,\psi({\bf x}),\,
	A^{T}_{a}({\bf x}),\,\psi^{T}_{a}({\bf x}),\,
	h^{TT}_{ab}({\bf x}),...).
\label{eqn:3.24}
	\end{equation}
Here the index $i$ is an element of condensed DeWitt notations which
we shall
intensively use throughout the paper. It includes discrete spin
labels of field
components and also continuous labels of spatial coordinates ${\bf
x}$. The
functions of spatial coordinates ${\bf x}$ in (\ref{eqn:3.24}) can be
decomposed as infinite series in the complete set of some spatial
harmonics,
which in view of the compactness of a spatial section is discrete and
countable. Then the continuous label ${\bf x}$ in $i$ will be
replaced by the
discrete quantum numbers enumerating these spatial harmonics. In both
cases,
however, the operations of respectively the integration over ${\bf
x}$ and
infinite summation over these numbers will be a part of a symbolic
contraction
of the condensed indices
\footnote
{It should be emphasized that, since we work in the noncovariant
canonical
formalism with the distinguished time variable, the condensed indices
do not
include time arguments of the fields. Correspondingly, these time
arguments and
such operations with them as integration will be, if necessary,
explicitly
written in all the formulae.}.

\subsection{The no-boundary regularity conditions for physical
variables}
\hspace{\parindent}
The no-boundary regularity conditions
(\ref{eqn:2.19})-(\ref{eqn:2.20}) must be
reformulated now in terms of physical variables (\ref{eqn:3.24}). To
begin
with, note that the gauge (\ref{eqn:3.23}) analytically continued to
the
Euclidean time, $\tau=\pi/2H(\!\varphi\!)+it$, gives the scale factor
$a_{E}(\tau)=\tau+O\,(\tau^{2})$ which automatically satisfies the
asymptotic
behaviour (\ref{eqn:2.20}) of the metric (this gauge picks up the
unit lapse
$N_{E}=1$ \cite{tunnelIII} and therefore guarantees that $\tau$
measures the
proper radial distance in the center of the Euclidean ball
$\fatg{\cal B}^{4}$
and does not produce the conical singularity for
$a_{E}(\tau)\sim\tau$ at
$\tau\rightarrow 0$). Thus, it remains to check that all the physical
fields
(\ref{eqn:3.24}) satisfy the condition of smooth regularity in the
center of
this Euclidean spacetime $\tau=0$. For spatially homogeneous modes
$\varphi=
\varphi\,(\tau)$ this condition implies that their radial derivative
should
vanish at this point $(d\varphi/d\tau)\,(0)=0$, while inhomogeneous
modes
should disappear themselves. This conclusion easily follows from the
fact that
the coordinates of a spatial section shrinking to a point $\tau=0$
play the
role of angular coordinates in a spherical coordinate system with the
origin at
this point. Therefore, the above properties of spatially homogeneous
and
inhomogeneous modes is a direct corollary of the direction
independent limit of
these modes or their derivatives when approaching the point $\tau=0$
\footnote
{Note that we require the continuity of at most the first-order
spacetime
derivatives of fields. The explanation for this restricted notion of
regularity
in the no-boundary proposal can follow from the fact that the
Lagrangian in the
Euclidean action (\ref{eqn:3.10}) contains at most the first-order
derivatives
of fields. Therefore the no-boundary proposal does not rule out the
integration
fields with discontinuous higher-order derivatives, because they do
not give
infinite contribution to the Euclidean action and are not
automatically
suppressed in the Euclidean path integral.}.

\section{The method of collective coordinates}
\hspace{\parindent}
In field-theoretical models the variables $q$ represent the
continuous
infinitude of modes (\ref{eqn:3.24}), so that their constructive
treatment can
be performed only in certain approximations. The idea of such
approximations
consists in disentangling from the set of $q$ a certain finite subset
which
plays the most important role in the dynamics of the system and
exactly or
approximately decouple from the rest of degrees of freedom. Then
these
distinguished variables, which usually describe the collective
behaviour of the
system as a whole, are treated exactly, while the rest of the modes
are either
supposed to be frozen out or considered perturbatively. A typical
example of
such an approach is a minisuperspace quantum cosmology with spatially
homogeneous modes of the gravitational and matter fields as the only
modes
subject to dynamics and quantization \cite{Halliwell:bibliography}.

Freezing out the dynamical degrees of freedom is not, however, a
rigorous
procedure beyond the tree-level approximation, because in the quantum
domain
all the modes have zero-point fluctuations and, therefore, can
significantly
contribute to the effective dynamics of the collective variables.
That is why,
we consider the approach, when neither of the modes are completely
frozen out,
but the variables complimentary to collective degrees of freedom are
treated
perturbatively. Such a method was developed in mid seventies in the
context of
matter field models \cite{Banks-Bender-Wu} and in recent years was
intensively
applied in classical and quantum cosmology
\cite{Halliwell-Hawking,Brandenberger-M-F,Salopek} for the purpose of
studies
in the theory of the early inflationary Universe and the formation of
the
large-scale cosmological structure. It consists in separating all the
fields
into the macroscopic collective variables $\varphi$ (mainly it is the
so-called
inflaton scalar field) which drive the quasi-DeSitter dynamics of a
spatially
homogeneous cosmological background and a set of all inhomogeneous
modes $f$
describing the particle excitations. In \cite{tunnelIII} we shall
consider in
much detail the application of such a method to the Hawking model of
chaotic
inflation, while here we deal with the general theory of this method.

To begin with, consider the one-loop wavefunction (\ref{eqn:3.18})
with the
argument $q_{+}$ and make its splitting of the above type into the
set of
collective variables $\varphi$ and the rest of degrees of freedom
$f$
    \begin{equation}
    q_{+}=\left[\begin{array}{c}
    \varphi\\f\end{array}\right]=
    \left[\begin{array}{c}\varphi\\0\end{array}\right]
    +\eta,\,\,\eta=
    \left[\begin{array}{c}0\\f\end{array}\right].
\label{eqn:4.1}
    \end{equation}
Here, in general, $\varphi$ can be regarded as subcolumn of $q$ of
finite
dimensionality (like, for example, a finite set of physical variables
responsible for the scale factor and anisotropy parameters in
homogeneous
Bianchi models). On the contrary, $f$ is an infinite-dimensional
vector
(representing, in the same example, all spatially inhomogeneous field
harmonics
on a symmetric background).

Let us suppose that $\phi\,(\tau)$ is a solution of the classical
Euclidean
equations (\ref{eqn:3.19}) - (\ref{eqn:3.20})) corresponding to the
unperturbed
boundary condition at $\tau_{+}, \,q_{+}=(\varphi,0)$ and determined
entirely
by the collective variables $\varphi$. Similarly, we shall denote the
solution
of (\ref{eqn:3.19}) - (\ref{eqn:3.20})) with the perturbed boundary
conditions
(\ref{eqn:4.1}) as $\phi\,(\tau)+\eta\,(\tau)$, whence it follows
that
$\eta\,(\tau)$ satisfies up to quadratic terms the linearized
Euclidean
equations of motion subject to the condition of regularity at
$\tau=0$
    \begin{eqnarray}
     &&\fatg{F}\,(d/d\tau)\,\eta\,(\tau)=O\,
     (\,\eta^{2}),
\label{eqn:4.2}\\
     &&\eta\,(\tau_{+})=
     \left[\begin{array}{c}0\\f\end{array}\right],\,\,
     \eta\,(0)={\rm reg}.
\label{eqn:4.3}
     \end{eqnarray}
Thus the perturbation $\eta\,(\tau_{+})$ of the boundary conditions
$q=q\,(\tau_{+})$ gives rise to to the perturbation $\eta\,(\tau)$ of
the
classical extremal $\phi\,(\tau)$ and the corresponding perturbation
expansion
of the Euclidean action:
     \begin{equation}
     I\,[\,\phi+\eta\,]=I\,[\,\phi\,]+\delta I
     +\frac{1}{2}\,\delta^{2}I+O\,(\,\eta^{3}\,),
\label{eqn:4.4}
     \end{equation}
where the first-order variation of $I$, after integration by parts
with respect
to $\tau$, takes the following form
     \begin{equation}
     \delta I=\int_{0}^{\tau_{+}}d\tau\,
     \frac{\delta I}{\delta \phi\,(\tau)}\,\eta\,(\tau)+
     \left.\frac{\partial{\cal{L}}_{E}}
     {\partial\dot\phi}\,\eta\;\right|_{\tau_{+}}.
\label{eqn:4.5}
     \end{equation}
This integration by parts yields the integral (volume) term
containing the
left-hand side of Euclidean equations of motion and the surface term
-- the
contribution of the boundary at $\tau=\tau_{+}$ and, generally, the
contribution  of the lower integration limit $\tau_{-}=0$. But in
view of the
no-boundary prescription the latter is vanishing because of the
regularity
conditions at the center of the Euclidean spacetime ball -- the zero
value of
the radial coordinate $\tau_{-}=0$. Depending on the type of the mode
of the
field $\phi\,(\tau)$ this takes place either because
$\partial{\cal{L}}_{E}
/\partial\dot\phi(\tau_{-})=0$ (in case of spatially homogeneous
mode) or
because of $\eta\,(\tau_{-})=0$ (for spatially inhomogeneous ones).

The second-order variation in (\ref{eqn:4.4})
     \begin{equation}
     \delta^{2}I=\int_{0}^{\tau_{+}}d\tau\,
     \eta^{T}(\fatg{F}\eta)+
     \left.\eta^{T}(\fatg{W}\eta)\,
     \right|_{\tau_{+}}
\label{eqn:4.6}
     \end{equation}
with $T$ denoting the transposition of the column $\eta=\eta^{i}$,
follows from
varying the equation (\ref{eqn:4.5}) on account of the variational
relations
     \begin{eqnarray}
     \delta\frac{\delta I}{\delta\phi\,(\tau)}=
     \fatg{F}\,(d/d\tau\!)\,\eta(\tau),
\label{eqn:4.7}\\
     \delta\frac{\partial{\cal{L}}_{E}}
     {\partial\dot\phi}
     =\fatg{W}(d/d\tau\!)\,\eta\,(\tau).
\label{eqn:4.8}
     \end{eqnarray}
Here $\fatg{F}=\fatg{F}\,(d/d\tau)$ is, obviously, the differential
operator of
linearized Euclidean equations (\ref{eqn:3.21}) - (\ref{eqn:3.22}),
while
$\fatg{ W}=\fatg{W}(d/d\tau)$ we shall call the {\it Wronskian}
operator which
enters the following relation
	\begin{equation}
	\varphi^{T}_{1}\,(\fatg{F}\varphi_{2}\!)-
	(\fatg{F}\varphi_{1}\!)^{T}\varphi_{2}=
	-\frac{d}{d\tau}\left[\,\varphi^{T}_{1}\,
	(\fatg{W}\varphi_{2}\!)-(\fatg{W}\varphi_{1}\!)^{T}
	\varphi_{2}\,\right]
\label{eqn:6.6}
	\end{equation}
valid for arbitrary test functions $\varphi_{1}$ and $\varphi_{2}$
and usually
used for the construction of the conserved inner product for linear
modes of
$\fatg{F}\,(d/d\tau)$. For $\fatg{F}\,(d/d\tau)$ of the form
(\ref{eqn:3.22})
the Wronskian operator equals
     \begin{equation}
     \fatg{W}(d/d\tau)=a\,\frac{d}{d\tau}+b.
\label{eqn:4.9}
     \end{equation}

\subsection{Basis functions of linearized field modes}
\hspace{\parindent}
The perturbation $\eta\,(\tau)$ satisfies the equations of motion
(\ref{eqn:4.2}) - (\ref{eqn:4.3}) which can be solved by iterations
in
$\eta\,(\tau_{+})=(0,f)$. In the linear approximation this solution
can be
represented in terms of regular basis functions of the Euclidean
"wave"
operator $\fatg{F}\,(d/d\tau)$. They form the full set $\fatg{
u}(\tau)$ of
solutions of the homogeneous differential equation
     \begin{eqnarray}
     \fatg{F}\,(d/d\tau)\,\fatg{u}(\tau)=0,\;\;\;
     \fatg{u}\,(0)={\rm reg},
\label{eqn:4.10}
     \end{eqnarray}
which are regular in the Euclidean spacetime ball
$0\leq\tau\leq\tau_{+}$. But
before discussing the regularity properties of these basis functions,
let us
consider their general spin-tensor structure and the form it takes in
terms of
condensed DeWitt notations.

In view of functional matrix nature of the operator
$\fatg{F}\,(d/d\tau)=\fatg{F}_{ik}\,(d/d\tau)$ its basis functions
also form a
matrix $\fatg{u}(\tau)=\fatg{u}^{\,k}_{\,A}(\tau)$, in which the
condensed
upper index $k$ (acted upon by indices of the matrix operator) labels
the
components of a given basis function including its dependence on
spatial
coordinates, while the lower condensed index $A$ enumerates the basis
functions
themselves. The infinite ranges and the (discrete or continuous)
nature of
these indices $k$ and $A$ can be different depending on the
parametrization of
basic physical variables (\ref{eqn:3.24}) and their possible
decomposition in
spatial harmonics. What is, however, in common to all field
parametrizations
is that there is a one to one map between the sets $\{k\}$ and
$\{A\}$, so that
the matrix $\fatg{u}^{\,k}_{\,A}$ can be regarded as non-symmetric
but {\it
quadratic} and {\it invertible} matrix (parametrically depending on
$\tau$)
having with respect to its infinite-dimensional indices the inverse
$\fatg{u}^{-1}(\tau)=(\fatg{u}^{-1} )^{\,A}_{\,i}$:
      \begin{equation}
      \fatg{u}^{i}_{A}\,(\fatg{u}^{-1}\!)^{\,A}_{\,k}=
      \delta^{i}_{k}\,.
\label{eqn:4.12}
      \end{equation}

To illustrate the use of condensed DeWitt indices in the functional
matrix
$\fatg{u}(\tau)$ of the above type, let us consider a simple example
of a
scalar field $q^{i}=\phi\,({\bf x})$ in flat spacetime, when the
condensed
index does not contain any discrete labels and reduces to the set of
continuous
spatial coordinates $i={\bf x}$. The linear equation of motion
(\ref{eqn:4.10})
in this case is the Euclidean Klein-Gordon equation which has, as a
set of
basis functions regular at past infinity $\tau\rightarrow -\infty$,
the
following solutions
     \begin{eqnarray}
     &&\fatg{u}^{i}_{A}\,(\tau)\equiv
     \fatg{u}_{\bf k}({\bf x},\tau),\,\,\,\,
     i={\bf x},A={\bf k},
\label{eqn:4.13}\\
     &&\fatg{u}_{\bf k}({\bf x},\tau)=
     {e}^{\,\omega({\bf k}\!)\,\tau+
     i{\bf k}{\bf x}},\,\,\,\,
     \omega\,({\bf k})=\sqrt{\,{\bf k}^{2}+m^{2}},
\label{eqn:4.14}
     \end{eqnarray}
enumerated by the continuous set of spatial momentum vectors ${\bf
k}$. Every
square-integrable function of spatial coordinates $h^{i}=h({\bf x})$
can be
decomposed in plane waves of the above type in the form
     \begin{equation}
     h\,({\bf x})=\int d^{3}{\bf k}\,
     {\rm e}^{\,\omega\,({\bf k}\!)\,\tau+
     i{\bf k}{\bf x}}\,h_{\bf k},
\label{eqn:4.15}
     \end{equation}
which can be rewritten in condensed notations as
     \begin{equation}
     h^{i}=\fatg{u}^{i}_{A}h^{A},\,\,\,
     h^{A}\equiv h_{\bf k}
\label{eqn:4.16}
     \end{equation}
(we omit for brevity the time argument here, for it enters this
transformation
only as a parameter). This means that this equation provides a linear
one to
one map between $h^{i}$ and $h^{A}$. The transformation inverse to
(\ref{eqn:4.15})
      \begin{equation}
      h_{\bf k}=\frac{1}{(2\pi)^{3}}\int d^{3}{\bf x}\,
      {\rm e}^{-\omega\,({\bf k}\!)\tau-
      i{\bf k}{\bf x}}\,h\,({\bf x}),
\label{eqn:4.17}
      \end{equation}
in condensed indices has a simple form
      \begin{equation}
      h^{A}=(\fatg{u}^{-1}\!)^{\,A}_{\,i}\,h^{i}
\label{eqn:4.18}
      \end{equation}
where $(\fatg{u}^{-1}\!)^{\,A}_{\,i}$ denotes the inverse of
$\fatg{u}^{i}_{A}$
(\ref{eqn:4.12}) with a kernel given by eq.(\ref{eqn:4.17}).

It is also possible to decompose a scalar field in spherical-wave
basis
functions of the Klein-Gordon equation, enumerated instead of a
spatial
momentum vector by its continuous norm $k=|{\bf k}|$ and discrete
orbital
$l=0,1,2,...$ and azimuthal $m,\, -l\leq m\leq l,$ quantum numbers,
in which
case the condensed label $A=(k,l,m)$ will be of mixed
continuous-discrete
nature. In spatially closed cosmology the spatial section of
spacetime is
compact and, therefore, the corresponding set of spatial harmonics in
the
decomposition (\ref{eqn:4.16}) is discrete and countable. For a
general set of
linearized physical fields (\ref{eqn:3.24}), the condensed index $A$
of basis
functions includes three discrete quantum numbers and the
corresponding spin
label $A=(n,\,l,\,m,\,{\rm spin})$ which are again in one to one
correspondence
with $i=({\bf x},\,a T,\,ab TT,...)$, and so on. But the above
peculiarities of
"fine" structure of various field models can always be encoded in
universal
relations (\ref{eqn:4.12}), (\ref{eqn:4.16}) and (\ref{eqn:4.18})
written in
DeWitt's notations which we shall imply throughout the paper.

In view of the general decomposition (\ref{eqn:4.1}) of $q^{i}$, the
full set
of basis functions $\fatg{u}\,(\tau)$ contains the modes of both the
collective
variable $\varphi$ and the rest of degrees of freedom $f$. Generally
the
collective variables of the system interact very nonlinearly with its
microscopic modes. But physically the decomposition (\ref{eqn:4.1})
makes sense
when they decouple at least in the linearized approximation, which
means that
in the basis (\ref{eqn:4.1}) of $\varphi$ and $f$ the differential
operator
$\fatg{F}\,(d/d\tau)$ has a block-diagonal structure
     \\
     \begin{equation}
     \fatg{F}\,(d/d\tau)=\left[\!\begin{array}{cc}
      F_{\varphi}\,(d/d\tau)&\!\!\!\!0\\
      0&\!\!\!\!F\,(d/d\tau)\end{array}\!\right],
\label{eqn:4.19}
     \end{equation}
     \\
with $F_{\varphi}\,(d/d\tau)$ and $F\,(d/d\tau)$ acting respectively
in
subspaces of $\varphi$ and $f$. This structure of
$\fatg{F}\,(d/d\tau)$ has a
simple illustration in the case when collective variables $\varphi$
represent a
spatially homogeneous background for inhomogeneous modes $f$. On such
symmetric
background the variables $f$ are decomposed into series of spatially
inhomogeneous harmonics which are orthogonal to the homogeneous
linear modes of
$\varphi$: their bilinear combinations give zero in the cross
$\varphi\!-\!f$
terms when integrated over a compact spatial section in the quadratic
part of
the Euclidean action (\ref{eqn:4.6}) and, thus, provide the
block-diagonal form
(\ref{eqn:4.19}).

The block-diagonal structure (\ref{eqn:4.19}) implies a similar form
of all the
matrix coefficients of the operator $\fatg{F}\,(d/d\tau)$, its
Wronskian
operator (\ref{eqn:4.9})
      \begin{equation}
      \fatg{W}(d/d\tau)=\left[\!\begin{array}{cc}
      W_{\varphi}\,(d/d\tau)&\!\!\!\!0\\
      0&\!\!\!\!W(d/d\tau)\end{array}\!\right]
\label{eqn:4.20}
      \end{equation}
and also allows one to choose the matrix $\fatg{u}\,(\tau)$ in the
block-diagonal form with the basis functions $u_{\varphi}(\tau)$ and
$u(\tau)$
of the linearized modes of $\varphi$ and $f$ respectively
      \begin{eqnarray}
      &&\fatg{u}\,(\tau)=\left[\!\begin{array}{cc}
      u_{\varphi}\,(\tau)&\!\!\!\!0\\
      0&\!\!\!\!u\,(\tau)
      \end{array}\!\right],\\
\label{eqn:4.21}
      \nonumber\\
      &&F_{\varphi}\,(d/d\tau\!)\,u_{\varphi}\,(\tau)=0,\,\,\,
      F\,(d/d\tau\!)\,u\,(\tau)=0.
\label{eqn:4.22}
      \end{eqnarray}

Finally let us consider the conditions which select the regular basis
functions
of the operator $\fatg{F}$. Due to the no-boundary nature of the
underlying
manifold $\fatg{M}$, its point of vanishing coordinate radius
$\tau=\tau_{-}\equiv 0$ is a singular point of the radial part of
$\fatg{F}(d/d\tau)$. Indeed, for physical fields (\ref{eqn:3.24}) of
all
possible spins, $s=0,1/2,1,3/2,2,...$, the coefficient $a=a_{ik}$ in
$\fatg{F}\,(d/d\tau\!)$ (the Euclidean version of eq.(\ref{eqn:3.4}))
can be
collectively written as
	\begin{eqnarray}
	a_{ik}=(^{4}g)^{1/2}g^{\tau\tau}g^{a_{1}a_{2}}...
	g^{a_{s}a_{2s}}\,
	\delta\,({\bf x}_{i}-{\bf x}_{k}),
\label{eqn:6.2.1}
	\end{eqnarray}
$i=(a_{1},...a_{s},{\bf x}_{i}),\,k=(a_{2},...a_{2s},{\bf x}_{k}),$
and in the
regular metric (\ref{eqn:2.20}) has the following behaviour
	\begin{equation}
	a=a_{0}\,\tau^{k}+O\,(\,\tau^{k+1}\,),\,\,
	k=3-2s,\,\,\tau\rightarrow\tau_{-}=0,
\label{eqn:6.2.2}
	\end{equation}
where $a_{0}$ is defined by eq.(\ref{eqn:6.2.1}) with respect to the
round
metric $c_{ab}$ on a 3-sphere of the unit radius and the unit lapse
$g^{\tau\tau}=N^{-2}=1$
\footnote
{Even though the expression (\ref{eqn:6.2.1}) is formally valid only
for
integer-spin fields, this behaviour with the parameter $k=3-2s$ also
holds for
half-integer spins, because every  gamma matrix substituting the
corresponding
metric coefficient in (\ref{eqn:6.2.1}) contributes one power of
$\tau^{-1}$.}
{}.
Therefore the equations (\ref{eqn:4.10}) for basis functions have the
form
	\begin{equation}
	\left(\frac{d^2}{d\tau^2}+f\frac{d}{d\tau}+
	g\right)\fatg{u}(\tau)=0,                   \label{eqn:6.2.3}
	\end{equation}
with the coefficients $f$ and $g$ having the following asymptotic
behaviour
	\begin{equation}
	f=\frac{k}{\tau}\,\fatg{I}+
	O\,(\,\tau^{0}\,),\,\,\,
	g=\frac{g_{\,0}}{\tau^2}+
	O\,(\,\tau^{-1}\,).                   \label{eqn:6.2.4}
	\end{equation}
Here the leading singularity in the potential term $g$ originates
from the
spatial Laplacian $g^{ab}\nabla_{a}\nabla_{b}$ entering the operator
$\fatg{F}$, which scales in the metric (\ref{eqn:2.20}) as
$1/\tau^{2}$, and
the leading term of $f$ is always a multiple of the unity matrix
\fatg{I} with
the same parameter $k=3-2s$ as in (\ref{eqn:6.2.2}). In the
representation of
spatial harmonics, the eigenfunctions of a spatial Laplacian, the
(functional)
matrix $g_{\,0}$ can be also diagonalized, $g_{\,0}={\rm
diag}\{-\omega^2_{i}\}$ , so that, without the loss of generality,
the both
singularities in (\ref{eqn:6.2.3}) can be characterised by simple
numbers $k$
and $\omega^2=\omega^2_{i}$ for every component of
$\fatg{u}=\fatg{u}^{i}$.

As it follows from the theory of differential equations with singular
points
\cite{Olver}, in this case there are two types of solutions
$\fatg{u}_{-}(\tau)$ and $\fatg{u}_{+}(\tau)$ differing by their
behaviour near
$\tau_{-}=0$:
	\begin{eqnarray}
	&&\fatg{F}\fatg{u_{\pm}}=0,              \label{eqn:6.9}\\
	&&\fatg{u}_{-}(\tau)=\fatg{U}_{\!-}\,
	\tau^{\mu_{-}}
	+O\,(\,\tau^{1+\mu_{-}}),
\label{eqn:6.2.5}\\
	&&\fatg{u}_{+}(\tau)=\fatg{V}_{\!+}\,
	\tau^{\mu_{+}}
	+O\,(\;\tau^{1+\mu_{+}}),
\label{eqn:6.2.6}
	\end{eqnarray}
where $\mu_{\pm}$ are the roots of the quadratic equation involving
only the
coefficients of leading singularities
$\mu^{2}+(k-1)\,\mu-\omega^{2}=0$.
    In view of non-negativity of $\omega^2$ (the eigenvalue of
$-c^{ab}\nabla_{a} \nabla_{b}$) these roots are of opposite signs,
$\mu_{-}\mu_{+}= -\omega^2\leq0$, and we can choose $\mu_{-}$ to be
non-negative in order to have $\fatg{u}(\tau)=\fatg{u}_{-}(\tau)$ as
a set of
regular basis functions at $\tau_{-}=0$, the remaining part of them
$\fatg{u}_{+}(\tau)$ beeing singular. By our assumption the operator
$\fatg{F}$
does not have zero eigenvalues on the Euclidean spacetime of the
no-boundary
type (otherwise, its functional determinant and the one-loop
prefactor of the
wavefunction are not defined). Therefore, there are no basis
functions which
are simultaneously regular at $\tau_{-}=0$ and vanishing for positive
$\tau\leq\tau_{+}$, and their matrix can be considered invertible
everywhere in
this range of $\tau$ except the origin $\tau_{-}=0$
\footnote
{This property holds until the first caustic of solutions of
classical
equations for {\it physical} variables at which {\normalsize
$\fatg{u}_{-}(\tau)=0$} (because {\normalsize $\fatg{u}_{-}(\tau)$}
is a
derivative of the parametric family of these solutions with respect
to their
parameter, this relation is just an equation for their enveloping
curve).
However, this is not the caustic of Einstein equations in {\it
superspace} of
the theory, responsible for the transition from the Euclidean to the
Lorentzian
regime. As it follows from the Hawking model of chaotic inflation
\cite{tunnelIII}, the latter is generated by zeros of the
Faddeev-Popov
determinant of the ADM reduction procedure, rather than by zeros of
{\normalsize $\fatg{u}_{-}(\tau)$}. Degeneration of the Faddeev-Popov
matrix
indicates the presence of Gribov copies in the quantization procedure
and
serves as a strong motivation for the third quantization of gravity
\cite{BarvU,BKr,Operd}, therefore, as it could have been expected on
physical
ground, the Euclidean-Lorentzian tunnelling phenomena are directly
related to
the physics of baby-universe production.
}.
In what follows we shall denote the regular basis functions either by
$\fatg{u}(\tau)$ or by $\fatg{u}_{-}(\tau)$, when we prefer to
emphasize their
regularity in the "center" of the Euclidean ball $\tau_{-}\equiv 0$.

\subsection{Perturbation theory in microscopic variables and $\;\;\;$
$\hbar$-expansion}
\hspace{\parindent}
The basis functions of the above type will serve us as a technical
tool for two
purposes: the perturbation theory in microscopic variables $f$ and
the
reduction method for the functional determinants in the one-loop
prefactor. Let
us here, first, develope this perturbation theory in powers of
$\eta\,(\tau)=O\,(f)$ and show how it actually reduces to the
expansion in
$\hbar$.

To begin with, note that in virtue of the invertibility of
$\fatg{u}_{-}(\tau)$ the linearized solution of the boundary-value
problem
(\ref{eqn:4.2}) - (\ref{eqn:4.3}) has the form:
      \begin{equation}
      \eta\,(\tau)=\fatg{u}(\tau\!)\,
      \fatg{u}^{-1}(\tau_{+}\!)\,\eta\,(\tau_{+})+
      O\,(\,\eta^3\,),                    \label{eqn:4.23}
      \end{equation}
Here we suppress the indices of functional matrices
$\fatg{u}(\tau)=\fatg{
u}^{\,i}_{\,A}(\tau),\,\, \fatg{u}^{-1}(\tau_{+})=[\,\fatg{u}^{-1}
(\tau_{+}\!)\,]^{\,A}_{\,i}$ and columns $\eta(\tau)=\eta^{i}(\tau)$
implying
again the DeWitt rule of summation-integration over supercondensed
labels.
Substituting this solution into the linear and quadratic terms of the
perturbed
Euclidean action (\ref{eqn:4.4}) one can see that the volume
contributions
vanish in the quadratic approximation due to the background $\delta
I/\delta\phi(\tau)=0$ and linearized (\ref{eqn:4.2}) equations of
motion. The
remainig surface terms at the boundary $\tau=\tau_{+}$ give
      \begin{eqnarray}
      I\,[\,\phi+\eta\,]=I\,[\,\phi\,]\,+
      \,\left[\,\frac{\partial{\cal{L}}_{E}}
      {\partial\dot\phi}\,\eta\,+\,
      \frac{1}{2}\eta^{T}\,(\fatg{W}\!
      \fatg{u})\,\fatg{u}^{-1}\eta\,\right]
      _{\tau_{+}}
      +\,O\,(\,\eta^{3}\,).                    \label{eqn:4.24}
      \end{eqnarray}
In view of the form of the boundary-value perturbation
(\ref{eqn:4.3}) only the
$f$-component of the Euclidean momentum contributes to the right-hand
side of
this equation, but since it is computed at the field background
$\phi(\tau)=(\varphi(\tau),0)$ with identically vanishing $f(\tau)$
this
momentum component $\partial{\cal{L}}_{E}/\partial\dot f(\tau_{+})$
also
vanishes and only the quadratic form in $\eta$ survives in
(\ref{eqn:4.24}).
This form in its turn can be completely  rewritten in terms of the
quantities
on the $f$-subspace due to (\ref{eqn:4.3}) and the block-diagonal
structure of
basis functions and the Wronskian operator:
       \begin{eqnarray}
       &&I\,[\,\phi+\eta\,]=I\,[\,\phi\,]+
       \frac{1}{2}f^{T}D\,(\tau_{+}\!)\,f
       +O\,(\,f^{3}),
\label{eqn:4.25}
       \\
       &&D\,(\tau_{+}\!)=[\, W(d/d\tau_{+}\!)\,
       u\,(\tau_{+}\!)\,]\,\,u^{-1}(\tau_{+}\!).
\label{eqn:4.26}
       \end{eqnarray}
This expression can be used in the equation (\ref{eqn:3.18}) for the
wavefunction together with the preexponential factor $(\,{\rm
Det}\,\fatg{
F}/{\rm Det}\,\fatg{a}\,)^{-1/2}$ calculated at the perturbed
classical
background $\phi(\tau)+\eta(\tau),\, \eta(\tau)=O\,(\,f\,),$
\\
        \begin{equation}
	\left(\,\frac{{\rm
	Det}\,\fatg{F}\,[\,\phi+\eta\,]}
	{{\rm Det}\,\fatg{a}\,[\,\phi+\eta\,]}\,
	\right)^{-1/2}=
	\left(\,\frac{{\rm Det}\,\fatg{F}\,
	[\,\phi\,]}{{\rm Det}\,\fatg{a}\,
	[\,\phi\,]}\,\right)^{-1/2}+O\,(\,f\,).   \label{eqn:4.27}
	\end{equation}

Our purpose now will be to show that respectively cubic
$O\,(\,f^{3})$ and
linear $O\,(\,f\,)$ corrections in eqs.(\ref{eqn:4.25}) and
(\ref{eqn:4.27})
give the contributions to the wavefunction of one and the same order
$O\,(\,\hbar^{1/2}\,)$, belonging to the two-loop approximation of a
semiclassical expansion. Substituting (\ref{eqn:4.25}) and
(\ref{eqn:4.27})
into (\ref{eqn:3.18}) and reexpanding the exponential in terms of
$O\,(\,f^{3}\,)$ we get
       	\begin{eqnarray}
	\Psi\,(q_{+},\tau_{+}\!)=\left(\,\frac{{\rm
	Det}\,\fatg{F}\,[\,\phi\,]}{{\rm
	Det}\,\fatg{a}\,[\,\phi\,]}\,\right)^{-1/2}\!
	{\rm exp}\left\{-\frac{1}{\hbar}I\,[\,\phi\,]-
        \frac{1}{2\hbar}f^{T}D(\tau_{+}\!)\,f \right\}
	\nonumber \\ \nonumber\\
	\times\,\left[\;1+O\,(\,f\,)+
	O\, (\,f^{3}/\hbar\, )\;\right].          \label{eqn:4.28}
        \end{eqnarray}

Thus, up to corrections of the above type the wavefunction is a
gaussian state
of microscopic variables $f$. The gaussian exponent suppresses the
states with
large $f$, because of the positive definiteness of the matrix
$D\,(\tau_{+}\!)$, which is a direct corollary of the boundedness of
the
Euclidean action from below. If this property is satisfied and the
extremal
$\phi\,(\tau)$ realizes at least a local minimum of the action, then
its
quadratic perturbation (\ref{eqn:4.6}) is positive-definite for {\it
arbitrary}
$\eta\,(\tau),\,\delta^{2}I>0$. On solutions of linearized equations
of motion
with the boundary data (\ref{eqn:4.3}) specified by $f$ it reduces to
        \begin{equation}
	\delta^{2}I=f^{T}D\,(\tau_{+}\!)\,f>0
\label{eqn:4.29}
	\end{equation}
and, thus, provides the negative definiteness of the quadratic form
in the
exponential of (\ref{eqn:4.28}). Therefore one can use the asymptotic
bound
        \\
        \begin{equation}
	{\large{\rm e}}^{\!\!\phantom {0}^{\textstyle
	-\frac{1}{2\hbar}f^{T}\!Df}}f^{n}=
	O\,(\,\hbar^{n/2}),\;\;\;
	\hbar\rightarrow 0,
\label{eqn:4.30}
	\end{equation}
which can be rigorously proved under certain assumptions of uniform
regularity
of $D$ for a wide class of positive definite quadratic functionals on
the space
of functions $f$ of many variables \cite{Maslov}. In view of this
bound the
linear $O\,(\,f\,)$ and cubic $O\,(\,f^{3}/\hbar\,)$ corrections in
(\ref{eqn:4.28}) are both $O\,(\hbar^{1/2})$, and therefore go beyond
the
one-loop approximation considered in this paper. Actually, such terms
of half
integer power in $\hbar$ belong to the two-loop approximation
$O\,(\hbar)$
because in quantum averages with the gaussian quantum state only even
powers of
$f$ will give a nonvanishing contribution and, therefore, the
corrections
$O\,(\hbar^{1/2})$ acquire at least one extra power of $\hbar^{1/2}$.

\subsection{The basis functions algorithm for the one-loop
preexponential
factor}
\hspace{\parindent}
As is shown in \cite{tunnelII} the regular basis functions
$\fatg{u}_{-}(\tau)$
can be used for the calculation of the one-loop preexponential factor
of the
wavefunction (\ref{eqn:4.28}). The nature of this procedure consists
in the
reduction which allows to obtain the functional determinant ${\rm
Det}\,\fatg{F}$ in terms of the quantity of the lower functional
dimensionality
-- the determinant of the non-degenerate matrix of regular basis
functions
$\fatg{u}^{i}_{A}(\tau)$ taken with respect to its condensed indices
\footnote
{This reduction method is actually a particular case of the Pauli-Van
Vleck-Morette formula \cite{Morette} for the one-loop kernel of the
heat
equation in the proper-time interval $[\tau_{-},\tau_{+}]$, adjusted
to the
case of the no-boundary type, when $\tau_{-}$ is a singular point of
the
dynamical equations having a special behaviour (\ref{eqn:6.2.5}) of
the
linearized modes \cite{tunnelII}. The corresponding algorithm
(\ref{eqn:6.2.11}) was also obtained in the authors' paper \cite{BKK}
by a
special technique of the $\zeta$-functional regularization for
operators with
the explicitly unknown spectra on manifolds with a boundary.
}.
These basis functions have the behaviour (\ref{eqn:6.2.5}) and are
defined up
to linear $\tau$-independent recombinations. The latter can be used
to make the
coefficient $\fatg{U}_{-}$ completely independent of the background
fields
$\phi$ on $\fatg{M}$ and, without loss of generality, equal the
functional
matrix unity $\fatg{I}$. Then this algorithm for a one-loop prefactor
takes the
form \cite{tunnelII}
	\begin{eqnarray}
	&&\left(\,\frac{{\rm Det}\,\fatg{F}}{{\rm Det}\,
	\fatg{a}}\,\right)^{-1/2}=
	{\rm Const}\,\left[\;{\rm det}\,
	\fatg{u}_{-}(\tau_{+}\!)\;\right]^{-1/2},
\label{eqn:6.2.11}\\
	\nonumber\\
	&&\,\,\fatg{u}_{-}(\tau)=
	\fatg{I}\;\tau^{\mu_{-}}+
	O\,(\,\tau^{1+\mu_{-}}\!),\;\;\;\tau\rightarrow
0.\label{eqn:6.2.12}
	\end{eqnarray}

Thus, combining eqs.(\ref{eqn:4.28}) and (\ref{eqn:4.30}) with this
reduction
algorithm, we get the one-loop Euclidean wavefunction
$\Psi\,(q_{+},\tau_{+}\!)= \Psi\,(\varphi,f,\,\tau_{+}\!)$ of
physical
variables $q_{+}= (\varphi,\,f)$ in the form
        \begin{eqnarray}
	&&\!\!\!\!\!\!\!\!\!\!\!\!\!\!\!\!\!\!\!\!
	\Psi\,(\varphi,f,\tau_{+}\!)=
	\left.\Psi_{[\,\phi\,]}\,(f,\tau_{+}\!)
	\;\right|_{\,\phi=
	\phi\,(\tau|\,\varphi,\,\tau_{+}\!)},\\
\label{eqn:4.31}
	\nonumber \\
        &&\!\!\!\!\!\!\!\!\!\!\!\!\!\!\!\!\!\!\!\!
	\Psi_{[\,\phi\,]}\,(f,\tau_{+}\!)\equiv
	{\rm Const}\,\left(\,{\rm det}\,
	\fatg{u}_{-}(\tau_{+}\!)_{\,[\,\phi\,]}\right)^{-1/2}\!
	{\rm exp}\left \{-\frac{1}
	{\hbar}\,I\,[\,\phi\,]-
	\frac{1}{2\hbar}f^{T}D\,(\tau_{+}\!)\,f\right \}
	\nonumber\\
        &&\;\;\;\;\;\;\;\;\;\;\;\;\;\;\;\;\;\;\;\;\;
	\;\;\;\;\;\;\;\;\;\;\;\;\;\;
	\;\;\;\;\;\;\;\;\;\;\;\;\;\;\;\;\;\;\;\;\;
	\;\;\;\;\;\times\left [\,\,1+
	O\,(\hbar^{1/2})\,\,\right].
\label{eqn:4.32}
        \end{eqnarray}

In the equation (\ref{eqn:4.31}) we clearly separated the dependence
of the
wavefunction on the collective variables from that on the microscopic
ones. In
contrast to a simple quadratic dependence on $f$, the variables
$\varphi$ enter
$\Psi(\varphi,f,\,\tau_{+})$ through the functional argument
$\phi(\tau)$ of
$\Psi_{[\,\phi\,]}\,(f,\tau_{+}\!)$, for they parametrize the
extremal
$\phi(\tau)= \phi(\tau |\,\varphi,\tau_{+}\!)$ of the Euclidean
equations of
motion (\ref{eqn:3.19})-(\ref{eqn:3.20}) with the boundary data
$q_{+}=(\varphi,0)$. The argument $\phi$, in its turn, enters the the
auxiliary
functional (\ref{eqn:4.32}) through the Euclidean action
$I\,[\,\phi\,]$,
one-loop preexponential factor $\left(\,{\rm det}\,
\fatg{u}_{-}(\tau_{+}\!)_{\,[\,\phi\,]}\right)^{-1/2}$ (subscript
$[\,\phi\,]$
indicating the functional dependence on the field background) and the
matrix of
quantum dispersions $D\,(\tau_{+}\!)=
D\,(\tau_{+}\!)_{\,[\,\phi\,]}$.

\section{The method of complex extremals}
\hspace{\parindent}
According to the discussion of Sect.3 the "Lorentzian" wavefunction
can be
obtained from the one-loop expression (\ref{eqn:4.31}) above by the
analytic
continuation (\ref{eqn:3.17}) into the complex plane of the Euclidean
time:\vspace{1mm}
        \begin{eqnarray}
	\Psi_{L}\,(\varphi,f,t_{+}\!)=
	\left.\Psi_{[\,\Phi\,]}\,(f,z_{+}\!)
	\;\right|_{\,\Phi=
	\Phi\,(z|\,\varphi,\,z_{+}\!)}.
\label{eqn:5.1}
	\end{eqnarray}
\vspace{-7mm}

\noindent When this analytic continuation proceeds along the contour
$C_{+}=C_{E}\cup C_{L}$ of Sect.2, the extremal field
$\Phi(z)\,|_{\!\!\!\!
\textstyle \phantom{0}_{C_{+}}}=\Phi(z\,|\,\varphi,z_{+}\!)$ on its
Euclidean
and Lorentzian segments will not generally represent real functions
$\phi(\tau)$ and $q(t)$. The class of models in which all the fields
can be
real on the analytically matched Lorentzian and Euclidean sections of
spacetime
is very limited and comprises the so-called real tunnelling
geometries
\cite{Gibbons-Hartle}. In this section we present the semiclassical
technique
that allows to handle the general case of complex extremals by
reducuing the
formalism to real-valued solutions of both Euclidean and Lorentzian
equations
of motion.

\subsection{Matching conditions between the Euclidean and Loren-
tzian
spacetimes}
\hspace{\parindent}
To begin with, we shall reserve the notations $q\,(t)$ and
$\phi\,(\tau)$ for
real parts of the complex field $\Phi\,(z)$ on respectively
Lorentzian and
Euclidean segments of the contour $C_{+}$, denote their imaginary
parts by
$h(t)$ and $\eta(\tau)$ and also introduce the notation $Q\,(t)$ for
the full
complex field on $C_{L}$:
         \begin{eqnarray}
	 &&\Phi(\tau)=\phi\,(\tau)+i\eta\,(\tau),\;\;
	 Q\,(t)=q\,(t)+ih\,(t),
\label{eqn:5.2}\\
	 &&Q(t)\equiv\Phi\,(z)\,|_{\!\!\!\!
	 \textstyle\phantom{0}_{C_{L}}}=\Phi\,(\tau_{B}+it).
\label{eqn:5.3}
	 \end{eqnarray}
Then the complex action (\ref{eqn:3.7}) on this contour takes the
following
form
         \begin{equation}
	 {\cal{I}}\,[\,\Phi(z)\,]\,
	 |_{\!\!\!\!\textstyle\phantom{0}_{C_{+}}}=
	 I\,[\,\Phi(\tau)\,]-iS\,[\,Q(t)\,],
\label{eqn:5.4}
	 \end{equation}
where $I\,[\,\Phi(\tau)\,]$ and $S\,[\,Q(t)\,]$ are the Euclidean and
Lorentzian actions (\ref{eqn:3.11}) and (\ref{eqn:3.1}) as functions
of their
complex functional arguments.

Let us now consider the variational principle for this
Lorentzian-Euclidean
action which gives the saddle point of the path integral
(\ref{eqn:3.13}) for
the no-boundary wavefunction. The first-order variation of
(\ref{eqn:5.4}) can
be obtained by using the equation (\ref{eqn:4.5}) and its analogue
for the
Lorentzian action
     \begin{equation}
     \delta{\cal I}=\int_{0}^{\tau_{+}}d\tau\,
     \frac{\delta I}{\delta \Phi}\,\delta\Phi+
     \left.\frac{\partial{\cal{L}}_{E}}
     {\partial\dot\Phi}\,\delta\Phi\,\right|_{\,\tau_{B}}
     -i\int_{0}^{t_{+}}dt\,
     \frac{\delta S}{\delta Q}\,\delta Q
     +i\left.\frac{\partial{\cal{L}}}
     {\partial\dot Q}\,\delta Q\,\right|_{\,t=0}.     \label{eqn:5.5}
     \end{equation}
\vspace{-6mm}

\noindent The fields and their variations satisfy the no-boundary
regularity
conditions at $\tau=0$ and fixed boundary conditions (\ref{eqn:3.15})
at
$t=t_{+}$, $Q(t_{+})=q_{+}$. As a result, $\delta Q(t_{+})=0$ and the
total-derivative term vanishes at the boundary of spacetime $t=t_{+}$
as well
as in its regular center $\tau=0$. Moreover, in view of the analytic
continuation (\ref{eqn:5.3}), the Euclidean and Lorentzian fields
satisfy the
matching conditions
      \begin{equation}
      \Phi\,(\tau_{B})=Q\,(0),
\label{eqn:5.6}
      \end{equation}
which means that $\delta\Phi(\tau_{B})=\delta Q(0)$. Therefore,
equating to
zero separately the volume and surface terms of (\ref{eqn:5.5}) in
the
variational equation $\delta{\cal I}=0$, one can get the system of
Euclidean
and Lorentzian equations of motion
      \begin{equation}
      \frac{\delta I}{\delta \Phi}=0,\;\;
      \frac{\delta S}{\delta Q}=0                     \label{eqn:5.7}
      \end{equation}
for fields subject to special matching conditions at the nucleation
point
$\tau=\tau_{B}\;(t=0)$
     \begin{equation}
     \left.\frac{\partial{\cal{L}}_{E}}
     {\partial\dot\Phi}\,\right|_{\,\tau_{+}}
     \!\!+\,i\left.\frac{\partial{\cal{L}}}
     {\partial\dot Q}\,\right|_{\,t=0}=0.
\label{eqn:5.8}
     \end{equation}

These matching conditions show that the tunnelling geometries with
real
physical fields exist only in case when both the Euclidean
$\partial{\cal{L}}_{E}/\partial\dot\Phi$ and Lorentzian
$\partial{\cal{L}}/
\partial\dot Q$ momenta separately vanish at the nucleation point.
The
covariant version of this statement in the gravitational sector of
all fields
sounds as a vanishing of the extrinsic curvature $K_{ab}$ of the
nucleation
surface \cite{Hal-Hartle,Gibbons-Hartle,Gibbons-Pohle}. This
corresponds to the
fact that the canonical gravitational momentum $\fatg{p}$ is linear
in the
second fundamental form of a hypersurface in the slicing of spacetime
associated with the Hamiltonian formalism of the theory, and the
above surface
of the Euclidean-Lorentzian transition is supposed to be a member of
such a
slicing. In the example of the DeSitter Universe generated by the
inert
cosmological constant this surface coincides with the equator of the
Euclidean
four-dimensional sphere at which the time derivative of the scale
factor
(\ref{eqn:1.4}) vanishes when approaching both from inside the
Euclidean
hemisphere and from the inflationary Lorentzian spacetime .

In the general case, when the momenta are nonzero at $\tau=\tau_{B}$,
the
fields $\Phi\,(\tau)$ and $Q\,(t)$ become complex, and the very
notion of the
Euclidean-Lorentzian transition becomes questionable, because complex
physical
fields generate complex-valued metric tensors which can hardly be
ascribed to
spacetimes of either Euclidean or Lorentzian signature. We shall
show, however,
that the Euclidean-Lorentzian decomposition of the full quantum
dynamics still
makes sense within the $\hbar$-expansion. To show this we shall
develope the
perturbation expansion in the imaginary parts $\varepsilon=(\eta,h)$
of the
complex fields (\ref{eqn:5.2}) and demonstrate that it corresponds to
the
asymptotic expansion in $\hbar^{1/2}$.

\subsection{Perturbation theory in the imaginary corrections and the
$\hbar$-expansion}
\hspace{\parindent}
This perturbation expansion begins with substituting the expressions
(\ref{eqn:5.2}) into the classical equations and the matching
conditions
(\ref{eqn:5.6}) - (\ref{eqn:5.8}) and expanding the result in powers
of
$\varepsilon=(\eta,h)$. The separation of the real and imaginary
parts in
(\ref{eqn:5.7}) then immeadiately leads to to the following system of
equations
for the real part of the Euclidean extremal and its imaginary part
treated as a
perturbation
       \begin{equation}
       \frac{\delta I\,[\,\phi\,]}
       {\delta\phi\,(\tau)}=O\,(\,\eta^{2}),\;\;
       \fatg{F}(d/d\tau\!)
       \,\eta(\tau)=O\,(\,\eta^{3}).
\label{eqn:5.9}
       \end{equation}
Similar equations hold for the Lorentzian fields
       \begin{equation}
       \frac{\delta S\,[\,q\,]}
       {\delta q\,(t)}=O\,(\,h^{2}),\;\;
       \fatg{F}_{\!L}(d/dt)
       \,h(t)=O\,(\,h^{3}),
\label{eqn:5.10}
       \end{equation}
where $\fatg{F}_{\!L}(d/dt)$ is the Lorentzian wave operator of
linearized
equations (at the background of $q(t)$), analogous to its Euclidean
version
(\ref{eqn:3.21})
        \begin{equation}
	\fatg{F}_{\!L}\equiv\fatg{F}_{\!L}(d/dt)\,
        \delta(t-t^{\prime})=
        \frac{\delta^{2}S\,[\,q\,]}
        {\delta q(t)\,
        \delta q(t^{\prime})}\,.
\label{eqn:5.11}
        \end{equation}

The expansion of the matching condition (\ref{eqn:5.8}) up to linear
terms in
$\varepsilon$ can be performed with the aid of the variational
equation
(\ref{eqn:4.8}) and its Lorentzian version
         \begin{equation}
	 \delta\,\frac{\partial{\cal{L}}_{L}}
	 {\partial\dot q}
         =\fatg{W}_{\!L}(d/dt)\;\delta q(t),
\label{eqn:5.12}
	 \end{equation}
which serves as a definition of the{\it Wronskian} operator
$\fatg{W}_{\!L}=
\fatg{W}_{\!L}(d/dt)$ for (\ref{eqn:5.11}). The separation of the
real and
imaginary parts of eq.(\ref{eqn:5.8}) then yields the system of
matching
conditions coupling the dynamics of Euclidean and Lorentzian
variables
          \begin{eqnarray}
	  \left.\frac{\partial{\cal{L}}_{E}}
         {\partial\dot\phi}\;\right|_{\,\,\tau_{+}}\!
	 &=&\!\!\!\!\left.
	 \phantom{\frac{\cal L}{\dot\phi}}
	 \fatg{W}_{\!L}\,h\;\right|_{\,t=0}
	 \!\!+\,O\,(\,\varepsilon^{2}),
\label{eqn:5.13}\\
	 \left.\frac{\partial{\cal{L}}_{L}}
	 {\partial\dot q}\;\right|_{\,t=0}\!\!
	 &=&\!\!\!\!\!\!\!\!\!\!\left.
	 \phantom{\frac{\cal L}{\dot\phi}}
	 -\,\fatg{W}\,\eta\;\right|_{\,\,\tau_{+}}
	 \!+\;O\,(\,\varepsilon^{2}).
\label{eqn:5.14}
          \end{eqnarray}

Now we can calculate up to quadratic terms in $\varepsilon$ the
complex action
(\ref{eqn:5.4}) at its complex extremal. Linear terms of the action
follow from
(\ref{eqn:5.5}), while the quadratic terms can be obtained by using
the
equation (\ref{eqn:4.6}) for the Euclidean action and its obvious
Lorentzian
analogue involving the operators $\fatg{F}_{L}$ and $\fatg{W}_{L}$.
In virtue
of the equations (\ref{eqn:5.9}) - (\ref{eqn:5.10}) all the volume
terms turn
to be $O\,(\,\varepsilon^{3})$, so that the contribution of imaginary
corrections in the quadratic approximation reduces to the sum of
surface terms
at the nucleation point $\tau=\tau_{+}\,(t=0)$
         \begin{eqnarray}
	 {\cal I}[\,\Phi\,]\!\!&=&\!\!I[\,\phi\,]-i\,S[\,q\,]
	 +i\left.\frac{\partial{\cal{L}}_{E}}
         {\partial\dot\phi}\,\eta\;\right|_{\,\tau_{+}}\!\!
	 -\left.\frac{\partial{\cal{L}}_{L}}
         {\partial\dot q}\,h\;\right|_{\,t=0} \nonumber
	 \\ \nonumber\\
	 &\phantom{=}&\!\!-\left.\frac{1}{2}\;\eta^{T}
	 (\fatg{W}\eta)\;\right|_{\,\tau_{+}}\!\!
	 -\left.\frac{i}{2}\;h^{T}
	 (\fatg{W}_{\!L}h)\;\right|_{\,t=0}
	 \!+\;O\,(\,\varepsilon^{3}).
\label{eqn:5.15}
	 \end{eqnarray}
\vspace{-6mm}

\noindent Here we took into account the reality of
$q_{+}\,(h(t_{+})=0)$ and
the no-boundary regularity conditions leading to vanishing surface
terms at
$t=0$ and $\tau=0$. Now notice that in virtue of (\ref{eqn:5.6})
$h(0)=\eta(\tau_{+})$. Then the use of equations (\ref{eqn:5.13}) -
(\ref{eqn:5.14}) allows to rewrite the above expression as
         \begin{eqnarray}
	 {\cal I}[\,\Phi\,]=I\,[\,\phi\,]-i\,S\,[\,q\,]
	 +\frac{1}{2}\;\varepsilon^{T}(\fatg{\cal W}
	 \varepsilon)+O\,(\,\varepsilon^{3}),
\label{eqn:5.16}
	 \end{eqnarray}
where $\varepsilon^{T}(\fatg{\cal W}\varepsilon)$ denotes the full
resulting
quadratic form in the variables $\varepsilon\equiv{\rm
Im}\,\Phi(z)=(h(t),\eta(\tau))$
        \begin{equation}
	\varepsilon^{T}(\fatg{\cal W}\varepsilon)=
        \left.\eta^{T}(\fatg{W}\eta)\;\right|_{\,\tau_{+}}
	+i\left.h^{T}(\fatg{W}_
	{\!L}\,h)\;\right|_{\,t=0}.     \label{eqn:5.17}
	\end{equation}

The crucial point of our derivations in this section is that the net
effect of
the Lorentzian-Euclidean matching conditions (\ref{eqn:5.13})-
(\ref{eqn:5.14})
and linear terms in the expression (\ref{eqn:5.15}) consists in
changing the
overall sign of the quadratic form in $\eta$ and $h$. This has a
drastic
consequence for the asymptotic $\hbar$-expansion of the wavefunction
(\ref{eqn:5.1}) with the complex extremal $\Phi(z\,|\,
\varphi,z_{+}\!)$.
Indeed, substituting the expression (\ref{eqn:5.16}) into  the
functional
$\Psi_{\textstyle [\,\Phi\,]}(f,z_{+}\!)$ given by (\ref{eqn:4.32})
and
reexpanding everything, except this exponentiated quadratic form, in
powers of
$\varepsilon$, one has
         \begin{equation}
	 \Psi_{\!L}\,(\varphi,f,t_{+}\!)=
	  {\rm e}^{\!\!\phantom {0}^{\textstyle
	 -\frac{1}{2\hbar}\varepsilon^{T}(\fatg{\cal W}\varepsilon)}}
	 \left\{\,\Psi_{\textstyle
	 [\,{\rm Re}\,\Phi\,]}(f,z_{+}\!)
	 +O\,(\varepsilon)+O\,
	 (\varepsilon^{3}/\hbar)\,\right\}\!,
\label{eqn:5.18}
	 \end{equation}
Since the variables $\eta(\tau)$ satisfy up to higher order terms the
linearized equations of motion (\ref{eqn:5.9}) and no-boundary
regularity
conditions, the real part of this quadratic form coincides with the
part of the
Euclidean action quadratic in the field disturbancies $\eta$
         \begin{equation}
	 {\rm Re}\,\left[\,\varepsilon^{T}
	 (\fatg{\cal W}\varepsilon)\,\right]
	 =\delta^{2}_{\eta}\,I+O\,(\,\eta^{4})
\label{eqn:5.19}
	 \end{equation}
and is positive definite by the assumption that the extremal realizes
the local
minimum of the Euclidean action.

Therefore, one can use the analogue of the asymptotic bound
(\ref{eqn:4.30}) to
show that
          \begin{equation}
	  {\rm e}^{\!\!\phantom {0}^{\textstyle
	  -\frac{1}{2\hbar}\varepsilon^{T}
	  (\fatg{\cal W}\varepsilon)}}\varepsilon^{n}
	  =O\,(\,\hbar^{n/2}),
	  \,\,\,\hbar\rightarrow 0,
\label{eqn:5.20}
	  \end{equation}
whence it follows that all the perturbation corrections of the
eq.(\ref{eqn:5.18}) in powers of $\varepsilon$ actually belong to
higher orders
of a semiclassical expansion. In contrast to (\ref{eqn:4.30}) the
exponentiated
quadratic form here has a kernel which is a differential operator
$\fatg{\cal
W}$ with respect to $t$ and $\tau$. This generalization, however,
does not
break the validity of the bound, because, according to \cite{Maslov},
the
statements like (\ref{eqn:4.30}) or (\ref{eqn:5.20}) are valid for
quadratic
functionals of integro-differential nature provided sufficient
smoothness of
their functional arguments. Another potential difficulty with the
above bound
is that the real part of our quadratic form involves only $\eta$
variables, and
one would think that the powers of the Lorentzian field $h(t)$ are
not
exponentially suppressed in (\ref{eqn:5.20}). This is not, however,
the case
because the fields $\eta(\tau)$ and $h(t)$ are not independent, for
they
satisfy the linearised differential equations
(\ref{eqn:5.9})-(\ref{eqn:5.10})
on the Euclidean and Lorentzian segments of $C_{+}$ with common
boundary
conditions at the nucleation point $\eta(\tau_{+}\!)=h(0)$. Therefore
the
Lorentzian imaginary disturbancies $h(t)$ can be parametrized in
terms of
$\eta(\tau)$ by certain integral operation, which provides the
asymptotic bound
(\ref{eqn:5.20}) for the whole set of $\varepsilon$
\footnote
{This statement can be proved more rigorously by decomposing
$\eta(\tau)$ and
$h(t)$ in the basis functions of respectively Euclidean and
Lorentzian wave
operators subject to no-boundary regularity conditions at $\tau=0$
and
Dirichlet boundary conditions at $t=t_{+}$ (remember that
$h(t_{+}\!)=0$). Then
the quadratic form and perturbative corrections become expressed
entirely in
terms of the independent set of variables
$f\equiv\eta(\tau_{B}\!)=h(0)$, and
one can apply the asymptotic bound (\ref{eqn:4.30}) with some
effective complex
kernel $D$ having a positive definite real part.}.

Thus, despite the complex nature of classical extremals $\Phi\,(z)$,
the
semiclassical expansion for tunnelling systems can still be performed
on the
real-valued background ${\rm Re}\,\Phi(z)\,|_{\!\!\!\!
\phantom{0}_{\,C_{+}}}=(\phi(\tau),q(t)),$ and with the corresponding
elements
of the Feynman diagrammatic technique -- the inverse propagator, its
basis
functions and the matrix (\ref{eqn:4.26}) of quantum dispersions for
microscopic variables:
	\begin{eqnarray}
	&&\fatg{F}=\fatg{F}\,[\,{\rm Re}\,\Phi\,],
\label{eqn:5.23}\\
	&&D_{L}\,(t)\equiv D\,(\tau_{B}+it)
	_{\;[\,{\rm Re}\,\Phi\,]}.
\label{eqn:5.24}
	\end{eqnarray}
Imaginary corrections everywhere except the quadratic form of the
action
(\ref{eqn:5.16}) can be treated by perturbations generating in higher
orders
additional set of Feynman diagrams.

One should emphasize a crucial role played by the boundedness from
below of the
Euclidean action, which provides the positivity of the form
(\ref{eqn:5.19}).
In Einstein gravity theory this property is violated in the sector of
the
conformal mode which, as is widely believed, enters the set of
physical
variables in spatially closed quantum cosmology
\footnote
{In asymptotically flat spacetime this mode is unphysical due to the
general
coordinate invariance of the theory and the gravitational constraints
-- the
basis of powerful positive-energy and positive-action theorems
\cite{Schoen-Y,Witten}.
}.
The only known procedure of handling this mode consists in the
rotation of its
integration contour in the path integral to the complex plane
\cite{Gibbons-H-P} so that to make the integration convergent due to
a reversal
of sign in the exponentiated quadratic form. In contrast to a
widespread
practice, this means that the same conformal rotation must be done in
the
argument of the wavefunction, which implies the complexification of
the
configuration-space point $q_{+}$ above, $h\,(t_{+})\not=0$, and the
corresponding modification of the formalism of complex extremals
\cite{tunnelIII}. We shall not consider this modification here, and
in what
follows assume good properties of the Euclidean action. In this paper
this will
be justified by isolating the conformal mode into the sector of
collective
variables and considering (see Sect.8) only the high-energy behaviour
of their
quantum distribution. This behaviour is unaffected by the tree-level
properties
of the classical action and is determined by the quantum anomalous
scaling of
the theory (see discussion in Sect.9). As concerns the microscopic
modes $f$,
for which the Euclidean action is supposed to have good positivity
properties,
we shall show the efficiency of the above technique for interpreting
their
quantum state in the Lorentzian world which nucleates from the
Euclidean
spacetime.

\section{Euclidean vacuum via nucleation of the Loren- tzian Universe
from the
Euclidean spacetime}
\hspace{\parindent}
Combining eqs.(\ref{eqn:4.32}), (\ref{eqn:5.18}) and (\ref{eqn:5.20})
one can
obtain the needed analytic continuation of the Euclidean wavefunction
into the
Lorentzian regime. Under this analytic continuation the originally
real
Euclidean basis functions $\fatg{u}_{-}(\tau)$ go over into complex
functions
$\fatg{u}_{-}(z)$ on the contour $C_{+}$ (\ref{eqn:2.22}). The
complex nature
of $\fatg{u}_{-}(z)\equiv \fatg{u}_{-}(z)_{\,\,[\,\Phi\,]}$
originates from the
complexity of both their time argument $z=\tau_{B}\!+it$ and the
functional
argument $\Phi\,(z)$ -- the complex classical background
(\ref{eqn:5.2}) at
which their wave operator (\ref{eqn:3.21}) is determined. Thus, if we
introduce
the following notation $(\fatg{v}(t),\,\fatg{v}^*(t))$ for the pair
of complex
conjugated functions originating from
$\fatg{u}_{-}(z)_{\,\,[\,\Phi\,]}$ at the
Lorentzian segment $C_{L},\,\,z=\tau_{B}+it,$
	\begin{equation}
	\fatg{v}\,(t)=\left(\fatg{u}_{-}(\tau_{B}\!+it)_{\,\,
	[\,\Phi\,]}\right)^*,\,\,\,
	\fatg{v}^*(t)=\fatg{u}_{-}(\tau_{B}\!+it)_{\,\,
	[\,\Phi\,]},
\label{eqn:7.1}
	\end{equation}
then the wavefunction of the Lorentzian Universe (\ref{eqn:5.1})
takes the form
	\begin{eqnarray}
	\Psi_{\!L}\,(\varphi,f,t_{+}\!)
	={\rm Const}\,\left[\;{\rm det}\,
	\fatg{v}^*(t_{+}\!)\;\right]^{-1/2}\!{\rm exp}
	\left\{-\frac{1}{2\hbar}f^{T}
	D_{L}(t_{+}\!)\,f\,\right \}    \nonumber \\
	\nonumber \\
        \times\,{\rm e}^{\!\!\phantom{0}^{\textstyle
        -\frac{1}{\hbar}{\cal I}\,[\,\Phi\,]}}
	\left [\,\,1+O\,(\hbar^{1/2})\,\,\right],
\label{eqn:7.2}
        \end{eqnarray}
where we have reabsorbed the quadratic form in $\varepsilon$ into the
full
complex classical action (\ref{eqn:5.16}).

In virtue of the relation (\ref{eqn:7.1}) the functions
$(\fatg{v}(t),\,
\fatg{v}^*(t))$ satisfy the complex conjugated equations
	\begin{eqnarray}
	\fatg{F}_{[\,\Phi\,]}\,(d/idt)\,
	\fatg{v}^*(t)=0,\,\,\,
	\left[\fatg{F}_{[\,\Phi\,]}\,
	(d/idt)\,\right]^*\fatg{v}\,(t)=0,
\label{eqn:7.3}
	\end{eqnarray}
which are a direct corollary of the equation for
$\fatg{u}_{-}(z)_{\,\,[\,\Phi\,]}$. However, the method of complex
extremals
allows us to treat the imaginary part $\varepsilon\equiv {\rm
Im}\,\Phi\,(z)$
by perturbations and consider, instead of the complex operator
$\fatg{F}_{[\,\Phi\,]}\,(d/dz)$ acting on the contour
$C_{+}=C_{E}\cup C_{L}$,
the real Euclidean $\fatg{F}\,(d/d\tau)$ and Lorentzian operators
$\fatg{F}_{L}\,(d/dt)$ acting on the corresponding segments $C_{E}$
and $C_{L}$
and related to $\fatg{F}_{[\,\Phi\,]}\,(d/dz)$ by
	\begin{eqnarray}
	&&\left.\fatg{F}_{[\,\Phi\,]}
	\,(d/dz)\;\right|_{\,C_{E}}=
	\fatg{F}\,(d/d\tau)+O\,(\,\varepsilon\,),
\label{eqn:7.4}\\
	&&\left.\fatg{F}_{[\,\Phi\,]}
	\,(d/dz)\;\right|_{\,C_{L}}=
	-\fatg{F}_{\!L}\,(d/dt)+O\,(\,\varepsilon\,).
\label{eqn:7.5}
	\end{eqnarray}
Here the Euclidean operator is defined by eq.(\ref{eqn:5.23}), while
the
Lorentzian operator
	\begin{equation}
	\fatg{F}_{\!L}\,(d/dt)=
	-\left.\fatg{F}_{[\,{\rm Re}\,\Phi\,]}
	\,(d/idt)\,\,\right|_{\,C_{L}}.
\label{eqn:7.6}
	\end{equation}
coincides with the hyperbolic wave operator (\ref{eqn:5.11}) of the
Lorentzian
field theory, calculated at the real-valued background $q\,(t)={\rm
Re}\,Q\,(t)$, and has the form analogous to (\ref{eqn:3.22})
	\begin{eqnarray}
	\fatg{F}_{\!L}\,(d/dt)=-\frac{d}{dt}\,
        a\,\frac{d}{dt}-\frac{d}{d\tau}\;b_{L}
        +b^{T}_{L}\,\frac{d}{d\tau}-c.
\label{eqn:7.7}
        \end{eqnarray}
Here the coefficients $a$ and $c$ are trivially related to their
Euclidean
versions (and, therefore, have the same notation), while the
coefficient
$b_{\!L}$ represents a Wick rotation of its Euclidean counterpart:
$b_{\!L}=ib$.

Obviously, in view of eq.(\ref{eqn:7.3})-(\ref{eqn:7.5}) the
functions
(\ref{eqn:7.1}) satisfy the inhomogeneous equations
$\fatg{F}_{\!L}\fatg{v}
=O\,(\,\varepsilon\,)$,
$\fatg{F}_{\!L}\fatg{v}^*=O\,(\,\varepsilon\,)$, but
their right-hand sides $O\,(\,\varepsilon\,)=O\,(\,\hbar^{1/2})$ can
be again
discarded in the one-loop approximation due to the semiclassical
technique of
complex extremals. In what follows we shall work with this one-loop
accuracy
and, therefore, assume the following real-valued differential
equation for the
complex Lorentzian basis functions
\footnote
{Alternatively, one can view {\normalsize$\fatg{v}$} and
{\normalsize$\fatg{v}^*$} to be {\it exact} solutions of
eq.(\ref{eqn:7.8}) as
well as {\normalsize$\fatg{u}_{-}(\tau)\equiv
\fatg{u}_{-}(\tau)_{\,\, [\,{\rm
Re}\,\Phi\,]}$} to be the {\it exact} basis functions of the operator
{\normalsize$\fatg{F}(d/d\tau)$} (\ref{eqn:5.23}). But these
{\normalsize$\fatg{v}$} and {\normalsize $\fatg{u}_{-}$} are not
exactly
related by (\ref{eqn:7.1}) because the collection {\normalsize
$(\fatg{F}(d/d\tau),\,\fatg{F}_{\!L}\,(d/dt))$} cannot be regarded as
a smooth
differential operator {\normalsize $\fatg{F}(d/dz)$} on {\normalsize
$C_{+}=C_{E}\!\cup\! C_{L}$}\,: discarding {\normalsize ${\rm
Im}\,\Phi$} in
(\ref{eqn:5.23}) and (\ref{eqn:7.6}) means that the first-order
derivatives of
{\normalsize ${\rm Re}\,\Phi\,(z)$} and the operator coefficients
{\normalsize
$a$, $b$ and $c$} are discontinuous at {\normalsize $z=\tau_{B}$} and
result in
the discontinuity of the second-order derivatives of {\normalsize
$\fatg{v}$}
and {\normalsize $\fatg{u}_{-}$}. Therefore, for such definition of
{\normalsize $\fatg{v}$} and {\normalsize $\fatg{u}_{-}$}, one can at
most
demand matching their zeroth and first-order derivatives at
{\normalsize
$z=\tau_{B}$}, which implies the validity of (\ref{eqn:7.1}) as well
as the
relation {\normalsize $d\fatg{u}_{-}/d\tau=id\fatg{v}^*/dt$} {\it
only} at
{\normalsize $z=\tau_{B}\,\,(t=0).$}
}
	\begin{equation}
	\fatg{F}_{\!L}\fatg{v}=0,\,\,\,\,
	\fatg{F}_{\!L}\fatg{v}^*=0.
\label{eqn:7.8}
	\end{equation}

The Wronskian operator $\fatg{W}_{\!L}\,(d/dt)$ corresponding to
(\ref{eqn:7.7}) generates the variational equation (\ref{eqn:5.12})
for the
Lorentzian canonical momentum and also enters the Lorentzian analogue
of the
Wronskian relation (\ref{eqn:6.6}) valid for arbitrary test functions
$h_{1}(t)$ and $h_{2}(t)$
	\begin{eqnarray}
	 &&h^{T}_{1}\,(\fatg{F}_{\!L}\,h_{2}\!)-
	 (\fatg{F}_{\!L}\,h_{1}\!)^{T}h_{2}=
	 -\frac{d}{dt}\left[\,h^{T}_{1}\,
	 (\fatg{W}_{\!\!L}\,h_{2}\!)-
	 (\fatg{W}_{\!\!L}\,h_{1}\!)^{T}
	 h_{2}\,\right],
\label{eqn:7.9}\\
	&&\fatg{W}_{\!\!L}\,(d/dt)=
	a\frac{d}{dt}+b_{\!L},\;\;\;
	\fatg{W}_{\!\!L}\,(d/dt)=\left.
	i\fatg{W}_{[\,{\rm Re}\,\Phi\,]}\,(d/idt)
	\,\,\right|_{\,C_{L}}.
\label{eqn:7.10}
	\end{eqnarray}

A very important property of the above Lorentzian wave and Wronskian
operators
is their reality
	\begin{equation}
	\fatg{F}_{\!L}^*\,(d/dt)=
	\fatg{F}_{\!L}\,(d/dt), \,\,\,
	\fatg{W}_{\!\!L}^*\,(d/dt)=
	\fatg{W}_{\!\!L}\,(d/dt),
\label{eqn:7.11}
	\end{equation}
which is crucial for establishing the conventional complex structure
on the
space of classical solutions and gives rise in the wording of
\cite{Gibbons-Pohle} to the "beginning of time". Let us emphasize
here, that,
in its turn, this property follows from the following two features of
the above
formalism. Firstly, it relies on the reality of both the Lorentzian
and
Euclidean Lagrangians of physical variables, related by the Wick
rotation
(\ref{eqn:3.8}), when they are calculated at real-valued fields
$q(t)$,
$\phi(\tau)$ (cf. eqs.(\ref{eqn:3.9}) - (\ref{eqn:3.10})). As it was
discussed
in Sect.3, this condition must be granted by a proper choice of gauge
for
physical variables, which should not introduce into the Lagrangian
(\ref{eqn:3.8}) the explicit time dependence generating its imaginary
part on
the complex contour $C_{+}$. And, secondly, it is based on the
semiclassical
method of Sect.5 which always reduces the calculations and the
corresponding
elements of the diagrammatic technique to those of the real field
background
${\rm Re}\,\Phi(z)=(q(t), \,\,\phi(\tau))$.

For the second-order differential equation $\fatg{F}_{\!L}\,h=0$ with
real
coefficients, the complex linear space of solutions $h\,(t)$ can be
equipped
with the indefinite inner product conserved in time
	\begin{eqnarray}
	<h_1,h_2>=i\left[\,h^{\dagger}_1\,
	(\fatg{W}_{\!\!L}^{\phantom{*}}\,h_2^{\phantom{*}})-
	(\fatg{W}_{\!\!L}^{\phantom{*}}\,
	h_1^{\phantom{*}})^{\dagger}\,h_2\,\right],
\label{eqn:7.12}
	\end{eqnarray}
where $h^{\dagger}\equiv (h^*)^{T}$ is a Hermitian conjugation
involving both
the transposition (of vectors and matrices) and the complex
conjugation. When
the functions $h\,(t)$ are related by the analytic continuation to
their
Euclidean counterparts $\varphi\,(\tau)$, this inner product can be
expressed
in terms of the Wronskian construction of the Euclidean operator
$\fatg{F}$
	\begin{eqnarray}
	\varphi^{T}_{1}\,
	(\fatg{W}\!\varphi_{2})-
	(\fatg{W}\!\varphi_{1})^{T}\varphi_{2}=
	-<h_1^*,h_2>,\;\;\;
	h_{1,2}(t)=\varphi_{1,2}(\tau_{B}+it).
\label{eqn:7.13}
	\end{eqnarray}
Now, if we take as $\varphi_{1,2}(\tau)$ two regular basis functions
of the
Euclidean operator $\fatg{u}_{-}(\tau)$, which have a vanishing
Wronskian
	\begin{equation}
	\fatg{u}_{-}^{T}\,(\fatg{W}\!\fatg{u}_{-}\!)
	-(\fatg{W}\!\fatg{u}_{-}\!)^{T}
	\fatg{u}_{-}=0
\label{eqn:7.14}
	\end{equation}
resulting from their regular behaviour at $\tau=0$
(\ref{eqn:6.2.12}), then it
follows that the complex conjugated Lorentzian basis functions
(\ref{eqn:7.1})
satisfy the following orthogonality relation
	\begin{equation}
	<\fatg{v}^*,\fatg{v}>=0.
\label{eqn:7.15}
	\end{equation}

On the other hand, Lorentzian basis functions of one " positive
frequency" have
the conserved matrix of inner products
	\begin{equation}
	\fatg{\Delta}=<\fatg{v},\fatg{v}>, \;\;\;\;
	\fatg{\Delta}\equiv\fatg{\Delta}_{AB},
\label{eqn:7.16}
	\end{equation}
which can be calculated at the point of nucleation $t=0$
$(\tau=\tau_{B})$
where the following matching conditions hold between the Lorentzian
and
Euclidean modes: $\fatg{u}_{-}(\tau_{B})=\fatg{v}(0)= \fatg{v}^*(0)$,
$(\fatg{W}\!\fatg{u}_{-}\!)\,(\tau_{B})=i\fatg{W}_{\!L}\,\fatg{v}^*(0
)$
\footnote
{According to the discussion above, the Euclidean modes are real only
up to
negligible terms $O\,(\,\varepsilon\,)=O\,(\,\hbar^{1/2}\,)$ or,
alternatively
(see the previous footnote), can be regarded {\it exactly} real, in
which case
these matching conditions, replacing (\ref{eqn:7.1}), serve for the
continuation of the Euclidean modes into Lorentzian ones as exact
solutions of
the real wave equations.
}.
In virtue of these matching conditions this matrix equals the
following real
symmetric matrix
	\begin{equation}
	\fatg{\Delta}=
	2\,\left.\fatg{u}_{-}^{T}
	(\fatg{W}\!\fatg{u}_{-}\!)\;
	\right|_{\,\tau_{B}},\;\;\;\;
	\fatg{\Delta}=
	\fatg{\Delta}^{T},\;\;\;\;
	\fatg{\Delta}=
	\fatg{\Delta}^{*},	  \label{eqn:7.17}
	\end{equation}
which coincides with the kernel of the positive definite quadratic
part of the
Euclidean action in field disturbances $\delta\phi\,(\tau)=
\fatg{u}(\tau)\,\eta$ satisfying linearized equations of motion
(cf.eq.(\ref{eqn:4.24}))
	\begin{equation}
	\delta^2 I=\eta^{T}\left[\left.\fatg{u}^{T}_{-}
	(\fatg{W}\!\fatg{u}_{-}\!)\,
	\right|_{\,\tau_{B}}\right]\eta=
	\frac{1}{2}\,\eta^{T}\fatg{\Delta}\,\eta\, >\, 0.
\label{eqn:7.18}
	\end{equation}
Therefore, $\fatg{\Delta}$ is a real positive-definite symmetric
matrix, which
together with the orthogonality relations (\ref{eqn:7.15}) implies
that the
Lorentzian modes $\fatg{v}\,(t)$ and $\fatg{v}^*(t)$ can be regarded
as a set
of positive and negative frequency modes of the Lorentzian wave
operator.

Thus far we have considered the Lorentzian linearized modes of all
physical
variables $q$. According to their decomposition (\ref{eqn:4.1}) into
colective
variables $\varphi$ and the microscopic variables $f$ of Sect.4 and
the
corresponding block-diagonal structure of all the relevant Euclidean
operators
and modes (\ref{eqn:4.19}) - (\ref{eqn:4.22}), a similar
decomposition
properties hold for their Lorentzian counterparts
	\begin{eqnarray}
	&&\fatg{F}_{\!L}={\rm diag}\;(F_{\!\varphi\;
L},\,F_{L}),\,\,\,
	\fatg{W}_{\!L}={\rm diag}\;(W_{\!\varphi\; L},
	\,W_{\!L}),                                   \nonumber\\
	&&\fatg{v}\,(t)={\rm diag}\;
	(v_{\varphi}\,(t),\,v\,(t)),\,\,\,
	\fatg{\Delta}={\rm diag}\;
	(\Delta_{\varphi},\,\Delta).
\label{eqn:7.19}
	\end{eqnarray}
Therefore all the Wronskian orthogonality relations of the above type
hold
separately in the sector of collective variables and the sector of
microscopic
ones.

The operators $F_{\!\varphi\; L},\,\,W_{\!\varphi\; L}$ and their
modes
$v_{\varphi}\,(t)$ determine the quantum properties of the collective
variable
$\varphi$ and of the corresponding field background $\Phi\,(z)$ in
the
definition of the wavefunction (\ref{eqn:7.2}). Let us recall that,
in contrast
to the usual definitions of the classical background, our background
has a
quantum nature, for it is parametrized by the collective quantum
variable
$\varphi$, $\Phi\,(z)= \Phi\,(z|\varphi,\,z_{+}\!)$, as an extremal
of
equations of motion subject to the boundary conditions depending on
$\varphi$
(cf. eq.(\ref{eqn:4.1})).

The quantum properties of the lineraized microscopic modes $f$ in the
wavefunctions (\ref{eqn:7.2}) are determined by the matrix of quantum
dispersions $D_{L}$ defined, according to (\ref{eqn:4.26}) and
(\ref{eqn:5.24}), entirely in terms of $W_{L}(d/dt)$ and $v^*(t)$:
	\begin{equation}
	D_{L}\,(t)=-i\,\left[\,W_{\!L}\,(d/dt)\,
	v^*(t)\,\right]\,[\,v^*(t)\,]^{-1}.
\label{eqn:7.20}
	\end{equation}
In virtue of eq.(\ref{eqn:7.15}) this complex matrix is symmetric and
has a
positive-definite real part derivable from the corollary
$W_{L}v^*=(v^{\dagger})^{-1}(W_{L}v)^{\dagger}v^* $ of
eq.(\ref{eqn:7.15}):
	\begin{eqnarray}
	&&D_{L}^{T}-D_{L}=
	(v^{\dagger})^{-1}<v,\,v^*>\,(v^*)^{-1}=0,
\label{eqn:7.21}\\
	&&D_{L}^{*}+D_{L}=
	(v^{\dagger})^{-1}\Delta\,v^{-1}.
\label{eqn:7.22}
	\end{eqnarray}

These properties of $D_{L}$ finally allow us to show that the
gaussian state
(\ref{eqn:7.2}) is a vacuum of linearized modes $f$ relative to the
positive-negative frequency decomposition with respect to Lorentzian
basis
functions (\ref{eqn:7.1}). Indeed, consider the Hermitian (in the
quantum
Hilbert space, but not in the space of complex matrices) Heisenberg
operator of
the linear quantum field $\hat f(t)$ decomposed into a set of
$(v(t),\,v^*(t))$
	\begin{equation}
	\hat f(t)=v\,(t)\,\hat a+v^*(t)\,\hat a^*\equiv
	v_{A}(t)\,\hat a^{A}+v^*_{A}(t)\,\hat a^{*A}
\label{eqn:7.23}
	\end{equation}
with the operatorial Hermitian-conjugated coefficients $\hat a=\hat
a^{A},$ and
$\hat a^*=\hat a^{*A}$. The canonically conjugated momentum $\hat
p\,(t)$ for
this linearized field can be obtained in virtue of (\ref{eqn:5.12})
by acting
on $\hat f(t)$ with the Wronskian operator
	\begin{equation}
	\hat p\,(t)=W_{L}(d/dt)\,\hat f(t)=
	(W_{L}v)\,(t)\;\hat a+(W_{L}v)^*(t)\;\hat a^*.
\label{eqn:7.24}
	\end{equation}
As a corollary of the orthogonality (\ref{eqn:7.15}) the following
matrix
relation holds
      \begin{equation}
      \left[\begin{array}{cc}
      (W_{L}v)^{\dagger}&-v^{\dagger}\\
      (W_{L}v)^{T}&-v^{T}\end{array}\right]
      \left[\begin{array}{cc}
      v&v^*\\
      W_{L}v& W_{L}v^*\end{array}\right]=
      \left[\begin{array}{cc}
      i\,\Delta&0\\
      0&-i\,\Delta\end{array}\right]
\label{eqn:7.25}
      \end{equation}
which allows one immeadiately to solve the system of equations
(\ref{eqn:7.23})-(\ref{eqn:7.24}) for the operators $(\hat a,\,\hat
a^*)$:
	\begin{equation}
	\hat a=i\,\Delta^{-1}v^{\dagger}\hat p-i\,
	\Delta^{-1}(W_{L}v)^{\dagger}\hat f,\;\;\;
	\hat a^*=-i\,\Delta^{-1}v^{T}\hat p+
	i\,\Delta^{-1}(W_{L}v)^{T}\hat f.
\label{eqn:7.26}
	\end{equation}
In view of the standard equal-time canonical commutation relations
for the
phase-space operators $[\,\hat f,\,\hat p^{T}\,]=i\,\hbar\,I$ ($I$ is
a unit
matrix in the $f$-sector of the full space of fields, $\fatg{I}={\rm
diag}\,(I_{\varphi},I)$, and all the other commutators vanish), $\hat
a$ and
$\hat a^*$ have the following only nonvanishing commutator
\footnote
{To avoid confusion with the matrix notations of this equation, we
have written
the indices of $\hat a$ and $\hat a^*$ explicitly. The correct matrix
form of
these commutation relation would be $[\;\hat a,\,a^{*T}]=\Delta$
reflecting the
fact that $[\;\hat a,\,a^{*T}]$ is a direct product of two vectors,
but not the
scalar contraction of their indices. For the same reason, the correct
form of
the commutation relations for $\hat f$ and $\hat p$ above also
involves the
transposed quantities: $[\;\hat f,\,\hat p^{T}\,]=-[\;\hat p,\,\hat
f^{T}\,]=i\,\hbar\,I$.
}
	\begin{equation}
        \left[\;\hat a^{A},\hat a^{*B}\right]
	=\hbar\;(\Delta^{-1}\!)^{\,AB}.
\label{eqn:7.27}
	\end{equation}

Since $\Delta$ is a real positive definite matrix, it can be
diagonalized by
linear transformations of positive-frequency basis functions $v\,(t)$
making
their set orthonormal
	\begin{equation}
	<v_{A},\,v_{B}>=\Delta_{AB}=\delta_{AB},
\label{eqn:7.28}
	\end{equation}
so that $\hat a$ and $\hat a^*$ become respectively the usual
annihilation and
creation operators. In particular, the operator $\hat a$ rewritten in
the
coordinate representation of the phase space operators, $\hat f=f$,
$\hat
p=\hbar\partial/i\partial f$,
	\begin{equation}
	\hat a\,(f,\partial/\partial f)
	=\hbar\,v^{\dagger}\,\frac{\partial}{\partial f}
	-i\,(W_{L}v)^{\dagger}f
\label{eqn:7.29}
	\end{equation}
annihilates the gaussian quantum state (\ref{eqn:7.2}) of lineraized
quantum
perturbations in the Lorentzian Universe
	\begin{equation}
	\hat a\,(f,\partial/\partial f)
	\,\Psi_{\!L}\,(\varphi,f,t_{+}\!)=0,
\label{eqn:7.30}
	\end{equation}
which can be easily verified by using eq.(\ref{eqn:7.15}).

Thus, in the no-boundary prescription of Hartle and Hawking the
Lorentzian
Universe nucleates from the Euclidean spacetime with the vacuum state
of
linearized physical modes, corresponding to the field decomposition
in special
positive and negative frequency basis functions. They originate by
the analytic
continuation (\ref{eqn:7.1}) from regular linear  modes in the
Euclidean ball.
This decomposition and, therefore, the definition of the vacuum is
unique,
because the only admissible freedom in the choice of basis functions
(\ref{eqn:7.1}), which does not violate the regularity condition,
consists in
the unitary rotations of the positive-frequency subset of $v\,(t)$,
preserving
(\ref{eqn:7.28}) and not mixing $v\,(t)$ and $v^*\,(t)$.

It is worth emphasizing here again a crucial role played by the
boundedness of
the Euclidean action from below. In view of the positivity  of the
quadratic
form (\ref{eqn:7.18}) it guarantees the positivity of $\Delta$ and of
the real
part of $D_{L}$. Therefore, it makes the gaussian state
(\ref{eqn:7.2})
normalizable and exponentially damping large quantum fluctuation. As
we see,
this is the same property which underlies the perturbation theory in
the
imaginary part of complex extremals and allows the construction of a
special
positive-negative frequency decomposition and vacuum state for
complex
tunnelling geometries.

The emergence of the vacuum state for quantum inhomogeneties from the
Hartle-Hawking wavefunction has been observed by a number of authors
\cite{Halliwell-Hawking,Wada} in the context of different models with
the
real-valued classical DeSitter and quasi-DeSitter background and, in
particular, has been most transparently demonstrated in the language
of the
path integral approach in \cite{Laflamme}. In DeSitter models this
special
vacuum state coincides with the so-called Euclidean vacuum
\cite{Euclideanvac}
which exhibits a number of remarkable properties \cite{Allen}
\footnote
{In particular, it is DeSitter invariant \cite{Allen} and has a
Hadamard
singularity of two-point Green's functions, thus providing the
covariant
renormalization of ultraviolet infinities with local counterterms
\cite{DW:Dynamical,DW:LesH}
}
and has important applications in the theory of the inflationary
Universe
\cite{large-scale}, because it provides the spectrum of density
fluctuations
responsible for the formation of the large scale structure of the
observable
Universe. Here we have extended these conclusions by showing that in
the
framework of $\hbar$-expansion a similar unique vacuum is selected by
the
no-boundary proposal even for complex-valued tunnelling geometries
which
certainly constitute the most general case.

\section{Quantum distribution of tunnelling universes}
\hspace{\parindent}
The Lorentzian wavefunction of the Universe (\ref{eqn:7.2}) carries
the
information about all dynamical variables of the system. However, in
practical
applications, when only a part of variables are available to the
observer at
the experimental level, it suffices to have the corresponding density
matrix
which can be obtained from the wavefunction by tracing out the
unmeasured
degrees of freedom. This procedure of tracing out certain variables
can
drastically change the form of the initial full quantum state as a
function of
the needed observables and, therefore, represents a nontrivial step
towards
quantities having a direct physical interpretation. Usually the
degrees of
freedom which carry the most important information about the system
are some
macroscopic collective variables of the type considered in Sect.4,
that is why
the matter of primary interest within the above method of collective
coordinates is their density matrix
	\begin{equation}
	\hat\rho\,(t)={\rm Tr}_{\phantom{0}_{\phantom{0}_
	{\!\!\!\!\!\!\!\!\!\!\!\!\!\!\!\!\!
	\textstyle\phantom{0}_{f}}}}\,\,
	|\,\Psi_{\!L}\,(t)\!><\!\Psi_{\!L}\,(t)\,|
\label{eqn:8.1}
	\end{equation}
obtained from the pure state density matrix by tracing out the
microscopic
degrees of freedom $f$. This object encodes all the correlation
functions of
the collective variables $\varphi$ generally nonlinearly interacting
with one
another and the rest of the physical modes. In particular, it
includes the
density of their probability distribution which is just the diagonal
element
$\rho\,(\varphi,\,t)=\rho\,(\varphi,\varphi\,|\,t)$ of
$\hat\rho\,(t)=\rho\,(\varphi,\varphi^{\prime}|\,t)$ in the
coordinate
representation of $\varphi$. Since we work with the wavefunction of
physical
variables in the Hilbert space with the trivial inner product
(\ref{eqn:1.6}),
this diagonal element equals
	\begin{equation}
	\rho\,(\varphi,\,t)=
	\int df\,|\Psi_{\!L}\,(\varphi,f,t)\,|^2.   \label{eqn:8.2}
	\end{equation}
The knowledge of the wavefunction (\ref{eqn:7.2}) easily allows us to
calculate
the full density matrix (\ref{eqn:8.1}), but here we shall mainly
concentrate
on this quantity playing a very important role in quantum cosmology
of
tunnelling universes, for it determines their probability
distribution in the
space of such macroscopic variables as a Hubble constant, parameters
of
anisotropy, etc.

Substituting (\ref{eqn:7.2}) into (\ref{eqn:8.2}) and taking into
account the
equation (\ref{eqn:7.22}) for the real part of the matrix of quantum
dispersions, one immeadiately finds the following answer for the
gaussian
integral over $f$
	\begin{eqnarray}
	\rho\,(\varphi,\,t)=
	{\rm Const}\,\frac{\;\;({\rm det}\,\Delta_{\varphi})^{1/2}}
	{|\,{\rm det}\,v_{\varphi}\,(t)\,|}\;
	({\rm det}\,\fatg{\Delta})^{-1/2}\,
	{\rm e}^{\!\!\phantom{0}^{\textstyle
        -\frac{2}{\hbar}{\rm Re}\,{\cal I}\,[\,\Phi\,]}}
	\left [\,\,1+O\,(\hbar^{1/2})\,\,\right],
\label{eqn:8.3}
        \end{eqnarray}
where the real part of the complex action (\ref{eqn:5.16})
	\begin{equation}
	{\rm Re}\,{\cal I}[\,\Phi\,]=I\,[\,\phi\,]
	 +\frac{1}{2}\;\eta^{T}(\fatg{W}\!
	 \eta)\,(\tau_{B})+O\,(\,\hbar^{3/2}),
\label{eqn:8.4}
	 \end{equation}
must include the positive definite real part of the quadratic form
(\ref{eqn:5.17}) damping the contribution of the imaginary part of
the complex
extremal $\varepsilon=O\,(\hbar^{1/2})$. Here we took into account
the
block-diagonal form (\ref{eqn:7.19}) of the matrices $\fatg{v}$ and
$\fatg{\Delta}$
	\begin{equation}
	{\rm det}\,\fatg{v}\,(t)=
	{\rm det}\,v_{\varphi}\,(t)\,{\rm det}\,v\,(t),\;\;\;
	{\rm det}\,\fatg{\Delta}=
	{\rm det}\,\Delta_{\varphi}\,{\rm det}\,\Delta,
\label{eqn:8.5}
	\end{equation}
due to which the one-loop preexponential factor of (\ref{eqn:8.3})
features the
determinant of the {\it full} Wronskian matrix $\fatg{\Delta}$ and
the
determinant of the Lorentzian modes of the collective variables
$v_{\varphi}\,(t)$ normalized by $({\rm
det}\,\Delta_{\varphi})^{1/2}$ to unity
in the inner product (\ref{eqn:7.12}). The emergence of the full
Wronskian
matrix in this algorithm in conjunction with the reduction method for
functional determinants on closed spacetimes of \cite{tunnelII} make
us to
consider in the next section a special geometric interpretation of
the result
(\ref{eqn:8.3}).

\subsection{Doubling the Euclidean spacetime}
\hspace{\parindent}
Here we shall present a detailed geometrical construction of doubled
Euclidean
spacetime serving for both the interpretation and covariant
calculation of the
one-loop partition function (\ref{eqn:8.3}), which was proposed for
these
purposes in \cite{BKam:norm} and also used in \cite{Gibbons-Hartle}
for the
general tree-level analysis of real tunnelling geometries, amounting
to the
so-called unique conception theorem.

To begin with, note that the Wronskian matrix $\fatg{\Delta}$ in
(\ref{eqn:8.3}) can be represented by the equation (\ref{eqn:7.17})
in terms of
the regular basis functions on the Euclidean spacetime ball $\fatg{
M}\equiv\fatg{M}_{-}=\fatg{\cal B}^{4}$ with the "center" at
$\tau_{-}$. This
spacetime carries a real Euclidean metric and matter fields
characterized by
the real physical fields $\phi\,(\tau)={\rm Re}\,\Phi\,(\tau)$ and
has as a
boundary the spatial hypersurface $\Sigma_{B}$ of constant
$\tau=\tau_{B}$ at
which a semiclassical nucleation of the Lorentzian Universe from the
Euclidean
one takes place. Now consider its orientation reversed copy
$\fatg{M}_{+}$,
which can be regarded as mirror image of $\fatg{M}_{-}$ with respect
to this
boundary. One can now construct the doubled manifold ${\bf
2}\fatg{M}$ by
joining $\fatg{M}_{-}$ and $\fatg{M}_{+}$ across their common
boundary
$\Sigma_{B}$ (see Fig.4)
	\begin{equation}
	{\bf 2}\fatg{M}=\fatg{M}_{-}\!
	\cup\fatg{M}_{+}.
\label{eqn:8.7}
	\end {equation}
This doubled manifold is, obviously, closed, has a topology of a
four-dimensional sphere and admits an isometry $\theta$ mapping its
two halves
$\fatg{M}_{\pm}$ onto one another
	\begin{equation}
	\fatg{M}_{\pm}=
	\theta\fatg{M}_{\mp}:\;\;\;\;
	\{\,x\in\fatg{M}_{\pm},\,\,
	\theta\,x\in\fatg{M}_{\mp}\,\}.		\label{eqn:8.8}
	\end{equation}

The foliation of $\fatg{M}_{-}$ by three-dimensional compact surfaces
of
constant $\tau$ can be continuously extended by this isometry to the
whole of
${\bf 2}\fatg{M}$ with the parameter $\tau$ ranging between
$\tau_{-}$ and
$\tau_{+}$ -- the value corresponding to the "center" of
$\fatg{M}_{+}$
coinciding with the "north" pole of the sphere-like doubled manifold.
This
foliation can be represented by the continuous one-parameter family
of surfaces
$\Sigma\,(\tau)$ which expand from zero volume at the south cap of
${\bf
2}\fatg{M}$ and then again shrink to a point at the north cap after
passing the
equatorial section $\Sigma_{B}=\Sigma\,(\tau_{B})$ at
$\tau_{B}=(\tau_{-}\!\!+\tau_{+})/2$
	\begin{eqnarray}
	&\tau\,(x)=\tau,\;\;\;\;
	&{\bf x}\,(x)={\bf x},\;\;\;\;
	x\in\fatg{M}_{-},\;\;
	\tau_{-}\leq\tau\leq\tau_{B}, \nonumber \\
	&\tau\,(\theta x)=\tau_{+}\!+\tau_{-}\!-\tau,\;\;\;\;
	&{\bf x}\,(\theta x)={\bf x},\;\;\;\;
	\theta x\in\fatg{M}_{+}.
\label{eqn:8.9}
	\end{eqnarray}
This foliation explicitly demonstrates the reversal of Euclidean time
on
$\fatg{M}_{+}$ in contrast to spatial coordinates {\bf x} identically
related
on surfaces $\Sigma$ and $\theta\Sigma$. Obviously, the coordinate
$\tau$
parametrizing the whole of ${\bf 2}\fatg{M}$ plays the role of the
latitude
angle $\theta$ on the four-dimensional sphere homeomorphic to ${\bf
2}\fatg{M}$, ranging from $0$ to $\pi$, while ${\bf x}$ are the
"angular"
coordinates on quasispherical spatial sections $\Sigma$.

The four geometry and matter fields on ${\bf 2}\fatg{M}$ are also a
subject of
the isometry map (\ref{eqn:8.8}), which means that on $\fatg{M}_{+}$
	they are
defined as a reflection image of those on the original spacetime
$\fatg{M}_{-}=\fatg{M}$. In the foliation (\ref{eqn:8.9}) this fact
can be
easily represented as a following definition of the physical
variables
$\phi\,(\tau)$ for $\tau_{B}\leq\tau\leq\tau_{+}$ in terms of those
for
$\tau_{-}\leq\tau\leq\tau_{B}$:
	\begin{equation}
	\phi\,(\tau)=\phi\,(\tau_{+}\!+\tau_{-}\!-\tau).
\label{eqn:8.10}
	\end{equation}
Such fields are continuous but not generally analytic at the
"equatorial"
junction surface $\Sigma_{B}$ unless their normal derivative
$d\phi/d\tau(\tau_{B})$	vanishes there. In case of real tunnelling
geometries
considered in \cite{Gibbons-Hartle,Gibbons-Pohle} this condition is
satisfied
as a geometrically invariant requirement of vanishing extrinsic
curvature of
$\Sigma_{B}$
	\begin{equation}
	\left.K_{ab}\,\right|_{\,\Sigma_{B}}=0,
\label{eqn:8.11}
	\end{equation}
this fact providing the analytic matching of real Euclidean manifold
$\fatg{M}_{-}$ with its double $\fatg{M}_{+}$ and with the nucleating
real
Lorentzian spacetime $\fatg{M}_{L}$. For complex tunnelling fields,
we consider
here, this condition is, however, generally violated, because the
normal (time)
derivative of $\phi\,(\tau)$ is proportional in view of the matching
condition
(\ref{eqn:5.13}) to the imaginary part of the extremal:
$\dot\phi\,(\tau_{B})\sim\partial{\cal
L}_{E}/\partial\dot\phi\,(\tau_{B})
=O\,(\varepsilon)$.

The lack of smoothness of the background fields does not prevent,
however, from
extending the basis function $\fatg{u}_{-}(\tau)$ defined on
$\fatg{M}_{-}$ to
the whole of ${\bf 2}\fatg{M}$ as a solution of
	\begin{equation}
	\left.\phantom{A^0_0}\fatg{F}\fatg{u}_{-}
	\right|_{\,{\textstyle\bf 2}\!
	{\small{\fatg{M}}}}=0
\label{eqn:8.12}
	\end{equation}
with continuous zeroth and first order derivatives at $\Sigma_{B}$
(the second
order derivatives will generally jump at $\Sigma_{B}$, because the
coefficients
of the differential operator $\fatg{F}$ are discontinuous at this
surface).
These basis functions are regular at the south pole $\tau_{-}$ of the
doubled
manifold, but singular at $\tau_{+}$, because we assume that the
positive-definite Euclidean operator $\fatg{F}$ does not have zero
modes on
${\bf 2}\fatg{M}$.

On the other hand, we can consider the set of basis functions
$\fatg{u}_{+}$ on
the doubled manifold which are the reflection image of $\fatg{u}_{-}$
defined
in the foliation (\ref{eqn:8.9}) by the relation
\footnote
{
Generally, this relation, as well as the equation (\ref{eqn:8.10}),
should be
understood not for separate components of
{\normalsize$\fatg{u}_{\pm}$} and
{\normalsize$\phi$}, but for their tensor objects as a whole,
accounting for
the orientation reversing map of local basis on
{\normalsize$\fatg{M}_{-}$} and
{\normalsize$\fatg{M}_{+}$}. However, except for spinors, since the
physical
variables basically involve the spatial components of tensor fields,
the
relation (\ref{eqn:8.13}) literally holds in a special coordinate
system
(\ref{eqn:8.9}) which does not reverse the orientation of spatial
sections. The
more complicated case of spinors, accounting for the complex
structure of the
Dirac operator, was considered in \cite{Gibbons-Pohle} for Majorana,
Dirac and
Weyl spinor fields.
}
	\begin{equation}
	\fatg{u}_{+}(\tau)=
	\fatg{u}_{-}(\tau_{+}\!+\tau_{-}\!-\tau).
\label{eqn:8.13}
	\end{equation}
These basis functions are regular at $\tau_{+}$, singular at
$\tau_{-}$ and
satisfy the following matching conditions at the junction surface
$\Sigma_{B}$
	\begin{equation}
	\fatg{u}_{-}(\tau_{B})=\fatg{u}_{+}(\tau_{B}),\;\;\;
	\fatg{W}\!\fatg{u}_{-}(\tau_{B})=
	-\fatg{W}\!\fatg{u}_{+}(\tau_{B}).
\label{eqn:8.14}
	\end{equation}
Therefore, the set of inner products of Lorentzian linear modes
$\fatg{\Delta}$, given by eq.(\ref{eqn:7.17}), can be rewritten in
the form of
the Wronskian matrix $\fatg{\Delta}_{+-}=(\fatg{\Delta}_{+-})_{AB}$
of these
two sets of Euclidean basis functions on the doubled manifold
(independent of
$\tau$ in virtue of the relation (\ref{eqn:6.6}))
	\begin{eqnarray}
	&&\fatg{\Delta}=\fatg{\Delta}_{+-},
\label{eqn:8.15}\\
	&&\fatg{\Delta}_{+-}=\fatg{u^{T}_{+}}\,
	(\fatg{W}\fatg{u}_{-}\!)-
	(\fatg{W}\fatg{u}_{+}\!)^{T}\fatg{u}_{-}.
\label{eqn:6.11}
	\end{eqnarray}

This property serves as a ground for the following important
observation.
According to the reduction technique of \cite{tunnelII} for
functional
determinants on a closed compact spacetime of spherical topology
(which is just
the case of ${\bf 2}\fatg{M}$), the one-loop preexponential factor of
the
Euclidean quantum theory on such a spacetime can be generated by the
determinant of the Wronskian matrix (\ref{eqn:6.11}) of the two
complete sets
of basis functions $\fatg{u}_{\pm}(\tau)$ which, in the
$\tau$-foliation of the
above type, are regular respectively at $\tau_{+}$ and $\tau_{-}$ and
have the
asymptotic behaviour (cf. eq.(\ref{eqn:6.2.12})
	\begin{eqnarray}
	\fatg{u}_{-}(\tau)=\fatg{I}\,
	(\tau-\tau_{-}\!)^{\mu_{-}}\!
	+O\,[\,(\tau-\tau_{-}\!)^{1+\mu_{-}}\!],\,\,\,\,
	\tau\rightarrow\tau_{-}, \\
\label{eqn:6.3.9}
	\fatg{u}_{+}(\tau)=\fatg{I}\,
	(\tau_{+}-\tau)^{\mu_{-}}\!
	+O\,[\,(\tau_{+}-\tau)^{1+\mu_{-}}],\,\,\,\,
	\tau\rightarrow\tau_{+}
\label{eqn:6.3.10}
        \end{eqnarray}
with the field-independent coefficient -- the matrix inity
$\fatg{I}$. This
reduction algorithm reads
        \begin{equation}
	({\rm det}\,\fatg{\Delta}_{+-})^{-1/2}=
	{\rm Const}\,
	\left[\,
	{\rm Det}\,\fatg{F}_{\phantom{0}_{\phantom{0}_
	{\!\!\!\!\!\!\!\!\!\!\!\!\!\!\!\!\!\!\!\!\!\!\!
	\!\!\!\!\!\!\!\!\textstyle\phantom{0}_
	{2M}}}}\;\;\;\,
	/
	{\rm Det}\,\fatg{a}_{\phantom{0}_{\phantom{0}_
	{\!\!\!\!\!\!\!\!\!\!\!\!\!\!\!\!\!\!\!\!\!\!\!
	\!\!\!\!\!\!\!\!\textstyle\phantom{0}_
	{2M}}}}\;\;\;\;
	\right]^{-1/2}.             \label{eqn:8.16}
	\end{equation}
and implies that the one-loop preexponential factor of our partition
function
(\ref{eqn:8.3}) in the main boils down to the contribution of
functional
determinants on ${\bf 2}\fatg{M}$. Such determinants are calculated
on the
representation space of $\fatg{F}$ -- the space of functions regular
on closed
compact manifold ${\bf 2}\fatg{M}$ and constitute the one-loop part
of the {\it
effective} action of the Euclidean theory on this spacetime.

\subsection{Covariant distribution function: covariance versus
unitarity}
\hspace{\parindent}
Using the above relation in (\ref{eqn:8.3}) we arrive at the
following
fundamental algorithm for the one-loop partition function of
tunnelling
geometries
	\begin{eqnarray}
	\rho\,(\varphi,\,t)=
	{\rm Const}\,\frac{\;\;({\rm det}\,\Delta_{\varphi})^{1/2}}
	{|\,{\rm det}\,v_{\varphi}\,(t)\,|}
	\!\!\!\!\!\!&&\!\!\!\!\!\!
	{\rm exp}\!\left\{-\frac{1}{\hbar}\fatg{\Gamma}_
	{\rm 1-loop}\,[\,\phi\,]\,\right\}
\nonumber\\
	\nonumber\\
	&\times&\!\!\!\!
	\mbox{\large e}^{\!\!\phantom{0}
	^{\textstyle -\frac{1}{\hbar}
	\eta^{T}(\fatg{W}\!\eta)_{\,B}}}
	\left [\,\,1\!+\!O\,(\hbar^{1/2}\!)\,\,\right].
\label{eqn:8.2.1}
        \end{eqnarray}
Here $v_{\varphi}\,(t)$ is the set of linearized Lorentzian modes of
collective
variables $\varphi$ which has a matrix of inner products
(\ref{eqn:7.12})
	\begin{equation}
	F_{\varphi\,L}\,(d/dt)\,v_{\varphi}\,(t)
	=0,\;\;\;\;
	<v_{\varphi},v_{\varphi}>=\Delta_{\varphi},
\label{eqn:8.2.2}
	\end{equation}
$\eta^{T}(\fatg{W}\!\eta)_{\,B}$ is a doubled quadratic form of the
action
(\ref{eqn:8.4}) in the imaginary corrections to the classical
extremal at the
nucleation surface $\tau_{B}$ and
	\begin{equation}
	\fatg{\Gamma}_{\rm 1-loop}\,[\,\phi\,]=
	I_{\,2M}[\,\phi\,]+\frac{\hbar}{2}\;\,
	{\rm Tr}_{\phantom{0}_{\phantom{0}_
	{\!\!\!\!\!\!\!\!\!\!\!\!\!\!\!\!\!\!\!\!
	\textstyle\phantom{0}_{2M}}}}
	{\rm ln}\,\fatg{F}-\frac{\hbar}{2}\;\,
	{\rm Tr}_{\phantom{0}_{\phantom{0}_
	{\!\!\!\!\!\!\!\!\!\!\!\!\!\!\!\!\!\!\!\!
	\textstyle\phantom{0}_{2M}}}}
	{\rm ln}\,\fatg{a}
\label{eqn:8.2.3}
	\end{equation}
is the one-loop effective action of the theory on the doubled
Euclidean
spacetime ${\bf 2}\fatg{M}$ with the real background field $\phi\,
(\tau)$ -- the real part of the exact complex extremal
$\Phi\,(z)=\Phi\,(z|\varphi,\,t)$, parametrized by the boundary data
$(\varphi,\,t)$ which is the argument of the partition function
(\ref{eqn:8.2.1}):
	\begin{equation}
	\phi\,(\tau)={\rm Re}\,\Phi\,(\tau|\varphi,\,t),
	\;\;\;0\leq\tau\leq\tau_{B}.
\label{eqn:8.2.4}
	\end{equation}
The effective action (\ref{eqn:8.2.3}) includes the classical
Euclidean action
on ${\bf 2}\fatg{M}$
	\begin {equation}
	I_{\,2M}[\,\phi\,]=2\,I_{\,M}[\,\phi\,]
\label{eqn:8.2.5}
	\end{equation}
and the one-loop contribution given by the logarithm of
(\ref{eqn:8.16}).

{}From the viewpoint of practical applications, it would seem that
the new
algorithm (\ref{eqn:8.2.1}) does not have any advantages over the
original
expression (\ref{eqn:8.4}), for the replacement of ${\rm
det}\,\fatg{\Delta}$
by its representation (\ref{eqn:8.16}) with the determinant of higher
functional dimensionality actually complicates the calculations.
However, the
new form of $\rho\,(\varphi,\,t)$ has a very important property of
covariance,
because in contrast to ${\rm det}\,\fatg{\Delta}$, involving the
intrinsically
non-covariant ADM reduction to physical variables, selection of the
particular
time, construction of basis functions, etc., there exist very
powerful
manifestly covariant methods for the calculation of
$\fatg{\Gamma}_{\rm
1-loop}\,[\,\phi\,]$. Apart from this, the variety of methods for
differential
operators on compact spaces, which is just the case of ${\bf
2}\fatg{M}$, is
much richer than on manifolds with boundaries. Altogether this allows
one to
perform a covariant regularization and renormalization of the
generally
divergent partition function and obtain its nontrivial high-energy
behaviour,
which will be considered in the last section of this paper where we
shall
briefly dwell on these methods.

As is known, the requirement of manifest general covariance served as
one of
the basic motivations for the creation of the Euclidean quantum
gravity in the
works of Hawking, Hartle and the others. Even the Euclideanization,
as it is,
was essentially aimed at getting rid of any signs of inequivalence
between
different coordinates on spacetime manifold. However, the usual price
one pays
for manifest covariance is the loss of manifest unitarity. The root
of the
difficulty is that in the covariant quantization the sector of
physical fields
and the physical inner product are deeply hidden in the full space of
the
theory involving ghosts, zero and negative norm states, etc., and
usually it
requires very subtle methods to recover unitarity from the manifestly
covariant
formalism or, vice versa, to render the unitary theory a manifestly
covariant
form \cite{DW:Dynamical,Fr-V,FV:Bern,BF:Ann,BarvU}. A remarkable
feature of the
partition function (\ref{eqn:8.2.1}) is that, beeing formulated in
terms
physical degrees of freedom with a standard inner product, it
combines both of
the desired properties: the covariance of radiative corrections
accumulated in
the Euclidean effective action (\ref{eqn:8.2.3}) and its one-loop
unitarity
encoded in the preexponential factor. Let us first consider this
unitarity
property.

\subsection{Unitarity and partition function of gravitational
instantons}
\hspace{\parindent}
Basically the unitarity in application to the density of the
partition function
means the conservation in the Lorentzian time $t_{+}$ of the total
probabilty
	\begin{equation}
	\int d\varphi\,\rho\,(\varphi,t_{+}\!)
	={\rm Const}.
\label{eqn:8.2.6}
	\end{equation}
To prove this statement, we shall make a number of observations. To
begin with,
note that a set of basis functions $v_{\varphi}\,(t)$ featuring in
(\ref{eqn:8.2.1}) can be obtained from the family of classical
extremals
parametrized by certain variables (constants of integration of
classical
equations) by differentiating them with respect to these variables.
The
resulting functions satisfy the linearized equations of motion and
the same
regularity conditions as these classical extremals. In our case the
extremals
$\Phi\,(z|\varphi,\,t_{+}\!)$ are parametrized by their boundary
conditions
$\varphi$ at the final moment of Lorentzian time $t_{+}$, but for our
purposes
it will be more convenient to parametrize them by the value
$\phi_{B}$ of their
real part at the moment $\tau_{B}$ of the Lorentzian nucleation (we
assume that
$\phi_{B}$ and $\tau_{B}$ are in one-to-one correspondence with
$\varphi$ and
$t_{+}$):
	\begin{equation}
	\Phi\,(z)=\Phi\,(z,\phi_{B})=\phi\,(z,\phi_{B})
	+i\,\eta\,(z,\phi_{B}),\;\;\;\;
	\phi_{B}=\phi\,(\tau_{B},\phi_{B}).
\label{eqn:8.2.7}
	\end{equation}
Thus, the Lorentzian basis functions can be regarded as
	\begin{equation}
	v^*_{\varphi}\,(t)=\left.
	\frac{\partial\Phi\,(z,\phi_{B})}
	{\partial\phi_{B}}\;\right|_{\,z=\tau_{B}+it}.
\label{eqn:8.2.8}
	\end{equation}
Since $\Phi\,(z_{+},\phi_{B})=\varphi$, the matrix of the above basis
functions
is real at $t_{+}$, and its determinant coincides with the Jacobian
of
transformation from $\phi_{B}$ to $\varphi$
	\begin{eqnarray}
	\varphi\rightarrow\phi_{B},\;\;\;
	{\rm det}\,v_{\varphi}\,(t_{+}\!)=
	{\rm det}\,
	\left(\partial\varphi/
	\partial\phi_{B}\!\right).                 \label{eqn:8.2.9}
	\end{eqnarray}

The further proof is based on the method of complex extremals
according to
which their imaginary part $\varepsilon$ can be treated by
perturbations in all
the terms except the negative-definite quadratic form in the total
exponential
of (\ref{eqn:8.2.1}). When combined with the classical action
(\ref{eqn:8.2.5}), this form gives rise to the (doubled) imaginary
part of the
full complex action (\ref{eqn:5.16}), ${\cal S}\,[\,\Phi(z)\,]\equiv
i{\cal I}\,[\,\Phi(z)\,]=S\,(t_{+},\varphi)+iI(t_{+},\varphi)$
calculated at
the contour $C_{+}$
	\begin{eqnarray}
	I_{\,2M}[\,\phi\,]+
	\eta^{T}(\fatg{W}\!\eta)_{\,B}=
	2\,I(t_{+},\varphi)
	+O\,(\,\eta^{3}),
\label{eqn:8.2.10}
	\end{eqnarray}
which is a complex Hamilton-Jacobi function of the data
$(t_{+},\varphi)$ at
the end $z_{+}$ of the complex extremal, ${\cal
S}\,[\,\Phi(z)\,]={\cal
S}\,(t_{+},\varphi)$. With respect to this data it satisfies the
Hamilton-Jacobi equation with the physical Hamiltonian
$H\,(\varphi,p)$,
generating the following equation for its imaginary part
$I=I(t_{+},\varphi)$
	\begin{eqnarray}
	\frac{\partial I}{\partial t_{+}}+
	\left.\frac{\partial H}{\partial p}\,\right |
	_{p=\partial S/\partial\varphi}
	\frac{\partial I}{\partial\varphi}=
	O\left[\,\left(\frac{\partial I}
	{\partial\varphi}\right)^3\right].
\label{eqn:8.2.11a}
	\end{eqnarray}
In view of the Euclidean-Lorentzian matching conditions of Sect.5,
$\partial
I/\partial\varphi=O\,(\varepsilon)=O\,(\,\hbar^{1/2})$, so that the
solution of
(\ref{eqn:8.2.11a}), $I\,(t,\bar\varphi(t))=
I\,(0,\bar\varphi(0))+O\,(\,\hbar^{3/2})$, is practically a constant
\cite{Hal-Hartle} along the real-valued trajectory $\bar\varphi\,(t)$
evolving
according to
	\begin{eqnarray}
	\dot{\bar\varphi}=\left.\frac{\partial H\,(\bar\varphi,p)}
	{\partial p}\,\right |
	_{p=\partial S(t,\bar\varphi)/
	\partial\bar\varphi},\,\;
	\bar\varphi\,(t_{+})=\varphi,
\label{eqn:8.2.11b}
	\end{eqnarray}
and differing from the real part of the exact complex extremal at
most by
$O\,(\varepsilon)=O\,(\,\hbar^{1/2})$ terms, $\bar\varphi\,(t)=
\phi\,(\tau_{B}+it,\phi_{B})+O\,(\varepsilon)$.

Thus, the dependence on $t_{+}$ in (\ref{eqn:8.2.10}) can be, with
the needed
one-loop accuracy, completely absorbed into the redefinition of the
field
variable, $\varphi\rightarrow\varphi_{B}\equiv\bar\varphi\,(0)$.
Correspondingly, the tree-level part () can be regarded as the
Euclidean action
$I_{2M}[\,\bar\phi\,]$ on a new real classical background
$\bar\phi\,(\tau)$
with the boundary condition $\varphi_{B}$ at $\tau_{B}$, the latter
beeing
determined as a function of $(t_{+},\varphi)$ from the solution of
(\ref{eqn:8.2.11b}):
	\begin{eqnarray}
	\delta I\,[\,\bar\phi\,]/\delta\,\bar\phi\,(\tau)=0,\;\;
	\bar\phi\,(\tau_{B})=
	\varphi_{B}(t_{+},\varphi).
\label{eqn:8.2.11c}
	\end{eqnarray}
With the same accuracy the full one-loop exponential in
(\ref{eqn:8.2.1}) turns
out to be the effective action on this new background
$\bar\phi(\tau)$ or can
be regarded as a function of its boundary data
$\varphi_{B}=\phi_{B}+O\,(\,\hbar^{1/2})$ at the nucleation surface
	\begin{equation}
	\fatg{\Gamma}_{\rm 1-loop}\,[\,\phi\,]+
	\eta^{T}(\fatg{W}\!\eta)_{\,B}
	=\fatg{\Gamma}_{\rm 1-loop}\,[\,\bar\phi\,]
	=\fatg{\Gamma}_{\rm 1-loop}\,(\,\varphi_{B}).
\label{eqn:8.2.12}
	\end{equation}
On the other hand, by differentiating the asymptotic bound
(\ref{eqn:5.20})
with respect to $\phi_{B}$, one finds that not only
$\varepsilon\,(\tau)=O\,(\hbar^{1/2})$ but also $\partial\varepsilon/
\partial\phi_{B}=O\,(\hbar^{1/2})$, whence one can rewrite
(\ref{eqn:8.2.9}) as
a Jacobian of the following change of variables
	\begin{equation}
	\varphi\rightarrow\varphi_{B},\;\;\;
	[\,{\rm det}\,v_{\varphi}\,(t_{+}\!)\,]^{-1}=
	{\rm det}\left(\partial\varphi_{B}/
	\partial\varphi\right)+O\,(\hbar^{1/2}),
\label{eqn:8.2.14}
	\end{equation}
absorbing all the dependence of (\ref{eqn:8.2.1}) on the Lorentzian
time
$t_{+}$.

Therefore, the quantum distribution function for tunnelling
geometries
	\begin{equation}
	\rho\,(\varphi,t_{+})=\rho_{2M}\,(\varphi_{\!B})\,
	\left.{\rm det}\,\frac{\partial\varphi_{\!B}}
	{\!\partial\varphi}\;\right|
	_{\,\varphi_{B}=
	\varphi_{B}\,(\varphi,\,t_{+})},    \label{eqn:8.2.15}
	\end{equation}
reduces to the following partition function of the gravitational
instantons
${\bf 2}\fatg{M}$ with a special classical background field subject
to the
boundary conditions (\ref{eqn:8.2.11c}) at the junction surface
$\Sigma_{B}$:
	\begin{equation}
	\rho_{2M}\,(\varphi_{\!B})={\rm Const}\,
	[\,{\rm det}\,\Delta_{\varphi}\,(\varphi_{B}\!)\,]^{1/2}\,
	\mbox{\large e}^{\!\!\phantom{0}
	^{\textstyle -\frac{1}{\hbar}\fatg{\Gamma}_
	{\rm 1-loop}\,(\varphi_{B})}}
	\left[\,1+O\,(\hbar^{1/2})\;\right].
\label{eqn:8.2.16}
	\end{equation}
As a result one has
	\begin{eqnarray}
	\int d\varphi\,\rho\,(\varphi,t_{+}\!)
	=\int d\varphi_{B}\,\rho_{2M}(\varphi_{B})
	={\rm Const},
\label{eqn:8.2.17}
	\end{eqnarray}
which accomplishes the proof of unitarity for the distribution
function
\footnote
{Strictly speaking, both {\normalsize$\fatg{\Gamma}_{\rm
1-loop}\,(\varphi_{B})$} and {\normalsize${\rm
det}\,\Delta_{\varphi}\,(\varphi_{B})$} in (\ref{eqn:8.2.16}) depend
also on
the value of {\normalsize$\tau_{B}$} which is a function of $t_{+}$
(as a part
of boundary conditions {\normalsize$(\varphi,t_{+}\!)$} on the
complex
extremal). However, as it follows from the analyses of these
extremals
\cite{tunnelIII}, {\normalsize$
d\tau_{B}/dt_{+}=O\,(\varepsilon)=O\,(\hbar^{1/2}\!)$}, so that this
value is
defined up to higher-order loop corrections by the location of
caustic surfaces
of the Euclidean classical histories in superspace of three-metrics
and matter
fields {\normalsize$\fatg{q}$}. According to the discussion of
Sect.2, the
necessity of introducing the complex time for tunnelling geometries
originates
from extending the classical extremals of the theory beyond these
caustic
surfaces, the gauge properties of which will be considered in
\cite{tunnelIII}.
}.

The algorithms (\ref{eqn:8.2.1}) or
(\ref{eqn:8.2.15})-\ref{eqn:8.2.16}) have a
very good graphical illustration demonstrating their unitarity. The
partition
function, as an inner product of the wavefunction with itself, is
shown on
Fig.5 as a composition of the two spacetime manifolds combined of
Euclidean and
Lorentzian domains and associated respectively with
$\Psi_{\!L}\,(\varphi,f,t)$
and $\Psi_{\!L}^{*}\,(\varphi,f,t)$. Due to unitarity, which makes
sense only
in physical Lorentzian spacetime, the Lorentzian "brims" of these two
"hats"
cancel, because this portion of the spacetime is described by the
unitary
evolution operator. What remains is, in the main, the doubled
Euclidean
manifold ${\bf 2}\fatg{M}$ -- the compact gravitational instanton of
spherical
topology serving as a support for the Euclidean action
(\ref{eqn:8.2.3}). This
non-trivial remnant can be explained by the fact that the dynamical
"evolution"
on Euclidean spacetime is described by the non-unitary heat equation
rather
than the Schrodinger one.

The algorithms of analytic continuation from Euclidean spacetime
(Wick
rotation) for the matrix elements between different quantum states --
IN and
OUT asymptotic vacua -- are well known in asymptotically flat case
and usually
serve as a calculational basis for conventional scattering theory. A
similar
technique for expectation values, that is for matrix elements of
operators with
respect to one and the same state, is much less popular because of
the
difficulties related to a manifest breakdown of analyticity. In
contrast to the
analyticity of the wavefunction, the corresponding expectation values
can never
be analytic for they involve both $\Psi_{\!L}\,(q,t)$ and its complex
conjugate
$\Psi_{\!L}^*\,(q,t)$. However, in the context of a special quantum
state --
the standard asymptotic IN-vacuum associated with the plane-wave
decomposition
of field operators -- there exists a special technique relating the
expectation
values in Lorentzian spacetime to the Euclidean effective action
\cite{CPTI}.
Apparently, the algorithm (\ref{eqn:8.2.1}) is the first analogue of
this
technique in the cosmological case of spatially closed spacetime and
in the
context of the no-boundary quantum state of Hartle and Hawking.
Obviously, this
state plays the role of the standard IN-vacuum of asymptotically flat
worlds,
its regular Euclidean modes beeing the counterparts of the positive
energy
plane waves which under the Wick rotation go over into the modes
vanishing in
the remote Euclidean "past". The same analogy also transpires in the
realization of the reduction methods for functional determinants of
\cite{tunnelII} where the south and north poles of the compact
sphere-like
manifold were associated with the $\pm\infty$ of the asymptotically
flat
spacetime, while the corresponding regular modes $\fatg{u}_{\pm}$
were
associated with the basis functions of the IN and OUT asymptotic
vacua.

The above picture of unitarity on the Lorentzian portion of the
tunnelling
geometry takes the simplest form in the case of exactly real
classical
extremals, when the doubled manifold represents a completely smooth
gravitational instanton
\footnote
{
Although, in this case the linearized modes of the collective
variables get a
zero norm with respect to the Wronskian inner product and, thus,
acquire the
status of zero modes on the Euclidean instanton, the discussion of
which goes
beyond the scope of this paper.
}.
In the general case of theories with the Euclidean action bounded
from below,
the exact extremals are complex, and in the one-loop approximation
the effect
of their complexity boils down to the gaussian factor in
(\ref{eqn:8.2.1})
damping the contribution of their large imaginary part. Apparently,
this
property explains the lack of interest in literature to complex
instantons, the
contribution of which is always exponentially suppressed in
comparison with the
real tunnelling solutions. However, as it was discussed in
Introduction, there
are physically interesting problems lacking the real solutions, in
which cases
the technique of the above type becomes indispensible. One of the
fundamental
examples is the Hawking model of chaotic inflation driven by a
macroscopic
collective variable -- the inflaton scalar field. We shall consider
this model
in much detail in a forthcoming paper \cite{tunnelIII} and also use
it in the
next section to illustrate the last issue of the present work -- the
high
energy behaviour and normalizability of the Hartle-Hawking
wavefunction.

\section{High-energy behaviour of the Hartle-Hawking wavefunction}
\hspace{\parindent}
The validity of a semiclassical loop expansion in quantum gravity
essentially
depends on the energy scale of the problem. At Planckian energies it
is
expected to break down because of the nonrenormalizability of the
Einstein
theory, related to the dimensional nature of its coupling constant.
On the
other hand, the energy scale of the problem is determined by the
quantum state
of the system and the location of maxima of the corresponding
partition
function for those variables which are supposed to play the major
role in its
dynamics. Therefore, the validity of the loop expansion in quantum
gravity with
the Hartle-Hawking state can, at least heuristically, follow from the
behaviour
of the partition function of collective variables constructed above.

This partition function includes the quantum contribution of the
collective
variables themselves and also of the infinite set of microscopic
field modes.
Therefore, it suffers from the ultraviolet divergences and requires
regularization and renormalization. In principle, these procedures
must be done
at all stages of calculating the wavefunction and the corresponding
partition
function. Only in this case we would have the consistent and
consequitive {\it
operatorial} quantization. However, at the present state of art in
high-energy
physics, only in simple low dimensional field models such an approach
has been
realized and has a well-established status. In realistic field
theories we
still have to skip the operatorial stage of quantization at the
unregularized
level and make regularization and (if possible) renormalization only
in the
final algorithms for matrix elements, expectation values, etc.,
presented by
loop Feynman diagrams. For this reason we shall discuss here the
regularization
in the final algorithm for a partition function rather than in the
wavefunction
itself.

Even apart from this liberty, there still remains a problem of
whether the
properly regularized infinities can be renormalized by physically
sensible
procedure. We shall not discuss here this issue, which is a subject
of a vast
literature on the over-Planckian structure of fundamental
interactions.
Instead, we shall simply assume that, whatever physical origin of
this
procedure is (either it is a fundamental finite string theory
underlying its
low-energy effective limit or the inclusion of the infinite set of
counterterms), the correct procedure of renormalization consists in
the
subtraction of the covariantly regularized ultraviolet infinities. It
is
difficult to perform a regularization, respecting the four
dimensional general
covariance, in the manifestly non-covariant formalism of the unitary
ADM
quantization. However, as it was shown in the previous section, our
partition
function combines manifest unitarity with the Euclidean effective
action
accumulating the divergent quantum corrections, which can be rendered
covariant
form and, therefore, covariantly renormalized. Here we shall briefly
sketch
this procedure which eventually yields the high-energy behaviour of
the
partition function.

\subsection{Covariant renormalization and anomalous scaling}
\hspace{\parindent}
The basic step of converting the effective action (\ref{eqn:8.2.3}),
calculated
in terms of reduced (Euclidean) physical variables $\phi$, into a
covariant
form consists in its transformation to the original set of
gravitational
four-metric and matter fields $\fatg{g}=(\fatg{\phi},N_{E})$
($\fatg{\phi}$ is
a set of 3-metric coefficients and matter fields in spatial foliation
of the
Euclidean spacetime, and $N_{E}$ are the corresponding lapse and
shift
functions) taken in some gauge which has a form of local conditions
on
$\fatg{g}$ and its spacetime derivatives
	\begin{equation}
	\chi\,(\fatg{g},\dot{\fatg{g}})=0
\label{eqn:9.1}
	\end{equation}
(remember that, in the condensed canonical notations of our paper,
only time
derivatives with respect to a selected foliation of spacetime are
explicitly
written, while the spatial derivatives are encoded in the contraction
of
condensed indices). Such a transformation is identical for the
classical part
of the effective action $I\,[\,\phi\,]=\fatg{I}\,[\,\fatg{g}\,]$,
where the
boldfaced notation is used for the classical action in the initial
variables
(cf. Sect.2). For its one-loop part it is given by the one-loop
approximated
Faddeev-Popov ansatz
	\begin{eqnarray}
	\frac{\hbar}{2}\;
	{\rm Tr}\,{\rm ln}\,\fatg{F}-
	{\rm Tr}\,{\rm ln}\,\fatg{a}=
	\frac{\hbar}{2}\;
	{\rm Tr}\,{\rm ln}\,\fatg{\cal F}-
	\hbar\;{\rm Tr}\,{\rm ln}\,\fatg{\cal Q}+
	O\,(\delta\fatg{I}/\delta\fatg{g}).
\label{eqn:9.2}
	\end{eqnarray}
It involves the wave operator of the full set of fields $\fatg{g}$,
determined
by the total action $\fatg{I}_{\rm tot}[\,\fatg{g}\,]$ which includes
the gauge
breaking term with the gauge of the above type
	\begin{eqnarray}
	\fatg{\cal F}=\frac{\,\,\delta^2\!\fatg{I}_{\rm tot}}
	{\delta\fatg{g}\,\,\delta\fatg{g}},\;\;\;\;\;
	\fatg{I}_{\rm tot}[\,\fatg{g}\,]=
	\fatg{I}\,[\,\fatg{g}\,]+\!\int d\tau\,
	\chi^2\,(\fatg{g},\dot{\fatg{g}})
\label{eqn:9.3}
	\end{eqnarray}
and the corresponding ghost operator $\fatg{\cal Q}$ whose action on
a small
test function $f$ is determined by the infinitesimal coordinate gauge
transformation of gauge conditions (\ref{eqn:9.1}) generated by this
function,
$\fatg{\cal Q}\,f=\Delta^{f}\chi$
\footnote
{We disregard in the right hand side of (\ref{eqn:9.2}) the
contribution of the
local measure (the counterpart to {\normalsize{${\rm Tr}\,{\rm
ln}\,\fatg{a}$}}
in the left hand side) which is ultralocal and proportional to
{\normalsize
{$\delta^{4}(0)$}}. In the covariant regularizations disregarding the
strongest
volume divergences (which are cancelled by this measure
\cite{FV:Bern,tunnelII}) it is either identically vanishing
(dimensional
regularization) or reduces to the logarithmic renormalization
ambiguity
($\zeta$-functional regularization).
}.

On mass shell, that is on the solution of classical equations
$\delta\fatg{I}/\delta\fatg{g}=0$ for the background $\fatg{g}$,
which is just
the case of eqs.(\ref{eqn:2.18}) - (\ref{eqn:2.19}) and
(\ref{eqn:3.19}) -
(\ref{eqn:3.20}), the expression (\ref{eqn:9.2}) is gauge independent
\cite{BV:PhysRep} and exactly generates the one-loop effective action
in
physical variables
\footnote
{This is a particular case of the general statement on the gauge
independence
of the path integral in quantum gravity and local gauge theories,
which is
valid on shell beyond the one-loop approximation at least within the
perturbative loop expansion
\cite{DW:Dynamical,Faddeev,Fr-V,DW:LesH}.}.
This freedom in the choice of $\chi\,(\fatg{g},\dot{\fatg{g}})$
allows to
choose them as belonging to the class of the background covariant
gauge
conditions \cite{DW:Dynamical,DW:LesH,BV:PhysRep}, in which all the
traces of
the noncovariant (3+1)-splitting of the spacetime completely
disappear and
$\fatg{\cal F}$ and $\fatg{\cal Q}$ become local covariant
differential
operators of the second order. The functional determinants of such
operators
already admit the covariant regularization and have powerful
calculational
methods for their asymptotic scaling behaviour.

These methods, in the main, reduce to the combination of the
Schwinger-DeWitt
proper time technique \cite{DW:Dynamical,DW:LesH,BV:PhysRep} and the
dimensional or $\zeta$-functional regularization. The latter has a
technical
advantage that it automatically subtracts the divergences, that is
why we
choose it here to demonstrate the covariant renormalization of the
effective
action. In the $\zeta$-functional regularization \cite{zeta} one has
for a
second-order covariant differential operator $\fatg{\cal F}$
	\begin{equation}
	{\rm Tr}\,{\rm ln}\,\fatg{\cal F}
	=-\zeta^{\prime}(0)-\zeta\,(0)\;{\rm ln}\,\mu^{2},
\label{eqn:9.4}
	\end {equation}
where the $\zeta$-function $\zeta\,(p),\,p\rightarrow 0,$ is defined
by the
analytic continuation in $p$ from the domain of convergence of the
following
functional trace (or, equivalently, the sum over the eigenvalues
$\lambda$ of
$\fatg{\cal F}$)
	\begin{equation}
	\zeta\,(p)={\rm Tr}\,\fatg{\cal F}^{-p}
	=\sum_{\lambda}\lambda^{-p}
\label{eqn:9.5}
	\end{equation}
and $\mu^{2}$ is the mass parameter reflecting the logarithmic
renormalization
ambiguity in (\ref{eqn:9.4}). The coefficient of ${\rm ln}\,\mu^{2}$
in
(\ref{eqn:9.4}) as a local functional of the background fields
$\fatg{g}$
	\begin{equation}
	\zeta\,(0)=\frac{1}{(4\pi)^2}\,
	A_{2}[\,\fatg{g}\,]
\label{eqn:9.6}
	\end{equation}
determines the structure of (subtracted) ultraviolet divergences and
is defined
by the coefficient $A_{2}$ of the proper-time expansion for the
functional
trace of the heat kernel of the operator  $\fatg{\cal F}=\fatg{\cal
F}
[\,\fatg{g}\,]$ \cite{DW:Dynamical,Gilkey,DW:LesH}
	\begin{equation}
	{\rm Tr}\,{\rm e}^{-{\textstyle s}
	\fatg{\cal F}[\,\fatg{g}\,]}=
	\frac{1}{(4\pi s)^{\,2}}\,
	\sum_{n=0}^{\infty}A_{n/2}
	[\,\fatg{g}\,]\,s^{n/2},\;\;\;\;
	s\rightarrow 0.
\label{eqn:9.7}
	\end{equation}
These coefficients are determined by volume and surface integrals of
local
invariants over the spacetime manifold $\fatg{M}$ and its boundary
$\fatg{\partial M}$. Such invariants, in their turn, are constructed
out of the
coefficients of the operator $\fatg{\cal F}$, spacetime curvature and
the
extrinsic curvature of $\fatg{\partial}\fatg{M}$ and, therefore, are
easily
calculable for any spacetime and field background $\fatg{g}$.

The renormalization parameter $\mu^{2}$ introduces into the theory a
new scale.
The classical action has in the limit of small distances the
asymptotic scaling
invariance under the global conformal transformations of the 4-metric
and
matter fields $\fatg{g}=(g_{\mu\nu}(x),\phi\,(x))$
	\begin{eqnarray}
	&&\bar g_{\mu\nu}(x)=\Omega^{2}g_{\mu\nu}(x),\;\;\;
	\bar\phi\,(x)=\Omega^{c_{\phi}}\,\phi\,(x),
\label{eqn:9.8}\\
	&&\fatg{I}\,[\,\bar{\fatg{g}}\,]\simeq
	\fatg{I}\,[\,\fatg{g}\,],\;\;\;
	\Omega\rightarrow 0,
\label{eqn:9.9}
	\end{eqnarray}
where $c_{\phi}$ represents the set of conformal weights of fields
$\phi\,(x)$
\footnote
{This is a general case of the nonminimal coupling between the scalar
curvature
and the dilaton (or, in context of the iflationary Universe,
inflaton) scalar
field $\phi$ of the conformal weight $-1$, which generates the
effective
gravitational constant $k^2\sim 1/\phi^{2},\;\phi\rightarrow\infty$.
For the
minimal case different kinetic terms of the classical action scale
differently,
but this does not change essentially the scaling properties of
quantum
corrections considered below.
}.
On the contrary, quantum corrections are asymptotically scale
invariant only
under the simultaneous rescaling of fields and the renormalization
mass
parameter $\mu^{2}\rightarrow\bar\mu^{2}=\Omega^{-2}\mu^{2}$, and,
therefore,
they have the anomalous scaling behaviour defined by the coefficient
of ${\rm
ln}\,\mu^{2}$ in (\ref{eqn:9.4})
	\begin{equation}
	\int d^{4}x\,\left(2\bar g_{\mu\nu}\,
	\frac{\delta}{\delta \bar g_{\mu\nu}}+
	\sum_{\phi} c_{\phi}\,\bar\phi\frac{\delta}
	{\delta \bar\phi}\right)\,
	\frac{1}{2}\,{\rm Tr}\,{\rm ln}\,
	\fatg{F}[\,\bar{\fatg{g}}\,]\simeq
	-\zeta\,(0),\;\;\;\Omega\rightarrow 0.
\label{eqn:9.10}
	\end{equation}
Thus, in this limit the scaling behaviour of the full effective
action
	\begin{equation}
	\fatg{\Gamma}_{\rm 1-loop}\,
	[\,\bar{\fatg{g}}\,]\simeq
	-\zeta^{\rm tot}(0)\;{\rm ln}\,\Omega+
	\fatg{\Gamma}_{\rm 1-loop}\,[\,\fatg{g}\,],\;\;\;
	\Omega\rightarrow 0
\label{eqn:9.11}
	\end{equation}
is determined by the total $\zeta$-function including the
contributions of both
the gauge field $\fatg{\cal F}$ and ghost $\fatg{\cal Q}$ operators
in
(\ref{eqn:9.2})
\footnote
{For theories invariant under local Weyl transformations
(\ref{eqn:9.8}) with
local parameter $\Omega=\Omega\,(x)$ the relations (\ref{eqn:9.9})
and
(\ref{eqn:9.10}) - (\ref{eqn:9.11}) hold exactly, and the integrand
on the
left-hand side of (\ref{eqn:9.10}) represents the well-known
conformal anomaly
given by the volume density of $A_{2}$ -- the coincidence limit of
the
two-point DeWitt coefficient $a_{2}\,(x)=a_{2}\,(x,x)$
\cite{DW:Dynamical,DW:LesH}.}.

\subsection{Anomalous scaling on the gravitational instanton and the
normalizability of the Hartle-Hawking wavefunction}
\hspace{\parindent}
Application of the last equation with due regard for the algorithm
(\ref{eqn:8.2.1}) is of crucial importance in the model of the
quantum birth of
the chaotic inflationary Universe. This model within a wide class of
the field
Lagrangians will be considered in \cite{tunnelIII}, where we shall
demonstrate
all the peculiarities of the general theory of the above type.
However, there
is a very important issue raised at the beginning of Sect.8 which can
be
resolved on the ground of equations (\ref{eqn:8.2.1}) and
(\ref{eqn:9.11}) in
the universal and model-independent way. This is the question of the
high-energy behaviour of the partition function which serves as a
criterion for
the applicability of the WKB approximation and normalizability of the
wavefunctions at over-Planckian scales.

As is well-known \cite{H-Page,Vilenkin:tun-HH} the tree-level
Hartle-Hawking
wavefunction is not normalizable in this model. As a function of the
only
collective variable, the inflaton scalar field $\varphi$, it goes to
a constant
for $\varphi\rightarrow\infty$ and does not suppress the contribution
of the
over-Planckian energy scales. This can be clearly seen from the
algorithm
(\ref{eqn:8.2.15}) with the tree-level partition function of
gravitational
instantons
	\begin{equation}
	\rho_{\rm tree}\,(\varphi_{B})={\rm Const}\,
	\mbox{\large e}^{\!\!\phantom{0}
	^{\textstyle -\frac{1}{\hbar}
	I_{2M}\,(\varphi_{B}\!)}}.              \label{eqn:9.12}
	\end{equation}
Here the classical Euclidean action on the doubled manifold
(\ref{eqn:8.2.5})
is parametrized as in (\ref{eqn:8.2.12}) in terms of the boundary
conditions
for an auxiliary classical field (\ref{eqn:8.2.11}) at the nucleation
surface
$\Sigma_{B}$ and has the following asymptotic behaviour at large
$\varphi_{\!B}$
	\begin{equation}
	I_{2M}\,(\varphi_{\!B})=I_{0}+I_{1}/\varphi_{\!B}^{2}+
	O\,(\,1/\varphi_{\!B}^{4}),\;\;\;\;
	\varphi_{\!B}\rightarrow\infty.
\label{eqn:9.13}
	\end{equation}

This behaviour follows from the fact that in this model the Euclidean
segment
of a complex classical history $\phi\,(\tau)=\Phi(\tau|\varphi,t)$,
$0\leq\tau\leq\tau_{B}$, giving the dominant contribution to the
partition
function, has in the large-$\phi$ limit a simple form of a
practically constant
and real scalar field coinciding with its value at the nucleation
point
	\begin{equation}
	\phi\,(\tau)\simeq{\rm Const}=
	\phi_{B}(\varphi,t),\;\;\;
	\phi_{B}(\varphi,t)\simeq\varphi_{B}.
\label{eqn:9.14}
	\end{equation}
The value $\phi_{B}(\varphi,t)$ is parametrized in accordance with
the form of
Lorentzian extremal by its final point $(\varphi,t)$ and always
satisfies the
inequality $\phi_{B}>\varphi$ which follows from the fact that the
scalar field
slowly decreases with the growth of $t$ during the Lorentzian
inflationary
stage. Therefore the limit $\varphi\rightarrow\infty$ guarantees
large values
of $\varphi_{B}\rightarrow\infty$. The corresponding constant scalar
field
(\ref{eqn:9.14}) generates an effective cosmological constant
$\Lambda=3H^{2}\,(\varphi_{B})$. Its dependence on $\varphi_{B}$ is
determined
by the form of the Lagrangian of the system, but for all Lagrangians
viable
from the viewpoint of the inflationary scenario it has the property
of the
monotonic growth $H(\varphi_{B})\rightarrow\infty$ for
$\varphi_{B}\rightarrow\infty$. The corresponding regular (in the
no-boundary
sense) solution of the Euclidean Einstein equations is the metric of
the
Euclidean DeSitter space (\ref{eqn:1.3}) - (\ref{eqn:1.4}) -- the
four-dimensional hemisphere of radius $R=1/H$, which generates the
Lorentzian
DeSitter Universe by the nucleation at $\tau_{B}=\pi/2H$. The
analysis of the
matching conditions at the nucleation point \cite{tunnelIII} shows
that both
the deviation of the full Euclidean-Lorentzian extremal from the
exactly
DeSitter form and its imaginary corrections are vanishing for large
$H$ and,
therefore, in this high-energy limit the doubled manifold ${\bf
2}\fatg{M}$ is
given by a gravitational instanton -- a four-dimensional sphere
$\fatg{S}^4$
of vanishing radius $R$, carrying the 4-geometry (\ref{eqn:1.3}) -
(\ref{eqn:1.4}) and constant scalar field (\ref{eqn:9.14}) which we
shall
denote by $\fatg{g}_{R}= (g_{\mu\nu}^{DS},\,\varphi_{\!B})$
	\begin{equation}
	{\bf 2}\fatg{M}=\fatg{S}^4,\;\;\;
	\fatg{g}_{R}=(g_{\mu\nu}^{DS},\,\varphi_{\!B}),\;\;\;
	R=1/H(\varphi_{\!B})\rightarrow 0.
\label{eqn:9.15}
	\end{equation}

The classical Euclidean action calculated on this instanton has a
form
(\ref{eqn:9.13}) with the coefficients depending on the model of the
Lagrangian
for coupled gravitational and inflaton scalar field \cite{tunnelIII}
\footnote
{For example, in models with the minimal interaction of a scalar
field the
expansion (\ref{eqn:9.13}) starts only with the subleading term,
while for an
inflaton field non-minimally coupled to a scalar curvature it has
nonzero
$I_{0}$ \cite{tunnelIII}.
}.
Therefore, the tree-level partition function (\ref{eqn:9.12}) of the
DeSitter
gravitational instantons weighted by their action, does not suppress
the
contribution of over-Planckian energy scales
$\phi_{\!B}\rightarrow\infty$ and
yields unnormalizable Hartle-Hawking wavefunction. This basically
means the
inconsistency of the semiclassical approximation.

The situation drastically changes in the one-loop approximation, when
the
classical action must be replaced by the effective one
(\ref{eqn:8.2.3}). Since
the scale of the DeSitter instanton is determined by the only
dimensional
quantity $H(\varphi_{\!B})$, the corresponding asymptotic behaviour
of
$\fatg{\Gamma}_{\rm 1-loop}$ follows from the equation
(\ref{eqn:9.11}) with
the parameter $\Omega$ replaced by the dimensionless ratio
$\Omega=\mu^{2}/H^{2}(\varphi_{\!B})$
	\begin{equation}
	\fatg{\Gamma}_{\rm 1-loop}\,
	[\,\fatg{g}_{R}]\simeq
	\fatg{Z}\;{\rm ln}\,
	\frac{H^{\!2}(\!\varphi_{\!B}\!)}{\mu^{2}},\;\;\;
	H(\varphi_{\!B})\rightarrow\infty,
\label{eqn:9.16}
	\end{equation}
where $\fatg{Z}$ is a total anomalous scaling (\ref{eqn:9.6}) on the
DeSitter
instanton of vanishing size with the background metric and inflaton
fields
(\ref{eqn:9.15})
	\begin{equation}
	\fatg{Z}=\left.\frac{1}{(4\pi)^{\,2}}\,
	A_{2}^{\rm tot}[\,\fatg{g}_{R}\,]\;\right|_
	{\,H\,(\varphi_{B})\rightarrow\infty}.
\label{eqn:9.17}
	\end{equation}
Thus, the partition function $\rho\,(\varphi_{\!B})$
(\ref{eqn:8.2.16}) of the
DeSitter gravitational instantons with the effective Hubble constant
$H(\varphi_{\!B})$ has the following high-energy behaviour
	\begin{equation}
	\rho\,(\varphi_{\!B})\simeq{\rm Const}\,
	[\,H(\varphi_{\!B}\!)\,]^{-\fatg{Z}-1},\;\;\;
	H(\varphi_{B})\rightarrow\infty,
\label{eqn:9.19}
	\end{equation}
where one extra negative power of $H(\varphi_{\!B})$ comes from the
Wronskian
normalization coefficient $(\Delta_{\varphi})^{1/2}$ for the
Lorentzian
inflaton mode $v_{\varphi}\,(t)=[\partial\phi_{\!B}\,(\varphi,t)/
\partial\varphi]^{-1}$ \cite{tunnelIII}.

Therefore, depending on the value of the fundamental dimensionless
quantity
(\ref{eqn:9.17}), this partition function either suppresses the
contribution of
the over-Planckian energy scales or infinitely enhances it and, thus,
serves as
a criterion of applicability of the semiclassical expansion. In
particular, in
generic theories with nonminimally coupled inflaton scalar field
having quartic
selfinteraction, for which $H(\varphi_{\!B}\!)\sim\varphi_{\!B}$,
this
partition function implies the high-energy normalizability of the
no-boundary
wavefunction
	\begin{equation}
	\int^{\infty} d\varphi_{\!B}\,
	\rho\,(\varphi_{\!B})<\infty,               \label{eqn:9.20}
	\end{equation}
provided the anomalous scaling exponent satisfies the condition
\cite{BKam:norm}
	\begin{equation}
	\fatg{Z}>-1.
\label{eqn:9.21}
	\end{equation}
The value of $\fatg{Z}$ is determined from (\ref{eqn:9.17}) by the
complete
matter content of the Universe \cite{Allen1,Fr-Ts} and, thus, the
above
condition can serve for a selection of physically viable particle
models which
have a physically consistent normalizable quantum state of the
Universe
justifying the use of a semiclassical $\hbar$-expansion.

These conclusions are, certainly, restricted to the one-loop
approximation
which was multiply used for the calculation of the path integral, for
the
justification of the perturbation theory in the imaginary part of the
complex
extremals and for the perturbation expansion in microscopic
variables. But
general principles of unitarity and covariance, which we have just
verified by
direct calculations in this approximation, allow us to conjecture
that beyond
one loop, and even non-perturbatively, the basic algorithm
(\ref{eqn:8.2.15})
-- (\ref{eqn:8.2.16}) will still be valid with $\fatg{\Gamma}_{\rm
1-loop}$
replaced by the full effective action $\fatg{\Gamma}$ calculated at
the
instanton solution of the exact effective equations. Therefore, the
normalizability (that is, the quantum consistency) condition
(\ref{eqn:9.21})
will still hold with the {\it exact nonperturbative} anomalous
scaling
$\fatg{Z}$ replacing its simple one-loop counterpart
(\ref{eqn:9.17}).

As concerns the one-loop approximation, which definitely gives the
possibility
to improve the predictions of the intrinsically inconsistent
tree-level theory,
in the following paper of this series \cite{tunnelIII} we shall apply
the
presented technique to a generic model of the nonminimally coupled
inflaton
scalar field. In particular, we shall analyze the possible extrema of
the
obtained partition function which might be responsible for the most
probable
inflationary universes tunnelling according to the
Hartle-Hawking-Vilenkin
proposal.

\section{Discussion}
\hspace{\parindent}
Thus we have considered some elements of the general theory for
tunnelling
geometries in the no-boundary quantum state of Hartle and Hawking. We
have
shown that within the $\hbar$-expansion this theory can be extended
to complex
tunnelling solutions which are exponentially suppressed as compared
to real
ones, but can contribute to interesting physical phenomena,
especially, in
problems lacking real solutions matching the classically forbidden
Euclidean
regime with the Lorentzian one. The nucleation of the latter from the
former
can be described perturbatively entirely in terms of the Euclidean
and
Lorentzian spacetimes with real metrics and matter fields. In the
calculation
of the quantum distribution function for the collective physical
variables,
they naturally lead to the notion of the real gravitational instanton
--
topologically closed solution of covariant Euclidean equations of
motion.
Despite the underlying complexity, this distribution function
features
unitarity and, thus, demonstrates a subtle interplay between
unitarity,
analyticity and covariance, encoded  in a partition function of
gravitational
instantons weighted by their Euclidean effective action.

The price one pays for such a real-field description of complex
tunnelling
phenomena is that the resulting real fields are not completely
smooth: they
suffer a jump of the first-order derivatives at the nucleation
surface
$\Sigma_{B}$ and, consequently, generate a sharp edge at the junction
"equatorial" surface between $\fatg{M}_{-}$ and $\fatg{M}_{+}$ in the
geometry
of the instanton ${\bf 2}\fatg{M}$ (see Fig.4 and Fig.5). This means,
that one
cannot directly generalize, by using these reality properties, the
conclusions
of Gibbons and Hartle in \cite{Gibbons-Hartle} and, in particular,
their unique
conception theorem about the nucleation of the topologically
connected
Lorentzian spacetime. The same property implies also that, from the
viewpoint
of physical implications, the framework of problems on the signature
change in
general relativity \cite{Sakharov,signature} should be extended to
account for
the subtleties of the above analytic continuation. This framework
might include
the consideration of gravitational instantons even with the
discontinuous
metric and matter fields, which would naturally arise if one would
consider the
non-diagonal elements of the density matrix
$\rho\,(\varphi,\varphi^{\prime}|\,t)$ instead of the probability
distribution
$\rho\,(\varphi,\,t)=\rho\,(\varphi,\varphi\,|\,t)$.

The technique of this paper heavily relies on the boundedness from
below of the
Euclidean gravitational action of physical variables. In quantum
cosmology of
spatially closed worlds this presents an immeadiate difficulty in the
conformal
sector which gives a negative-definite contribution and, in contrast
to
asymptotically-flat spacetimes \cite{Schoen-Y,Witten} seems to be not
ruled out
from the sector of physical degrees of freedom. If there is no way to
consistently declare the conformal mode the unphysical one, one is
left with
the only option briefly discussed in Sect.5 -- to shift this variable
into a
complex plane both in {\it the integration contour} of the path
integral
\cite{Gibbons-H-P} and in {\it the argument} of the wavefunction or
its quantum
distribution $\rho\,(\varphi,\,t)$. As discussed in \cite{tunnelIII},
a first
step of this program might consist in isolating this mode
$\varphi^{\rm conf}$
into the sector of collective variables and modifying the
corresponding
perturbation theory in the the imaginary part of the classical
extremal for
$\varphi^{\rm conf}$. As is also shown in \cite{tunnelIII} this
difficulty is
not very serious in the analysis of the over-Planckian,
$H\,(\varphi^{\rm
conf})\rightarrow\infty$, behaviour of $\rho_{2M}\,(\varphi^{\rm
conf})$,
because in this limit ${\rm Im}\,\varphi^{\rm conf}=O\,
(1/H(\varphi^{\rm conf}))$, and one can use the usual perturbation
theory in
${\rm Im}\,\varphi^{\rm conf}$ unrelated to the asymptotic bound
(\ref{eqn:5.20}). This might be explained by the fact that the
possible
suppression of the over-Planckian energy scales in
$\rho_{2M}\,(\varphi^{\rm
conf})$ originates not from its good gaussian nature, but from the
anomalous
scaling of $\fatg{\Gamma}_{\rm 1-loop}(\varphi^{\rm conf})$, the
calculation of
which implies merely the conformal rotation in the Euclidean path
integral.

In \cite{tunnelIII} we shall consider in much detail the application
of the
general theory developed here and the peculiarities of the conformal
mode for
the model of chaotic inflation with the nonminimal selfinteracting
inflaton
field. This model, in which the inflaton field coincides with the
Brans-Dicke
scalar, plays an important role in the theory of the early Universe,
for it
provides a very efficient resolution \cite{SalopBB} of the known
difficulties
in the formation of the observable cosmological large-scale structure
\cite{Salopek}. Similarly to the discussion of the previous section,
the
analysis of this model is most simple for large $H(\varphi_{\!B})$,
because in
this limit a theory has a trivial behaviour of the caustic surfaces
in
superspace, responsible for the nucleation of the Lorentzian Universe
from the
Euclidean spacetime. On the contrary, the situation becomes extremely
complicated for small effective $H(\varphi_{\!B})$, when the
inflationary
ansatz is no longer efficient and the caustic surface disappears
below some
critical value of $H(\varphi_{\!B})$ \cite{GrisR}. On the other hand,
this
infrared domain plays an important role in the theory of wormholes,
baby
universes and the cosmological constant \cite{Gid-S,Coleman,Hawk-w}
-- the
first step towards the third quantization of gravity. For this
reason, we shall
also have to consider in \cite{tunnelIII} some general properties of
caustic
surfaces for solutions of Einstein equations and, in particular, show
their
relevance to the problem of Gribov copies in the manifestly unitary
quantization of gravity. According to the discussion in
\cite{BarvU,BKr,Operd},
the Gribov problem serves as a strong motivation for the third
quantization of
gravity, because due to this problem the secondary quantized gravity
seems to
be intrinsically inconsistent (just as a first quantized interacting
relativistic particle is inconsistent outside of the secondary
quantization
framework). This gives a hope that such an analysis of superspace
caustic
surfaces in gravity theory might pave a constructive perturbative
approach to
its third quantization.\\

\section*{Acknowledgements}
\hspace{\parindent}
The authors benefitted from helpful discussions with G.Gibbons,
J.B.Hartle, Don
N.Page and R.M.Wald and are also grateful to A.Umnikov for his help
in the
preparation of the pictures for this paper. One of the authors
(A.O.B.) is
grateful for the support of this work provided by CITA National
Fellowship and
NSERC grant at the University of Alberta.

\newpage
\section*{\centerline{Figure captions}}
{\bf Fig.1} Graphical representation of the Lorentzian spacetime
$\fatg{L}$
nucleating  at the bounce surface $\Sigma_{B}$ from the Euclidean
manifold
$\fatg{E}$ of the no-boundary type, having the topology of the
four-dimensional
ball.
\\
\\
{\bf Fig.2} Euclidean spacetime of the no-boundary type originating
from the
tube-like manifold $\Sigma\times[\,\tau_{-},\tau_{+}]$ by shrinking
one of its
boundaries $\Sigma_{-}$ to a point. It inherits the foliation with
slices of
constant Euclidean time $\tau$ in the form of quasi-spherical
surfaces of
"radius" $\tau$ with the center at $\tau_{-}=0$.
\\
\\
{\bf Fig.3} The contour $C_{+}=C_{E}\cup C_{L}$ of integration over
complex
time in the action, corresponding to the splitting of the whole
spacetime into
the combination of Euclidean ($C_{E}$) and Lorentzian ($C_{L}$)
domains matched
at the nucleation (bounce) point $\tau_{B}\;(t=0)$.
\\
\\
{\bf Fig.4} The doubling of the Euclidean manifold, which arises in
the
calculation of the quantum distribution function for Lorentzian
universes. The
Euclidean spacetime of the no-boundary type $\fatg{M}_{-}$ matched
across the
nucleation surface $\Sigma_{B}$ with its orientation reversed copy
$\fatg{M}_{+}$ gives rise to a closed manifold ${\bf
2}\fatg{M}=\fatg{M}_{-}\cup\fatg{M}_{+}$ -- the gravitational
instanton of
spherical topology. For generic complex tunnelling geometries the
matching of
$\fatg{M}_{-}$ with $\fatg{M}_{+}$ is not smooth, which is shown on
the picture
by the edge at $\Sigma_{B}$.
\\
\\
{\bf Fig.5} The graphical representation of calculating the quantum
distribution of tunnelling Lorentzian universes: a composition of the
combined
Euclidean-Lorentzian spacetime $\fatg{M}_{-}\cup\fatg{L}$ with its
orientation
reversed and complex conjugated copy $\fatg{M}_{+}\cup\fatg{L^*}$
results in
the doubled Euclidean manifold ${\bf 2}\fatg{M}$ -- the gravitational
instanton
carrying the Euclidean effective action of the theory. The
cancellation of the
Lorentzian domains $\fatg{L}$ and $\fatg{L^*}$ reflects the unitarity
of the
theory in the physical spacetime of Lorentzian signature.

\end{document}